\documentclass[12pt]{article}
\pdfoutput=1 
\usepackage[nosort]{cite}
\usepackage[height=8.85in,width=6.45in]{geometry}
\usepackage{pifont}

\usepackage{times}

\usepackage{booktabs}
\usepackage[utf8]{inputenc}
\usepackage{slashed}
\usepackage{braket}
\usepackage[svgnames,psnames]{xcolor}
\usepackage{url}
\usepackage[colorlinks,citecolor=DarkGreen,linkcolor=FireBrick,linktocpage,pagebackref]{hyperref}
\usepackage{multirow}
\usepackage{courier}
\usepackage{bm}

\usepackage{dashbox}
\usepackage{caption}
\usepackage{subcaption}
\usepackage{enumitem}

\usepackage{blkarray}
\usepackage{arydshln}

\usepackage{calc}

\usepackage{amssymb}
\usepackage{dsfont}
\usepackage[T1]{fontenc} 
\usepackage{physics}
\usepackage{comment}
\usepackage{amsmath}
\usepackage{mathtools}
\usepackage{graphicx}
\usepackage{array, makecell}
\usepackage{float}
\setcellgapes{3pt}
\usepackage[normalem]{ulem}
\usepackage{pdflscape}
\usepackage[export]{adjustbox}

\renewcommand{\mod}{\operatorname{mod}}

\definecolor{azure}{rgb}{0.0, 0.5, 1.0}
\definecolor{darkblue}{rgb}{0.15,0.35,0.7}
\definecolor{reddish}{rgb}{0.65, 0.2, 0.2}
\definecolor{brandeisblue}{rgb}{0.0, 0.44, 1.0}
\definecolor{ceruleanblue}{rgb}{0.16, 0.32, 0.75}
\definecolor{indigo(dye)}{rgb}{0.0, 0.25, 0.42}
\definecolor{grey}{rgb}{0.9,0.9,0.9}
\definecolor{dgrey}{rgb}{0.3,0.3,0.3}

\usepackage{cleveref}
\usepackage{bm}
\crefname{lem}{lemma}{lemmas}
\crefname{thm}{theorem}{theorems}
\crefname{cor}{corollary}{corollaries}
\crefname{rem}{remark}{remarks}
\crefname{prop}{proposition}{propositions}

\usepackage{colortbl}
\definecolor{dgreen}{rgb}{0, 0.55, 0}
\definecolor{llightyellow}{rgb}{1.0, 0.95, 0.7}
\definecolor{llightblue}{rgb}{0.7, 0.9, 1.0}
\definecolor{llightpink}{rgb}{1.0, 0.85, 0.95}
\definecolor{llightgreen}{rgb}{0.7, 1.0, 0.4}
\colorlet{lightyellow}{llightyellow!50!white}
\colorlet{lightblue}{llightblue!50!white}
\colorlet{lightgreen}{llightgreen!50!white}
\colorlet{lightpink}{llightpink!50!white}

\usepackage[framemethod=TikZ]{mdframed} 
\usepackage{tikz-cd} 
\usetikzlibrary{arrows,snakes,shapes.arrows,decorations.markings}
     \tikzset{>=triangle 90}
     \tikzstyle{bbc}=[draw,circle,fill=black,scale=.75]
     \tikzstyle{rc}=[circle,fill=red,scale=.6]
     \tikzstyle{wc}=[draw,circle,scale=.75]

\tikzset{snake it/.style={decorate, decoration=snake}}

\tikzset{
	on each segment/.style={
		decorate,
		decoration={
			show path construction,
			moveto code={},
			lineto code={
				\path [#1]
				(\tikzinputsegmentfirst) -- (\tikzinputsegmentlast);
			},
			curveto code={
				\path [#1] (\tikzinputsegmentfirst)
				.. controls
				(\tikzinputsegmentsupporta) and (\tikzinputsegmentsupportb)
				..
				(\tikzinputsegmentlast);
			},
			closepath code={
				\path [#1]
				(\tikzinputsegmentfirst) -- (\tikzinputsegmentlast);
			},
		},
	},
	mid arrow/.style={postaction={decorate,decoration={
				markings,
				mark=at position .5 with {\arrow[#1]{stealth}}
	}}},
}

\usetikzlibrary{decorations.markings}

\tikzset{line/.style={line width=0.25mm},
curve/.style={line,smooth,tension=1},
->-/.style={decoration={
  markings,
  mark=at position #1 with {\arrow[>=stealth]{>}}},postaction={decorate}},
-<-/.style={decoration={
  markings,
  mark=at position #1 with {\arrow[>=stealth]{<}}},postaction={decorate}},
}

\tikzset{bg/.style={opacity=.5}}

\tikzset{
    partial ellipse/.style args={#1:#2:#3}{
        insert path={+ (#1:#3) arc (#1:#2:#3)}
    }
}

\makeatletter
\renewcommand\section{\@startsection {section}{1}{\z@}%
                               {-3.5ex \@plus -1ex \@minus -.2ex}
                               {2.3ex \@plus.2ex}%
                               {\normalfont\large\bfseries}}
\renewcommand\subsection{\@startsection{subsection}{2}{\z@}%
                                 {-3.25ex\@plus -1ex \@minus -.2ex}%
                                 {1.5ex \@plus .2ex}%
                                 {\normalfont\bfseries}}
\makeatother

\newfont{\goth}{ygoth.tfm scaled 1200}                   

\numberwithin{equation}{section}

\newcommand{\be}{\begin{equation}}
\newcommand{\ee}{\end{equation}}
\newcommand{\bee}{\begin{equation} \begin{aligned}}
\newcommand{\eee}{\end{aligned} \end{equation}}

\newcommand{\CA}{\mathcal{A}}
\newcommand{\CC}{\mathcal{C}}
\newcommand{\CD}{\mathcal{D}}

\newcommand{\CZ}{\mathcal{Z}}
\newcommand{\CH}{\mathcal{H}}
\newcommand{\CL}{\mathcal{L}}
\newcommand{\CM}{\mathcal{M}}
\newcommand{\CN}{\mathcal{N}}

\newcommand\CT{\mathcal{T}}

\newcommand\IZ{\mathbb{Z}}

\newcommand\scD{\mathcal{D}}

\newcommand\scH{\mathcal{H}}

\newcommand\scN{\mathcal{N}}
\newcommand\scO{\mathcal{O}}

\newcommand\scU{\mathcal{U}}

\newcommand\doubleZ{\mathbb{Z}}

\newcommand{\VEC}{\mathsf{Vec}}
\newcommand{\Rep}{\mathsf{Rep}}
\newcommand{\TY}{\mathsf{TY}}

\newcommand\BrPic{\mathsf{BrPic}}
\newcommand\EqBr{\mathsf{EqBr}}
\newcommand\Ising{\mathsf{Ising}}

\newcommand{\dsi}{\mathds{1}}
\newcommand{\otau}{\overline{\tau}}

\newcommand{\oq}{\overline{q}}

\newcommand{\ii}{\mathsf{i}}

\newcommand\Hom{\operatorname{Hom}}

\newcommand{\Aut}{\operatorname{Aut}}

\newcommand{\eqnref}[1]{\eqref{#1}}

\newcommand{\tabref}[1]{Table\,\ref{#1}}

\newcommand{\appref}[1]{Appendix\,\ref{#1}}

\newcommand{\fsa}{\epsilon_{\CD}}
\newcommand{\fsb}{\kappa_{\CD}}

\renewcommand{\title}[1]{\vbox{\center\LARGE{#1}}\vspace{5mm}}
\renewcommand{\author}[1]{\vbox{\center#1}\vspace{5mm}}
\newcommand{\address}[1]{\vbox{\center\em#1}}

\begin{document} 

\begin{titlepage}
 	\hfill  	YITP-SB-2023-33
 	\\

\title{Self-duality under gauging a non-invertible symmetry}

\author{Yichul Choi${}^{1,2}$, Da-Chuan Lu${}^3$, and Zhengdi Sun${}^{3,4}$}

		\address{${}^{1}$C.\ N.\ Yang Institute for Theoretical Physics, Stony Brook University,\\
        Stony Brook, New York 11794, USA\\
        ${}^{2}$Simons Center for Geometry and Physics, Stony Brook University,\\
        Stony Brook, New York 11794, USA\\
		${}^{3}$Department of Physics, University of California San Diego, CA 92093, USA\\
        ${}^{4}$Mani L. Bhaumik Institute for Theoretical Physics, Department of Physics and Astronomy, University of California Los Angeles, CA 90095, USA

		}

\abstract

We discuss two-dimensional conformal field theories (CFTs) which are invariant under gauging a non-invertible global symmetry. At every point on the orbifold branch of $c=1$ CFTs, it is known that the theory is self-dual under gauging a $\mathbb{Z}_2\times \mathbb{Z}_2$ symmetry, and has $\mathsf{Rep}(H_8)$ and $\mathsf{Rep}(D_8)$ fusion category symmetries as a result. We find that gauging the entire $\mathsf{Rep}(H_8)$ fusion category symmetry maps the orbifold theory at radius $R$ to that at radius $2/R$. At $R=\sqrt{2}$, which corresponds to two decoupled Ising CFTs (Ising$^2$ in short), the theory is self-dual under gauging the $\mathsf{Rep}(H_8)$ symmetry. This implies the existence of a topological defect line in the Ising$^2$ CFT obtained from half-space gauging of the $\mathsf{Rep}(H_8)$ symmetry, which
commutes with the $c=1$ Virasoro algebra but does not preserve the fully extended chiral algebra. We bootstrap its action on the $c=1$ Virasoro primary operators, and find that there are no relevant or marginal operators preserving it. Mathematically, the new topological line combines with the $\mathsf{Rep}(H_8)$ symmetry to form a bigger fusion category which is a $\mathbb{Z}_2$-extension of $\mathsf{Rep}(H_8)$. We solve the pentagon equations including the additional topological line and find 8 solutions, where two of them are realized in the Ising$^2$ CFT. Finally, we show that the torus partition functions of the Monster$^2$ CFT and Ising$\times$Monster CFT are also invariant under gauging the $\mathsf{Rep}(H_8)$ symmetry.

\end{titlepage}

\eject

\tableofcontents

\section{Introduction}

Given a quantum field theory (QFT), it is a useful strategy to first analyze in detail its global symmetries and 't Hooft anomalies, before one asks more difficult dynamical questions (for instance, its long-distance behavior).
The understanding of the former can drastically constrain the possible answers to the latter \cite{tHooft:1979rat}.
The advent of generalized global symmetries \cite{Gaiotto:2014kfa} has increased the power of symmetries and one's ability to constrain possible answers to the dynamical questions in QFTs.
At the same time, a new challenging quest emerges, which is to effectively discover new generalized symmetries in familiar and important classes of QFTs. 
The task of finding all the generalized global symmetries of a given QFT turns out to be a difficult one, and to date there is no systematic way to achieve such a goal for generic QFTs.
See \cite{McGreevy:2022oyu,Cordova:2022ruw,Gomes:2023ahz,Brennan:2023mmt,Luo:2023ive,Schafer-Nameki:2023jdn,Bhardwaj:2023kri,Shao:2023gho} for recent reviews on generalized global symmetries.

This paper focuses on finite generalized symmetries in 1+1d CFTs described by fusion categories \cite{Bhardwaj:2017xup,Chang:2018iay}.\footnote{We focus on internal symmetries, and do not consider spacetime symmetries such as time-reversal. Also, we always work with unitary, compact, bosonic CFTs with a unique vacuum.}
Such symmetries are generated by topological defect lines, which commute with both the left- and right-moving Virasoro algebras (also known as totally transmissive defects).
Topological defect lines in 1+1d CFTs have been studied extensively, with various applications and perspectives.
See, for instance, \cite{Bhardwaj:2017xup,Chang:2018iay,Verlinde:1988sn,Oshikawa:1996ww,Oshikawa:1996dj,Petkova:2000ip,Frohlich:2004ef,Frohlich:2006ch,Frohlich:2009gb,Aasen:2016dop,Aasen:2020jwb,Moore:1988qv,Fuchs:2002cm,Carqueville:2012dk,Brunner:2014lua,Thorngren:2019iar,Cordova:2019wpi,Lin:2019hks,Pal:2020wwd,Komargodski:2020mxz,Chang:2020imq,Thorngren:2021yso,Huang:2021zvu,Burbano:2021loy,Kaidi:2022cpf,Lin:2022dhv,Chang:2022hud,Lu:2022ver,Kaidi:2023maf,Zhang:2023wlu,Lin:2023uvm,Choi:2023xjw,Bashmakov:2023kwo,Haghighat:2023sax,vanBeest:2023dbu,Duan:2023ykn,Chen:2023jht,Nagoya:2023zky,Bhardwaj:2023idu,Bhardwaj:2023fca,Huang:2023pyk}.

Other than topological quantum field theories (TQFTs), the only QFTs for which the full set of generalized symmetries is known are the Virasoro minimal model CFTs, whose central charge $c<1$.
Perhaps the next simplest class of QFTs for which one may hope to classify all the generalized symmetries would be the $c=1$ CFTs \cite{Ginsparg:1987eb}, which includes free compact boson and its orbifold.
They play an important role both in condensed matter physics and in string theory.
For instance, the orbifold branch of $c=1$ describes the critical line on the phase diagram of the Ashkin-Teller model on the lattice \cite{ashkin1943at,kohmoto1981at,saleur1987at,Ginsparg:1988ui}. 
The Ashkin-Teller model is widely used to describe 1+1d deconfined quantum critical points \cite{jiang2019dqcp,huang2019dqcp,zhang2023dqcp}, SPT transitions \cite{verresen2021spt,tantivasadakarn2023spt,prakash2023spt} and edges of 2+1d gauge theories \cite{scaffidi2017bdyto,zhu2019bdyto,somoza2021bdyto}.

Many topological defect lines are known for $c=1$ CFTs \cite{Thorngren:2021yso,Fuchs:2007tx,Bachas:2007td,Becker:2017zai,Ji:2019ugf}, and in particular, \cite{Thorngren:2021yso} provides a zoo of fusion category symmetries at $c=1$.
However, the full classification of topological defect lines at $c=1$ is not yet accomplished.
It is interesting to study the origin and physical consequences of these topological defect lines in concrete models, and also to discover new ones.

Below, we find a new topological defect line at $c=1$, at a point on the orbifold branch of the moduli space of $c=1$ CFTs where the theory is described by two decoupled Ising CFTs (Ising$^2$ in short).
Our finding is based on a generalized version of the ``half-space gauging,'' discussed in \cite{Thorngren:2021yso,Choi:2021kmx,Choi:2022zal} (see also \cite{Koide:2021zxj,Kaidi:2021xfk}).
Given a QFT in arbitrary spacetime dimensions which is self-dual under gauging a discrete symmetry, which we may call the ``parent'' symmetry, one can construct a codimension-1 topological defect by gauging the parent symmetry in only half of spacetime.
The resulting codimension-1 topological defect obeys a universal non-invertible fusion algebra, generating a ``child'' non-invertible symmetry.
The most standard example of such a construction is in the 1+1d Ising CFT, where the parent symmetry is the $\mathbb{Z}_2$ spin-flip symmetry, and the child non-invertible symmetry is generated by the Kramers-Wannier duality defect line \cite{Oshikawa:1996ww,Oshikawa:1996dj,Petkova:2000ip,Frohlich:2004ef,Frohlich:2006ch,Frohlich:2009gb,Aasen:2016dop,Aasen:2020jwb,Seiberg:2023cdc}.

In the known examples of half-space gauging, the parent symmetry is generally a discrete, invertible (higher-form) symmetry, whereas the child symmetry is non-invertible.
Below, we discuss a generalization of this, where the parent symmetry is also non-invertible, and described by a fusion category $\CC$.
Given a 1+1d QFT which is self-dual under gauging a fusion category symmetry $\CC$, we claim that one can construct a new topological defect line obtained by gauging $\CC$ in only half of spacetime.
We elaborate on this more in Section \ref{sec:gauging_noninv}, and briefly summarize it below.

Let $\{\CL_i\}$ be the set of simple topological lines in $\CC$.
By ``gauging $\CC$,'' we always mean that we gauge an algebra object of the form
\begin{equation} \label{eq:algebra_intro}
    \mathcal{A} = \bigoplus_{i} \langle \mathcal{L}_i\rangle  \mathcal{L}_i.
\end{equation}
where $\langle \CL_i \rangle$ is the quantum dimension of the line $\CL_i$.
There exists an algebra object of the form \eqref{eq:algebra_intro} which can be gauged if and only if the fusion category symmetry $\CC$ can be realized in a trivially gapped phase, namely if $\CC$ is anomaly-free \cite{Choi:2023xjw}.
This is a generalization of the familiar fact that 't Hooft anomalies are obstruction to gauging a global symmetry, to the case of fusion category symmetries.
For instance, recall that a necessary condition for the fusion category $\CC$ to be anomaly-free is that all the quantum dimensions $\langle \CL_i \rangle$ are non-negative integers \cite{Chang:2018iay}, and notice that \eqref{eq:algebra_intro} is a well-defined object in $\CC$ only when such a condition is satisfied.
Given an anomaly-free fusion category, there may exist more than one algebra object of the form \eqref{eq:algebra_intro} which can be gauged, and this generalizes the choice of discrete torsion.

Now, let $\CT$ be a 1+1d QFT with a non-anomalous fusion category symmetry $\CC$ and an algebra object $\CA$ of the form \eqref{eq:algebra_intro} which can be gauged.
The partition function of the gauged theory, denoted as $\CT / \CC$, is obtained from that of $\CT$ by inserting a fine mesh of the algebra object $\CA$ across the dual triangulation of the spacetime manifold \cite{Fuchs:2002cm,Bhardwaj:2017xup}.

Instead of gauging $\CC$ everywhere on the spacetime manifold, we may gauge it in only half of spacetime, i.e. by inserting a mesh of the algebra object $\CA$ in half of spacetime.
This ``half-space gauging'' of $\CC$ produces a topological interface between the two theories $\CT$ and $\CT/\CC$.

If the theory $\CT$ is self-dual under gauging $\CC$, then the half-space gauging results in a topological defect line of $\CT$, which we denote as $\CD$.
Such a topological line $\CD$ obeys a fusion algebra:
\begin{align}
\begin{split}
    \CD \otimes \overline{\CD} &= \CA = \bigoplus_{i} \langle \mathcal{L}_i\rangle  \CL_i = \dsi \oplus \cdots \,,\\
    \CD \otimes \CL_i &= \CL_i \otimes \CD = \langle \CL_i \rangle \CD \,.
\end{split}
\end{align}
Here, $\overline{\CD}$ is the orientation-reversal of $\CD$.
We note that the resulting line $\CD$ is non-invertible.
The quantum dimension of the line $\CD$ is determined by the total dimension of the parent fusion category $\CC$, namely
\begin{equation}
    \langle \CD \rangle = \sqrt{\sum_i \langle \CL_i \rangle^2 } \,.
\end{equation}

As a concrete example, we discuss $c=1$ CFTs as mentioned above.
In particular, we take the CFTs on the orbifold branch of the $c=1$ moduli space, and the ``parent'' fusion category symmetry to be the $\Rep(H_8)$ symmetry which was discussed in \cite{Thorngren:2021yso}. 
Here, $\Rep(H_8)$ denotes the representation category of a 8-dimensional Hopf algebra constructed by Kac and Paljutkin.
It is also one of the Tambara-Yamagami categories based on $\mathbb{Z}_2 \times \mathbb{Z}_2$.
The $\Rep(H_8)$ symmetry, which exists everywhere on the orbifold branch of $c=1$, is free of an anomaly, and admits a unique algebra object of the form \eqref{eq:algebra_intro} which can be gauged.
The simple lines of $\Rep(H_8)$ consist of four invertible lines $\dsi$, $a$, $b$, $ab$, generating a $\mathbb{Z}_2 \times \mathbb{Z}_2$ symmetry, and one non-invertible line $\CN$, satisfying $\CN \otimes \CN = \dsi \oplus a \oplus b \oplus ab$.

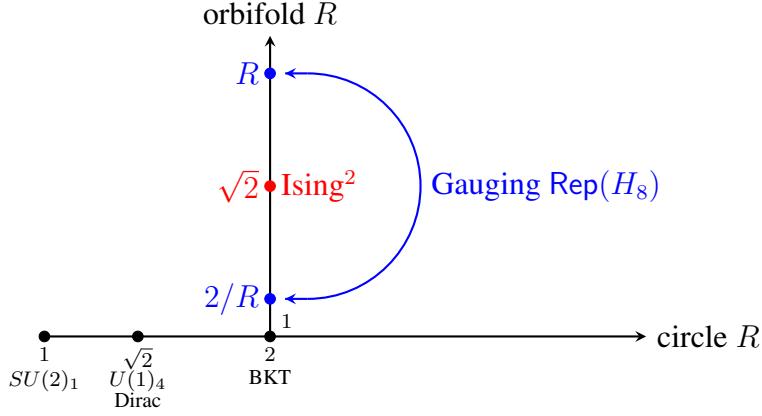
\begin{figure}[t]
    \centering
    \begin{tikzpicture}[scale=1.0]
	\draw[thick, black, -stealth] (-3,0) -- (5,0);
	\draw[thick, black, -stealth] (0,0) -- (0,4);
	\filldraw[black] (0,0) circle (2pt);
	\node[black, below] at (0,0) {\scriptsize $2$};
	\node[black, below] at (0,-0.3) {\scriptsize BKT};
	\node[black, right] at (5,0) {circle $R$};
	\node[black, above] at (0,4) {orbifold $R$};
	\filldraw[black] (-3,0) circle (2pt);
	\node[black, below] at (-3,0) {\scriptsize $1$};
	\node[black, below] at (-3,-0.3) {\scriptsize $SU(2)_1$};
	\filldraw[black] (-1.758,0) circle (2pt);
	\node[black, below] at (-1.758,0) {\scriptsize $\sqrt{2}$};
	\node[black, below] at (-1.758,-0.3) {\scriptsize $U(1)_4$};
	\node[black, below] at (-1.758,-0.6) {\scriptsize Dirac};
	\filldraw[red] (0,2) circle (2pt);
	\node[black, right] at (0,0.2) {\scriptsize $1$};
	\node[red, left] at (0,2) { $\sqrt{2}$};
	\node[red, right] at (0.0,2) {Ising$^2$};
        \draw[blue, thick] (0.5,3.5) arc(90:-90:1.5);
        \draw[blue, thick, ->-=1.0] (0.5,0.5) -- (0.2,0.5);
        \draw[blue, thick, ->-=1.0] (0.5,3.5) -- (0.2,3.5);
	\node[blue, right] at (2,2) {Gauging $\Rep(H_8)$};
	\node[blue, left] at (0,3.5) {$R$}; 
	\node[blue, left] at (0,0.5) {$2/R$};
	\filldraw[blue] (0,0.5) circle (2pt);
	\filldraw[blue] (0,3.5) circle (2pt);
\end{tikzpicture}	
    \caption{Moduli space of $c=1$ CFTs \cite{Ginsparg:1987eb}. The horizontal line is the circle branch consisting of free compact boson CFTs with radius $R$, and the vertical line is the orbifold branch obtained from gauging the charge conjugation symmetry of the circle branch theories. In addition, there are 3 isolated points (not shown). Along the orbifold branch, two theories at radii $R$ and $2/R$ are related by gauging the $\Rep(H_8)$ symmetry, and $R=\sqrt{2}$, corresponding to the Ising$^2$ CFT, is a fixed point under this gauging.}
    \label{fig:moduli_space}
\end{figure}

We find that gauging the algebra object $\CA = 1 \oplus a \oplus b \oplus ab \oplus 2\CN$ on the orbifold branch of $c=1$ maps the theory at radius $R$ to that at $2/R$, and vice versa (our convention is such that the $\mathsf{T}$-duality acts as $R \leftrightarrow 1/R$).
Thus, at a generic point, the orbifold CFT is not self-dual under gauging $\Rep(H_8)$, and instead such a gauging defines an order 2 operation on the orbfold branch.
However, the special point $R =\sqrt{2}$, namely the Ising$^2$ CFT, remains invariant under gauging $\Rep(H_8)$.
See Figure \ref{fig:moduli_space}.

This implies the existence of a new topological defect line $\CD$ in the Ising$^2$ CFT, coming from the half-space gauging of $\Rep(H_8)$.\footnote{This topological defect line $\CD$ of the Ising$^2$ CFT can be shown to be the product of a ``cosine'' line \cite{Chang:2020imq,Thorngren:2021yso} with a Kramers-Wannier duality line from one of the Ising factors \cite{Diatlyk:2023fwf}. A similar comment applies to the topological line $\CD'$ in Section \ref{sec:new_line} which obeys the same fusion algebra as $\CD$. 
We thank Yifan Wang for discussions on this point.}
Since gauging $\Rep(H_8)$ is an order 2 operation, we propose that $\CD$ is a self-dual line, namely $\overline{\CD} = \CD$.
This new topological line then obeys the following fusion algebra:
\begin{align} \label{eq:fusion_intro}
\begin{split}
    \CD \otimes \CD &= \dsi \oplus a \oplus b \oplus ab \oplus 2 \CN \,,\\
    \CD \otimes g &= g \otimes \CD = \CD \,, \\
    \CD \otimes \CN &= \CN \otimes \CD = 2\CD \,,
\end{split}
\end{align}
where $g \in \{ \dsi,a,b,ab\}$.
In particular, $\langle \CD \rangle = \sqrt{8} \notin \mathbb{Z}_{> 0}$, implying a nontrivial anomaly of the fusion category symmetry.

The Ising$^2$ CFT is rational with respect to the fully extended chiral algebra, namely two copies of the Ising chiral algebra.
The new topological defect line $\CD$ does not preserve this extended chiral algebra, and only a subalgebra of it (which includes the $c=1$ Virasoro algebra) is preserved.

We carefully analyze the action of this new topological line $\CD$ on the $c=1$ Virasoro primary operators of the Ising$^2$ CFT, by imposing several consistency conditions.
We find that there are no relevant or marginal operators preserving $\CD$.
Furthermore, by examining the spin selection rules derived from the explicit solutions to the pentagon identities based on the fusion algebra \eqref{eq:fusion_intro}, we determine the full fusion category structure formed by the line $\CD$ and the original $\Rep(H_8)$ symmetry.
Mathematically, the resulting fusion category is a $\mathbb{Z}_2$-extension of $\Rep(H_8)$.
We find that there are 8 such $\mathbb{Z}_2$-extensions satisfying the fusion algebra \eqref{eq:fusion_intro}, denoted as $\underline{\mathcal{E}}^{(i,\fsb,\fsa)}_{\doubleZ_2}\Rep(H_8)$ with $i,\fsb,\fsa =\pm $.
Among these, the ones corresponding to $\fsb = \fsa = +$ and $i=\pm$ are physically realized in the Ising$^2$ CFT (related by the $\mathbb{Z}_2$ symmetry that exchanges the two Ising factors).
Here, $\epsilon_{\CD}$ is the Frobenius-Schur indicator of $\CD$.

Finally, at the level of torus partition functions, we find that the Monster$^2$ and Ising$\times$Monster CFTs are also self-dual under gauging $\Rep(H_8)$. 
The Monster CFT is a $c=24$ holomorphic CFT with an extremely rich global symmetry group, the Monster group \cite{frenkel1984natural}. 
Similar to the Ising CFT, the Monster CFT is self-dual under gauging a $\IZ_{2}$ symmetry, and has the corresponding Kramers-Wannier-like duality defect line \cite{Lin:2019hks}. 
We leave for the future more detailed studies of potential new topological defect lines coming from half-space gauging of $\Rep(H_8)$ in these additional examples.

The rest of the paper is organized as follows.
In Section \ref{sec:review}, we begin by reviewing several properties of topological defect lines in 1+1d, finite invertible symmetries and their (half-space) gauging, as well as group extensions of fusion categories.
In Section \ref{sec:gauging_noninv}, we generalize the half-space gauging to the case where the parent symmetry is already non-invertible.
We also explain how to explicitly gauge the $\Rep(H_8)$ symmetry and compute the torus partition function of the gauged theory.
In Section \ref{sec:gauging_c=1}, we gauge the $\Rep(H_8)$ symmetry along the orbifold branch of $c=1$, and find that the Ising$^2$ CFT is self-dual under this gauging.
In Section \ref{sec:new_line}, we discuss in more detail several properties of the new topological defect line in the Ising$^2$ CFT, including its action on the Virasoro primary operators.
In Section \ref{sec:stacking}, we discuss various other 1+1d CFTs with the $\Rep(H_8)$ symmetry.
We find that the Monster$^2$ and Ising$\times$Monster CFTs are also self-dual at the level of torus partition functions.
We also provide non-examples, namely theories with the $\Rep(H_8)$ symmetry which are not self-dual under gauging.
Finally, in Section \ref{sec:F}, we explicitly solve the pentagon identities based on the fusion algebra \eqref{eq:fusion_intro}, and find 8 solutions.
We derive the spin selection rules, which allow us to physically distinguish the 8 fusion categories.
Two of them are realized in the Ising$^2$ CFT.

\section{Review} \label{sec:review}

We first briefly review several defining properties of topological defect lines in 1+1d CFTs \cite{Fuchs:2002cm,Bhardwaj:2017xup,Chang:2018iay} to set up the notations, as well as the half-space gauging for the case where the parent symmetry is an invertible symmetry.
Finally, group extension of fusion categories \cite{2009arXiv0909.3140E} is also reviewed.

\subsection{Topological defect lines in 1+1d}
Topological defect lines (denoted as $\mathcal{L}$) are line operators which commute with the stress-energy tensor, and various correlation functions are invariant under local deformations of them.
We focus on topological defect lines which satisfy the mathematical axioms of (unitary) fusion categories \cite{Fuchs:2002cm,Bhardwaj:2017xup,Chang:2018iay}.
Such topological lines generate finite (generalized) symmetries in 1+1d.\footnote{There are topological defect lines which go beyond the mathematical definition of fusion categories \cite{Chang:2020imq,Thorngren:2021yso}, the simplest example being just an ordinary continuous symmetry in 1+1d.}
For instance, we can fuse two topological lines $\mathcal{L}_a$ and $\mathcal{L}_b$ by putting them close to each other and generate a new topological line, which then in general decomposes into a finite sum of other topological lines,
\begin{equation}
    \mathcal{L}_a \otimes \mathcal{L}_b = \bigoplus_{c} N^{c}_{ab} \mathcal{L}_c, \quad N^c_{ab} \in \doubleZ_{\geq 0}.
\end{equation}
The \textit{simple} topological lines are those that cannot be written as a sum of at least two other lines.
We denote the trivial topological line as $\dsi$.

When $N_{ab}^c \ne 0$, two topological lines $\mathcal{L}_a$ and $\mathcal{L}_b$ can join each other locally and become the line $\mathcal{L}_c$ at a trivalent junction.
The set of topological junction operators form a vector space whose complex dimension is given by $N_{ab}^c$. 
We always fix a basis of the junction vector space and use the Greek letters $\mu,\nu,\cdots = 1,2,\cdots,N^{c}_{ab}$ to denote the corresponding basis vectors.\footnote{Later we will also use $\mu$ (and $\mu$ only) to denote the multiplication junction of an algebra object. 
The reader should be able to distinguish the two based on the context.}
\begin{equation}
    \begin{tikzpicture}[scale=0.8,baseline={([yshift=-.5ex]current bounding box.center)},vertex/.style={anchor=base,
    circle,fill=black!25,minimum size=18pt,inner sep=2pt},scale=0.50]
    \draw[thick, black] (-2,-2) -- (0,0);
    \draw[thick, black] (+2,-2) -- (0,0);
    \draw[thick, black] (0,0) -- (0,2);
    \draw[thick, black, -stealth] (-2,-2) -- (-1,-1);
    \draw[thick, black, -stealth] (+2,-2) -- (1,-1);
    \draw[thick, black, -stealth] (0,0) -- (0,1);
    \filldraw[thick, black] (0,0) circle (3pt);
    \node[black, right] at (0,0) {\footnotesize $\mu$};
    \node[black, below] at (-2,-2) {$\mathcal{L}_a$};
    \node[black, below] at (2,-2) {$\mathcal{L}_b$};
    \node[black, above] at (0,2) {$\mathcal{L}_c$};
    
\end{tikzpicture}, \quad \mu = 1,2,\cdots, N^{c}_{ab} \,.
\end{equation}
Given three topological defect lines, there are two possible ways they can fuse together. They are related by the so-called associativity map. 
In the explicitly chosen basis, the associativity map is characterized by the $F$-symbols,
\begin{equation}
\begin{tikzpicture}[baseline={([yshift=-1ex]current bounding box.center)},vertex/.style={anchor=base,
    circle,fill=black!25,minimum size=18pt,inner sep=2pt},scale=0.7]
	\draw[thick, black, -<-=0.5] (0,0) -- (-0.75,-0.75);
	\draw[thick, black, -<-=0.5] (0,0) -- (+0.75,-0.75);
	\draw[thick, black, ->-=0.5] (0,0) -- (+0.75,+0.75);
	\draw[thick, black, ->-=0.5] (+0.75,+0.75) -- (1.5,1.5);
	\draw[thick, black, -<-=0.5] (0.75,0.75) -- (2.25,-0.75);
	
	\node[below, black] at (-0.75,-0.75) {\scriptsize $\mathcal{L}_a$};
	\node[below, black] at (+0.75,-0.75) {\scriptsize $\mathcal{L}_b$};
	\node[below, black] at (+2.25,-0.75) {\scriptsize $\mathcal{L}_c$};
	\node[right, black] at (0.25,0.2) {\scriptsize $\mathcal{L}_e$};
 	\node[left, black] at (0,0) {\scriptsize $\mu$};
    \node[left, black] at (0.75,0.75) {\scriptsize $\nu$};
	\node[above, black] at (1.5,1.5) {\scriptsize $\mathcal{L}_d$};
	\filldraw[black] (0,0) circle (2pt);
	\filldraw[black] (0.75,0.75) circle (2pt);
	
	\node[black] at (5,0.25) {$\displaystyle = \sum\limits_{f,\rho,\sigma} \left[F^{abc}_d\right]_{(e,\mu,\nu),(f,\rho,\sigma)}$};
	
	\draw[thick, black,->-=0.5] (9-0.75,-0.75) -- (9+0.75,0.75);
	\draw[thick, black,->-=0.5] (9+0.75,+0.75) -- (9+1.5,1.5);
        \draw[thick, black,->-=0.5] (9+0.75,-0.75) -- (9+1.5,0.);
        \draw[thick, black,->-=0.5] (9+2.25,-0.75) -- (9+1.5,0.);
        \draw[thick, black,->-=0.5] (9+1.5,0.) -- (9+0.75,+0.75);
	
	\node[below, black] at (9-0.75,-0.75) {\scriptsize $\mathcal{L}_a$};
	\node[below, black] at (9+0.75,-0.75) {\scriptsize $\mathcal{L}_b$};
	\node[below, black] at (9+2.25,-0.75) {\scriptsize $\mathcal{L}_c$};
	\node[right, black] at (9+1.1,0.5) {\scriptsize $\mathcal{L}_f$};
	\node[above, black] at (9+1.5,1.5) {\scriptsize $\mathcal{L}_d$};

    \node[left, black] at (9+1.5,0) {\scriptsize $\rho$};
	\node[left, black] at (9+0.75,+0.75) {\scriptsize $\sigma$};
	\filldraw[black] (9+0.75,+0.75) circle (1.5pt);
	\filldraw[black] (9+1.5,0) circle (1.5pt);	
\end{tikzpicture}.
\end{equation}
The $F$-symbols satisfy the consistent conditions known as pentagon equations (our conventions follow \cite{Barkeshli:2014cna}), 
\begin{align}\label{eq:pentaequ}
&\sum_{\delta }\left[ F_{e}^{fcd}\right] _{\left( g,\beta ,\gamma \right)
\left( l,\nu, \delta \right) }\left[ F_{e}^{abl}\right]
_{\left( f,\alpha ,\delta \right) \left( k,\mu, \lambda \right) }\nonumber\\
=&\sum_{h,\sigma ,\psi ,\rho }\left[ F_{g}^{abc}\right] _{\left( f,\alpha
,\beta \right) \left( h,\psi,\sigma \right) }\left[ F_{e}^{ahd}\right]
_{\left( g,\sigma ,\gamma \right) \left( k,\rho, \lambda \right) }%
\left[ F_{k}^{bcd}\right] _{\left( h,\psi ,\rho \right) \left(
l,\nu ,\mu \right) }.
\end{align}

Fusion category symmetries are ubiquitous in 1+1d. 
For instance, the category $\VEC_G^\omega$ describes an ordinary finite $0$-form symmetry $G$ with an 't Hooft anomaly $[\omega] \in H^3(G,U(1))$.
The simple lines are labeled by group elements, $\mathcal{L}_g$ with $g \in G$, and their fusion rules are governed by the group multiplication law, $\mathcal{L}_g \otimes \mathcal{L}_h = \mathcal{L}_{gh}$. 
Below, we denote the line $\mathcal{L}_g$ by the corresponding group element $g$ for simplicity. 
Then, the $F$-symbols are given by
\begin{equation}
    \left[ F^{g,h,k}_{ghk} \right]_{gh,hk} = \omega(g,h,k).
\end{equation}
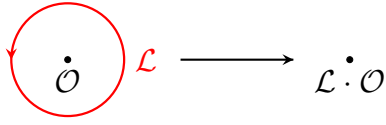
\begin{figure}[t]
    \centering
    \begin{tikzpicture}[baseline={([yshift=+.5ex]current bounding box.center)},vertex/.style={anchor=base,
    circle,fill=black!25,minimum size=18pt,inner sep=2pt},scale=0.75]
    \draw[red, ->-=0.5, line width = 0.3mm] (0,0) circle (1);
    \filldraw[black] (0,0) circle (1.5pt);
    \node[black, below] at (0,0) {$\mathcal{O}$};
    \node[red, right] at (+1,0) {$\mathcal{L}$};
    \draw[thick, ->-=1] (2,0) -- (4,0);
    \filldraw[black] (5,0) circle (1.5pt);
    \node[black, below] at (5,0) {$\mathcal{L}\cdot\mathcal{O}$};
    \end{tikzpicture}
    \caption{Action of a topological defect line $\mathcal{L}$ on a local operator $\mathcal{O}$. We start with the line $\mathcal{L}$ wrapping around the local operator $\mathcal{O}$. 
    After shrinking $\mathcal{L}$, $\mathcal{O}$ is transformed by $\mathcal{L}$ to another local operator $\mathcal{L}\cdot \mathcal{O}$.}
    \label{fig:Laction}
\end{figure}

The action of $G$ on a local operator is obtained by encircling the local operator with a closed loop of $g$.
Similarly, an arbitrary topological defect line $\mathcal{L}$ can act on a local operator $\mathcal{O}$, as shown in Figure \ref{fig:Laction}.
We denote such an action of a topological line $\CL$ on a local operator $\mathcal{O}$ as $\CL \cdot \mathcal{O}$.

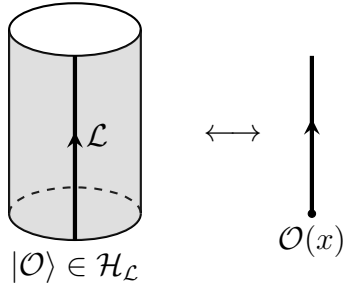
\begin{figure}[t]
    \centering
    \begin{tikzpicture}[scale=0.7]
        \draw [thick] (0,0) ellipse (1.25 and 0.5);
        \draw [thick] (-1.25,0) -- (-1.25,-3.5);
        \draw [thick] (-1.25,-3.5) arc (180:360:1.25 and 0.5);
        \draw [thick, dashed] (-1.25,-3.5) arc (180:360:1.25 and -0.5);
        \draw [thick] (1.25,-3.5) -- (1.25,0);
        \fill [gray,opacity=0.25] (-1.25,0) -- (-1.25,-3.5) arc (180:360:1.25 and 0.5) -- (1.25,0) arc (0:180:1.25 and -0.5);
        \draw [ultra thick, -<-=0.5] (0,-0.5) -- (0, -3.5-0.5);
        \node[black, below] at (0,-4) {$\ket{\scO}\in \CH_\CL$};
        \node[black, right] at (0,-2) {$\CL$};
        \node[black] at (3,-2) {$\longleftrightarrow$};
        \draw [ultra thick, -<-=0.5] (4.5,-0.5) -- (4.5, -3.5);
        \filldraw[black] (4.5,-3.5) circle (2pt);
        \node[black, below] at (4.5,-3.5) {$\scO(x)$};
        
    \end{tikzpicture}
    \caption{Under the state-operator correspondence, a state in the defect Hilbert space $\mathcal{H}_{\mathcal{L}}$ is mapped to a non-local operator $\mathcal{O}$ which is attached to $\mathcal{L}$.} \label{fig:non_local_operator_and_defect_Hilbert_space}
\end{figure}

A topological defect line $\mathcal{L}$ may end on a (non-local) operator $\mathcal{O}$, and under the state-operator map, such a non-local operator $\mathcal{O}$ is mapped to a state in the Hilbert space $\mathcal{H}_{\mathcal{L}}$ quantized on $S^1$ but with the boundary condition twisted by $\mathcal{L}$, see Figure \ref{fig:non_local_operator_and_defect_Hilbert_space}. $\mathcal{H}_{\mathcal{L}}$ is known as the defect (or twisted) Hilbert space.

\begin{figure}[t]
    \centering
    \begin{tikzpicture}[scale=0.8,baseline={(0,0.75)}]
    \node[above, red] at (0,3) {\scriptsize $\mathcal{L}_1$};
    \draw[thick, red, ->-=0.5] (0,0) -- (0,1);
    \draw[thick, red, ->-=0.5] (0,2) -- (0,3);
    \node[below] at (0,0) {\scriptsize$\mathcal{O}$};
    \draw [blue,thick,domain=-90:90, ->-=0] plot ({1*cos(\x)}, {1*sin(\x)});
    \draw [blue,thick,domain=90:270] plot ({1.5*cos(\x)}, {0.5+1.5*sin(\x)});
    \draw[thick, dgreen, ->-=0.5] (0,1) -- (0,2);
    \filldraw[] (0,0) circle (1.5pt);
    \filldraw[black] (0,1.0) circle (1.5pt);
    \filldraw[black] (0,2.0) circle (1.5pt);
    \node[black, left] at (0.,1.) {\scriptsize$\mu$};
    \node[black, right] at (0.,2.) {\scriptsize$\nu$};
    \node[blue, right] at (1.,0.) {\scriptsize $\mathcal{L}_2$};
    \node[dgreen, right] at (0.,1.5) {\scriptsize $\mathcal{L}_3$};
    \end{tikzpicture}
    \caption{The lasso diagram describing the action of a topological defect line $\mathcal{L}_2$ on a non-local operator $\mathcal{O}$ attached to $\mathcal{L}_1$. To fully determine the action, we need to also specify the line $\mathcal{L}_3$ as well as two junctions $\mu,\nu$.}
    \label{fig:lasso_action}
\end{figure}
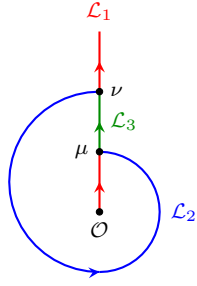

Topological defect lines can also act on non-local operators, via the \textit{lasso} diagram \cite{Chang:2018iay,Lin:2022dhv} in Figure \ref{fig:lasso_action}.
To define the action of $\mathcal{L}_2$ on a non-local operator $\mathcal{O}$ attached to $\mathcal{L}_1$, we must additionally specify a line $\mathcal{L}_3$ and two junctions $\mu,\nu$. 
Under the state-operator correspondence, this defines an operator $\widehat{\mathcal{L}_2}_{(\mathcal{L}_3,\mu,\nu)}$ acting on the defect Hilbert space $\mathcal{H}_{\mathcal{L}_1}$.\footnote{One may also define a more general map from $\mathcal{H}_{\mathcal{L}_1}$ to a different defect Hilbert space $\mathcal{H}_{\mathcal{L}_4}$, by a similar diagram as in  Figure \ref{fig:lasso_action}, with the topmost topological line replaced by $\CL_4$.}

The twisted partition function $\Tr_{\mathcal{H}_{\mathcal{L}_1}}\left(\widehat{\mathcal{L}_2}_{(\mathcal{L}_3,\mu,\nu)} q^{L_0 - c/24} \overline{q}^{\overline{L}_0 - c/24}\right)$ is an important observable of the 1+1d CFT. 
From the state-operator map applied to Figure \ref{fig:lasso_action}, we see that this twisted partition function corresponds to the torus partition function with the following network of topological defect lines inserted:
\begin{equation}
\Tr_{\mathcal{H}_{\mathcal{L}_1}}\left(\widehat{\mathcal{L}_2}_{(\mathcal{L}_3,\mu,\nu)} q^{L_0 - c/24} \overline{q}^{\overline{L}_0 - c/24}\right) \equiv Z[\mathcal{L}_1,\mathcal{L}_2,\mathcal{L}_3;\mu,\nu](\tau) = \begin{tikzpicture}[baseline={([yshift=+.5ex]current bounding box.center)},vertex/.style={anchor=base,
    circle,fill=black!25,minimum size=18pt,inner sep=2pt},scale=0.5]
    \filldraw[grey] (-2,-2) rectangle ++(4,4);
    \draw[thick, dgrey] (-2,-2) -- (-2,+2);
    \draw[thick, dgrey] (-2,-2) -- (+2,-2);
    \draw[thick, dgrey] (+2,+2) -- (+2,-2);
    \draw[thick, dgrey] (+2,+2) -- (-2,+2);
    \draw[thick, red, -stealth] (0,-2) -- (0.354,-1.354);
    \draw[thick, red] (0,-2) -- (0.707,-0.707);
    \draw[thick, blue, -stealth] (2,0) -- (1.354,-0.354);
    \draw[thick, blue] (2,0) -- (0.707,-0.707);
    \draw[thick, red, -stealth] (-0.707,0.707) -- (-0.354,1.354);
    \draw[thick, red] (0,2) -- (-0.707,0.707);
    \draw[thick, blue, -stealth] (-0.707,0.707) -- (-1.354,0.354);
    \draw[thick, blue] (-2,0) -- (-0.707,0.707);
    \draw[thick, dgreen, -stealth] (0.707,-0.707) -- (0,0);
    \draw[thick, dgreen] (0.707,-0.707) -- (-0.707,0.707);
    \filldraw[black] (0.707,-0.707) circle (2pt);
    \filldraw[black] (-0.707,0.707) circle (2pt);
    \node[red, below] at (0,-2) {\scriptsize $\mathcal{L}_1$};
    \node[blue, right] at (2,0) {\scriptsize $\mathcal{L}_2$};
    \node[dgreen, above] at (0.2,0) {\scriptsize $\mathcal{L}_3$};
    \node[black, below] at (0.9,-0.607) {\scriptsize $\mu$};
    \node[black, below] at (-0.707,0.707) {\scriptsize $\nu$};
\end{tikzpicture}.
\end{equation}
When every fusion coefficient $N_{ab}^c = 0,1$, we will simply drop the $\mu,\nu$ indices, and write $Z[\mathcal{L}_1,\mathcal{L}_2,\mathcal{L}_3]$. These twisted torus partition functions are constrained by the covariance under the modular transformations. For instance, under the $S$-transformation, we find
\begin{equation}
    Z[\mathcal{L}_1,\mathcal{L}_2,\mathcal{L}_3]\left(-\frac{1}{\tau}\right) = \sum_{\mathcal{L}_k} \left[F^{\mathcal{L}_1,\mathcal{L}_2,\overline{\mathcal{L}_1}}_{\mathcal{L}_2}\right]_{\mathcal{L}_3 \mathcal{L}_k} Z[\mathcal{L}_2,\overline{\mathcal{L}_1},\mathcal{L}_k](\tau).
\end{equation}

\subsection{Invertible symmetries and gauging}\label{sec:invert_sym_gauging}
Here, we review and rephrase the gauging of an anomaly-free finite group symmetry $G$ in terms of an algebra object $\mathcal{A}$, which generalizes naturally to the case of general non-invertible symmetries \cite{Fuchs:2002cm,Bhardwaj:2017xup}. 
For an ordinary global symmetry, an 't Hooft anomaly is characterized as an obstruction to gauge the symmetry, and its nontriviality implies that the low-energy phase cannot be trivially gapped while preserving the symmetry. 
For a (bosonic) theory in $d$ spacetime dimensions with a finite, internal $0$-form symmetry $G$, possible 't Hooft anomalies are classified by the elements of $H^{d+1}(G,U(1))$. 
In particular, the 't Hooft anomaly associated to a non-trivial class $[\omega(g,h,k)]\in H^3(G,U(1))$ obstructs orbifolding a 1+1d CFT with the $0$-form symmetry $G$. 
When the obstruction vanishes, $H^2(G,U(1))$ then parameterizes inequivalent ways of gauging $G$, known as the discrete torsion.

Let $\CC = \VEC_G^\omega$ below, where we keep $\omega$ to be arbitrary for now.
We will see that $\omega$ must be trivial for it to be possible to gauge $G$ consistently.
To gauge $G$ in a CFT $\mathcal{T}$ and to compute the corresponding orbifold partition function on a torus, we consider
\begin{equation}\label{eq:finite_G_gauging}
    Z_{\mathcal{T}/G} = \frac{1}{|G|}\sum_{\substack{g,h\in G \\ gh = hg}} \frac{\varphi(g,h)}{\varphi(h,g)} Z_{\mathcal{T}}[g,h,gh] = \frac{1}{|G|} \sum_{\substack{g,h\in G \\ gh = hg}} \frac{\varphi(g,h)}{\varphi(h,g)} \, \begin{tikzpicture}[baseline={([yshift=+.5ex]current bounding box.center)},vertex/.style={anchor=base,
    circle,fill=black!25,minimum size=18pt,inner sep=2pt},scale=0.5]
    \filldraw[grey] (-2,-2) rectangle ++(4,4);
    \draw[thick, dgrey] (-2,-2) -- (-2,+2);
    \draw[thick, dgrey] (-2,-2) -- (+2,-2);
    \draw[thick, dgrey] (+2,+2) -- (+2,-2);
    \draw[thick, dgrey] (+2,+2) -- (-2,+2);
    \draw[thick, black, -stealth] (0,-2) -- (0.354,-1.354);
    \draw[thick, black] (0,-2) -- (0.707,-0.707);
    \draw[thick, black, -stealth] (2,0) -- (1.354,-0.354);
    \draw[thick, black] (2,0) -- (0.707,-0.707);
    \draw[thick, black, -stealth] (-0.707,0.707) -- (-0.354,1.354);
    \draw[thick, black] (0,2) -- (-0.707,0.707);
    \draw[thick, black, -stealth] (-0.707,0.707) -- (-1.354,0.354);
    \draw[thick, black] (-2,0) -- (-0.707,0.707);
    \draw[thick, black, -stealth] (0.707,-0.707) -- (0,0);
    \draw[thick, black] (0.707,-0.707) -- (-0.707,0.707);

    \node[black, below] at (0,-2) {\scriptsize  $g$};
    \node[black, right] at (2,0) {\scriptsize $h$};
    \node[black, above] at (0.2,0) {\scriptsize $gh$};
\end{tikzpicture},
\end{equation}
where $[\varphi] \in H^2(G,U(1))$ is a choice of the discrete torsion. 
Each diagram with non-trivial $g$ and $h$ is a choice of a resolution of the 4-way junction diagram, and there exists another resolution 
\begin{equation}\label{eq:finite_G_resolution}
    \begin{tikzpicture}[baseline={([yshift=+.5ex]current bounding box.center)},vertex/.style={anchor=base,
    circle,fill=black!25,minimum size=18pt,inner sep=2pt},scale=0.5]
    \filldraw[grey] (-2,-2) rectangle ++(4,4);
    \draw[thick, dgrey] (-2,-2) -- (-2,+2);
    \draw[thick, dgrey] (-2,-2) -- (+2,-2);
    \draw[thick, dgrey] (+2,+2) -- (+2,-2);
    \draw[thick, dgrey] (+2,+2) -- (-2,+2);
    \draw[thick, black, ->-=.25] (0,-2) -- (0,2);
    \draw[thick, black, ->-=.25] (2,0) -- (-2,0);
    \node[black, below] at (0,-2) {\scriptsize $g$};
    \node[black, right] at (2,0) {\scriptsize $h$};
\end{tikzpicture} \longrightarrow \begin{tikzpicture}[baseline={([yshift=+.5ex]current bounding box.center)},vertex/.style={anchor=base,
    circle,fill=black!25,minimum size=18pt,inner sep=2pt},scale=0.5]
    \filldraw[grey] (-2,-2) rectangle ++(4,4);
    \draw[thick, dgrey] (-2,-2) -- (-2,+2);
    \draw[thick, dgrey] (-2,-2) -- (+2,-2);
    \draw[thick, dgrey] (+2,+2) -- (+2,-2);
    \draw[thick, dgrey] (+2,+2) -- (-2,+2);
    \draw[thick, black, -stealth] (0,-2) -- (0.354,-1.354);
    \draw[thick, black] (0,-2) -- (0.707,-0.707);
    \draw[thick, black, -stealth] (2,0) -- (1.354,-0.354);
    \draw[thick, black] (2,0) -- (0.707,-0.707);
    \draw[thick, black, -stealth] (-0.707,0.707) -- (-0.354,1.354);
    \draw[thick, black] (0,2) -- (-0.707,0.707);
    \draw[thick, black, -stealth] (-0.707,0.707) -- (-1.354,0.354);
    \draw[thick, black] (-2,0) -- (-0.707,0.707);
    \draw[thick, black, -stealth] (0.707,-0.707) -- (0,0);
    \draw[thick, black] (0.707,-0.707) -- (-0.707,0.707);

    \node[black, below] at (0,-2) {\scriptsize $g$};
    \node[black, right] at (2,0) {\scriptsize $h$};
    \node[black, above] at (0.2,0) {\scriptsize $gh$};
\end{tikzpicture} \quad \stackrel{?}{=} \quad \begin{tikzpicture}[baseline={([yshift=+.5ex]current bounding box.center)},vertex/.style={anchor=base,
    circle,fill=black!25,minimum size=18pt,inner sep=2pt},scale=0.5]
    \filldraw[grey] (-2,-2) rectangle ++(4,4);
    \draw[thick, dgrey] (-2,-2) -- (-2,+2);
    \draw[thick, dgrey] (-2,-2) -- (+2,-2);
    \draw[thick, dgrey] (+2,+2) -- (+2,-2);
    \draw[thick, dgrey] (+2,+2) -- (-2,+2);
    \draw[thick, black, ->-=0.5] (0,-2) -- (-0.707,-0.707);
    \draw[thick, black, ->-=0.5] (2,0) -- (0.707,0.707);
    \draw[thick, black, ->-=0.5] (-0.707,-0.707) -- (0.707,0.707);
    \draw[thick, black, ->-=0.5] (0,2) -- (0.707,0.707);
    \draw[thick, black, ->-=0.5] (-0.707,-0.707) -- (-2,0);

    \node[black, below] at (0,-2) {\scriptsize $g$};
    \node[black, right] at (2,0) {\scriptsize $h$};
    \node[black, above] at (-0.2,0) {\scriptsize $gh$};
\end{tikzpicture}.
\end{equation}
The two resolutions represent two different local trivializations of the same $G$-bundle, and they must agree in order for the gauging to be consistent. 
The 't Hooft anomaly $\omega$ exactly measures the ambiguity in the resolution of the twisted partition function.

Alternatively, we can also obtain the orbifold partition function \eqref{eq:finite_G_gauging} by inserting a single mesh of the non-simple topological defect line $\CA  = \bigoplus\limits_{g\in G} g$:
\begin{equation}\label{eq:finite_G_gauging_alg}
\mathcal{Z}_{\mathcal{T}/G} = \begin{tikzpicture}[baseline={([yshift= 0 ex]current bounding box.center)},vertex/.style={anchor=base,
    circle,fill=black!25,minimum size=18pt,inner sep=2pt},scale=0.5]
    \filldraw[grey] (-2,-2) rectangle ++(4,4);
    \draw[thick, dgrey] (-2,-2) rectangle ++(4,4);
    \draw[thick, red, -stealth] (0,-2) -- (0.354,-1.354);
    \draw[thick, red] (0,-2) -- (0.707,-0.707);
    \draw[thick, red, -stealth] (2,0) -- (1.354,-0.354);
    \draw[thick, red] (2,0) -- (0.707,-0.707);
    \draw[thick, red, -stealth] (-0.707,0.707) -- (-0.354,1.354);
    \draw[thick, red] (0,2) -- (-0.707,0.707);
    \draw[thick, red, -stealth] (-0.707,0.707) -- (-1.354,0.354);
    \draw[thick, red] (-2,0) -- (-0.707,0.707);
    \draw[thick, red, -stealth] (0.707,-0.707) -- (0,0);
    \draw[thick, red] (0.707,-0.707) -- (-0.707,0.707);
    
    \filldraw[red] (+0.707,-0.707) circle (2pt);
    \filldraw[red] (-0.707,+0.707) circle (2pt);
    
    \node[red, below] at (0,-2) {\scriptsize $\mathcal{A}$};
    \node[red, right] at (2,0) {\scriptsize $\mathcal{A}$};
    \node[red, above] at (0,2) {\scriptsize $\mathcal{A}$};
    \node[red, left] at (-2,0) {\scriptsize $\mathcal{A}$};
    \node[red, above] at (0,0) {\scriptsize $\mathcal{A}$};

    \node[red, below] at (+0.807,-0.707) {\scriptsize $\mu$};
    \node[red, above] at (-0.807,+0.707) {\scriptsize $\mu^\vee$};
\end{tikzpicture}.
\end{equation}
The consistency of the gauging requires the line $\CA$ to be an \emph{algebra object} in $\CC$ (with a few additional conditions) \cite{Fuchs:2002cm,Bhardwaj:2017xup}.
First, it requires the data of a fusion junction (multiplication) $\mu \in \Hom_{\mathcal{C}}(\mathcal{A}\otimes \mathcal{A},\mathcal{A})$ and a splitting junction (comultiplication) $\mu^\vee \in \Hom_{\mathcal{C}}(\mathcal{A}, \mathcal{A}\otimes \mathcal{A})$. 
In our case, they are given by
\begin{footnotesize}
\begin{equation}\label{eq:finite_G_alg}
\begin{tikzpicture}[scale=0.50,baseline = {(0,0)}]
    \draw[thick, red, ->-=.5] (-1.7,-1) -- (0,0);
    \draw[thick, red, -<-=.5] (0,0) -- (1.7,-1);
    \draw[thick, red, -<-=.5] (0,2) -- (0,0);
    \node[red, above] at (0,2) {\scriptsize$\mathcal{A}$};
    \node[red, below] at (-1.7,-1) {\scriptsize$\mathcal{A}$};
    \node[red, below] at (1.7,-1) {\scriptsize$\mathcal{A}$};
    \node[red, right] at (+0,0.2) {\scriptsize$\mu$};
    \filldraw[red] (0,0) circle (2pt);
\end{tikzpicture} = \frac{1}{\sqrt{|G|}}\sum_{g,h\in G} \varphi(g,h)\begin{tikzpicture}[scale=0.50,baseline = 0.]
    \draw[thick, black, ->-=.5] (-1.7,-1) -- (0,0);
    \draw[thick, black, -<-=.5] (0,0) -- (1.7,-1);
    \draw[thick, black, -<-=.5] (0,2) -- (0,0);
    \node[black, above] at (0,2) {\scriptsize$gh$};
    \node[black, below] at (-1.7,-1) {\scriptsize$g$};
    \node[black, below] at (1.7,-1) {\scriptsize$h$};
\end{tikzpicture}, \begin{tikzpicture}[scale=0.50,baseline = {(0,-0.5)}]
    \draw[thick, red, -<-=.5] (-1.7,1) -- (0,0);
    \draw[thick, red, ->-=.5] (0,0) -- (1.7,1);
    \draw[thick, red, ->-=.5] (0,-2) -- (0,0);
    \node[red, below] at (0,-2) {\scriptsize$\mathcal{A}$};
    \node[red, above] at (-1.7,1) {\scriptsize$\mathcal{A}$};
    \node[red, above] at (1.7,1) {\scriptsize$\mathcal{A}$};
    \node[red, right] at (+0,-0.2) {\scriptsize$\mu^\vee$};
    \filldraw[red] (0,0) circle (2pt);
\end{tikzpicture} = \frac{1}{\sqrt{|G|}} \sum_{g,h\in G} \varphi^\vee(g,h)\begin{tikzpicture}[scale=0.50,baseline = {(0,-0.5)}]
    \draw[thick, black, -<-=.5] (-1.7,1) -- (0,0);
    \draw[thick, black, ->-=.5] (0,0) -- (1.7,1);
    \draw[thick, black, ->-=.5] (0,-2) -- (0,0);
    \node[black, below] at (0,-2) {\scriptsize$gh$};
    \node[black, above] at (-1.7,1) {\scriptsize$g$};
    \node[black, above] at (1.7,1) {\scriptsize$h$};
\end{tikzpicture},
\end{equation}
\end{footnotesize}
where $\varphi^\vee(g,h) = \frac{1}{\varphi(g,h)}$. 
Expanding \eqref{eq:finite_G_gauging_alg} using \eqref{eq:finite_G_alg}, we indeed recover \eqref{eq:finite_G_gauging}:
\begin{equation}
\begin{aligned}
\begin{tikzpicture}[baseline={([yshift= 0 ex]current bounding box.center)},vertex/.style={anchor=base,
    circle,fill=black!25,minimum size=18pt,inner sep=2pt},scale=0.5]
    \filldraw[grey] (-2,-2) rectangle ++(4,4);
    \draw[thick, dgrey] (-2,-2) rectangle ++(4,4);
    \draw[thick, red, -stealth] (0,-2) -- (0.354,-1.354);
    \draw[thick, red] (0,-2) -- (0.707,-0.707);
    \draw[thick, red, -stealth] (2,0) -- (1.354,-0.354);
    \draw[thick, red] (2,0) -- (0.707,-0.707);
    \draw[thick, red, -stealth] (-0.707,0.707) -- (-0.354,1.354);
    \draw[thick, red] (0,2) -- (-0.707,0.707);
    \draw[thick, red, -stealth] (-0.707,0.707) -- (-1.354,0.354);
    \draw[thick, red] (-2,0) -- (-0.707,0.707);
    \draw[thick, red, -stealth] (0.707,-0.707) -- (0,0);
    \draw[thick, red] (0.707,-0.707) -- (-0.707,0.707);
    
    \filldraw[red] (+0.707,-0.707) circle (2pt);
    \filldraw[red] (-0.707,+0.707) circle (2pt);
    
    \node[red, below] at (0,-2) {\scriptsize $\mathcal{A}$};
    \node[red, right] at (2,0) {\scriptsize $\mathcal{A}$};
    \node[red, above] at (0,2) {\scriptsize $\mathcal{A}$};
    \node[red, left] at (-2,0) {\scriptsize $\mathcal{A}$};
    \node[red, above] at (0,0) {\scriptsize $\mathcal{A}$};

    \node[red, below] at (+0.807,-0.707) {\scriptsize $\mu$};
    \node[red, above] at (-0.807,+0.707) {\scriptsize $\mu^\vee$};
    
\end{tikzpicture} &= \frac{1}{|G|} \sum_{g,g',h,h'\in G} \delta_{g,g'} \delta_{h,h'} \delta_{gh,h'g'} \varphi(g,h)\varphi^\vee(h',g')
\begin{tikzpicture}[baseline={([yshift=-.5ex]current bounding box.center)},vertex/.style={anchor=base,
    circle,fill=black!25,minimum size=18pt,inner sep=2pt},scale=0.5]
    \filldraw[grey] (-2,-2) rectangle ++(4,4);
    \draw[thick, dgrey] (-2,-2) rectangle ++(4,4);
    \draw[thick, black, ->-=.5] (0,-2) -- (0.707,-0.707);
    \draw[thick, black, ->-=.5] (2,0) -- (0.707,-0.707);

    \draw[thick, black] (0,2) -- (-0.707,0.707);
    \draw[thick, blue, ->-=.5] (-0.707,0.707) -- (-0.177,1.676);
    
    \draw[thick, black] (-2,0) -- (-0.707,0.707);
    \draw[thick, blue, ->-=.5] (-0.707,0.707) -- (-1.677,0.177);
    
    \draw[thick, black, ->-=.5] (0.707,-0.707) -- (0,0);
    \draw[thick, blue, -<-=.5] (-0.707,0.707)-- (0,0);
    
    \filldraw[black] (-0.177,1.676) circle (2pt);
    \filldraw[black] (0,0) circle (2pt);
    \filldraw[black] (-1.677,0.177) circle (2pt);
    
    \node[black, below] at (0,-2) {\scriptsize $g$};
    \node[black, right] at (2,0) {\scriptsize $h$};
    \node[black, above] at (0,2) {\scriptsize $g$};
    \node[black, left] at (-2,0) {\scriptsize $h$};
    \node[black, above] at (0.7,-0.6) {\scriptsize $gh$};
    \node[blue, above] at (0.15,0.1) {\scriptsize $h'g'$};
    \node[blue, above] at (-1.1,0.3) {\scriptsize $h'$};
    \node[blue, above] at (-0.6,0.8) {\scriptsize $g'$};
    \end{tikzpicture} \\
    & = \frac{1}{|G|}\sum_{\substack{g,h \in G \\ gh = hg}} \frac{\varphi(g,h)}{\varphi(h,g)} \begin{tikzpicture}[baseline={([yshift=-.5ex]current bounding box.center)},vertex/.style={anchor=base,
    circle,fill=black!25,minimum size=18pt,inner sep=2pt},scale=0.5]
    \filldraw[grey] (-2,-2) rectangle ++(4,4);
    \draw[thick, dgrey] (-2,-2) rectangle ++(4,4);
    \draw[thick, black, ->-=.5] (0,-2) -- (0.707,-0.707);
    \draw[thick, black, ->-=.5] (2,0) -- (0.707,-0.707);
    \draw[thick, black, -<-=.5] (0,2) -- (-0.707,0.707);
    \draw[thick, black, -<-=.5] (-2,0) -- (-0.707,0.707);
    \draw[thick, black, ->-=.5] (0.707,-0.707) -- (-0.707,0.707);
    
    \node[black, below] at (0,-2) {\scriptsize $g$};
    \node[black, right] at (2,0) {\scriptsize $h$};
    \node[black, above] at (0,2) {\scriptsize $g$};
    \node[black, left] at (-2,0) {\scriptsize $h$};
    \node[black, above] at (0.2,0) {\scriptsize $gh$};

\end{tikzpicture}.
\end{aligned}
\end{equation}

The partition function \eqref{eq:finite_G_gauging_alg} admits an alternative resolution, similarly to \eqref{eq:finite_G_resolution}, and the different resolutions should agree with each other in order for the orbifold partition function to be well-defined. 
This requires the following condition on the junctions $\mu$, $\mu^\vee$:
\begin{equation}\label{eq:finite_G_alg_res}
    \begin{tikzpicture}[scale=0.50,baseline = {(0,0)}]
    \filldraw[grey] (-2,-2) rectangle ++(4,4);
    \draw[thick, dgrey] (-2,-2) rectangle ++(4,4);
    \draw[thick, red, -stealth] (0,-2) -- (0.354,-1.354);
    \draw[thick, red] (0,-2) -- (0.707,-0.707);
    \draw[thick, red, -stealth] (2,0) -- (1.354,-0.354);
    \draw[thick, red] (2,0) -- (0.707,-0.707);
    \draw[thick, red, -stealth] (-0.707,0.707) -- (-0.354,1.354);
    \draw[thick, red] (0,2) -- (-0.707,0.707);
    \draw[thick, red, -stealth] (-0.707,0.707) -- (-1.354,0.354);
    \draw[thick, red] (-2,0) -- (-0.707,0.707);
    \draw[thick, red, -stealth] (0.707,-0.707) -- (0,0);
    \draw[thick, red] (0.707,-0.707) -- (-0.707,0.707);
    
    \filldraw[red] (+0.707,-0.707) circle (2pt);
    \filldraw[red] (-0.707,+0.707) circle (2pt);
    
    \node[red, below] at (0,-2) {\scriptsize$\mathcal{A}$};
    \node[red, right] at (2,0) {\scriptsize$\mathcal{A}$};
    \node[red, above] at (0,2) {\scriptsize$\mathcal{A}$};
    \node[red, left] at (-2,0) {\scriptsize$\mathcal{A}$};
    \node[red, above] at (0,0) {\scriptsize$\mathcal{A}$};

    \node[red, below] at (+0.807,-0.707) {\scriptsize$\mu$};
    \node[red, above] at (-0.807,+0.707) {\scriptsize$\mu^\vee$};    
\end{tikzpicture} = \begin{tikzpicture}[scale=0.50,baseline = {(0,0)}]
    \filldraw[grey] (-2,-2) rectangle ++(4,4);
    \draw[thick, dgrey] (-2,-2) rectangle ++(4,4);
    \draw[thick, red, ->-=.5] (0,-2) -- (-0.707,-0.707);
    \draw[thick, red, ->-=.5] (2,0) -- (0.707,0.707);
    \draw[thick, red, -<-=.5] (0,2) -- (0.707,0.707);
    \draw[thick, red, -<-=.5] (-2,0) -- (-0.707,-0.707);
    \draw [line, red, ->-=.5] (-0.707,-0.707) -- (0.707,0.707);
    
    \filldraw[red] (-0.707,-0.707) circle (2pt);
    \filldraw[red] (0.707,+0.707) circle (2pt);
    
    \node[red, below] at (0,-2) {\scriptsize$\mathcal{A}$};
    \node[red, right] at (2,0) {\scriptsize$\mathcal{A}$};
    \node[red, above] at (0,2) {\scriptsize$\mathcal{A}$};
    \node[red, left] at (-2,0) {\scriptsize$\mathcal{A}$};
    \node[red, above] at (0,0) {\scriptsize$\mathcal{A}$};

    \node[red, below] at (-0.807,-0.707) {\scriptsize$\mu^\vee$};
    \node[red, above] at (+1.007,+0.607) {\scriptsize$\mu$};    
\end{tikzpicture} \longrightarrow \begin{tikzpicture}[scale=0.50,baseline = {(0,0)}]
    \draw[thick, red, -stealth] (0,-2) -- (0.354,-1.354);
    \draw[thick, red] (0,-2) -- (0.707,-0.707);
    \draw[thick, red, -stealth] (2,0) -- (1.354,-0.354);
    \draw[thick, red] (2,0) -- (0.707,-0.707);
    \draw[thick, red, -stealth] (-0.707,0.707) -- (-0.354,1.354);
    \draw[thick, red] (0,2) -- (-0.707,0.707);
    \draw[thick, red, -stealth] (-0.707,0.707) -- (-1.354,0.354);
    \draw[thick, red] (-2,0) -- (-0.707,0.707);
    \draw[thick, red, -stealth] (0.707,-0.707) -- (0,0);
    \draw[thick, red] (0.707,-0.707) -- (-0.707,0.707);
    
    \filldraw[red] (+0.707,-0.707) circle (2pt);
    \filldraw[red] (-0.707,+0.707) circle (2pt);
    
    \node[red, below] at (0,-2) {\scriptsize$\mathcal{A}$};
    \node[red, right] at (2,0) {\scriptsize$\mathcal{A}$};
    \node[red, above] at (0,2) {\scriptsize$\mathcal{A}$};
    \node[red, left] at (-2,0) {\scriptsize$\mathcal{A}$};
    \node[red, above] at (0,0) {\scriptsize$\mathcal{A}$};

    \node[red, below] at (+0.807,-0.707) {\scriptsize$\mu$};
    \node[red, above] at (-0.807,+0.707) {\scriptsize$\mu^\vee$};    
\end{tikzpicture} = 
\begin{tikzpicture}[scale=0.50,baseline = {(0,0)}]
    \draw[thick, red, ->-=.5] (0,-2) -- (-0.707,-0.707);
    \draw[thick, red, ->-=.5] (2,0) -- (0.707,0.707);
    \draw[thick, red, -<-=.5] (0,2) -- (0.707,0.707);
    \draw[thick, red, -<-=.5] (-2,0) -- (-0.707,-0.707);
    \draw [line, red, ->-=.5] (-0.707,-0.707) -- (0.707,0.707);
    
    \filldraw[red] (-0.707,-0.707) circle (2pt);
    \filldraw[red] (0.707,+0.707) circle (2pt);
    
    \node[red, below] at (0,-2) {\scriptsize$\mathcal{A}$};
    \node[red, right] at (2,0) {\scriptsize$\mathcal{A}$};
    \node[red, above] at (0,2) {\scriptsize$\mathcal{A}$};
    \node[red, left] at (-2,0) {\scriptsize$\mathcal{A}$};
    \node[red, above] at (0,0) {\scriptsize$\mathcal{A}$};

    \node[red, below] at (-0.807,-0.707) {\scriptsize$\mu^\vee$};
    \node[red, above] at (+1.007,+0.607) {\scriptsize$\mu$};    
\end{tikzpicture},
\end{equation}
which requires $\varphi^\vee(g,h) = \frac{1}{\varphi(g,h)}$ for the anomaly free ordinary symmetry. This is called the Frobenius condition.
The multiplication $\mu$ must satisfy the associativity condition (and there is a similar coassociativity condition on $\mu^\vee$):
\begin{equation}\label{eq:fintie_G_alg_constraint}
\begin{aligned}
\begin{tikzpicture}[baseline={([yshift=-.5ex]current bounding box.center)},vertex/.style={anchor=base,
    circle,fill=black!25,minimum size=18pt,inner sep=2pt},scale=0.5]
\draw[red, thick, ->-=0.5] (-1.5,-1.5) -- (-0.5,-0.5);
\draw[red, thick, ->-=0.5] (-0.5,-0.5) -- (0.5,0.5);
\draw[red, thick, ->-=0.5] (0.5,0.5) -- (1.5,1.5);

\draw[red, thick, ->-=0.5] (0.5,-1.5) -- (-0.5,-0.5);

\draw[red, thick, ->-=0.5] (2.5,-1.5) -- (0.5,0.5);

\filldraw[red] (-0.5,-0.5) circle (2pt);
\filldraw[red] (0.5,0.5) circle (2pt);

\node[red, above] at (-0.2,-0.2) {\scriptsize $\mathcal{A}$};

\node[red, below] at (-1.5,-1.5) {\scriptsize $\mathcal{A}$};
\node[red, above] at (1.5,1.5) {\scriptsize $\mathcal{A}$};
\node[red, below] at (2.5,-1.5) {\scriptsize $\mathcal{A}$};
\node[red, below] at (0.5,-1.5) {\scriptsize $\mathcal{A}$};
\node[red, left] at (-0.5,-0.5) {\scriptsize $\mu$};
\node[red, right] at (0.5,0.5) {\scriptsize $\mu$};

\end{tikzpicture} & = \hspace{0.1 in} \begin{tikzpicture}[baseline={([yshift=-.5ex]current bounding box.center)},vertex/.style={anchor=base,
    circle,fill=black!25,minimum size=18pt,inner sep=2pt},scale=0.5]
\draw[red, thick, ->-=0.5] (-1.5,-1.5) -- (0.5,0.5);
\draw[red, thick, ->-=0.5] (0.5,0.5) --  (1.5,1.5);
\draw[red, thick, ->-=0.5] (0.5,-1.5) -- (1.5,-0.5);

\draw[red, thick, ->-=0.5] (2.5,-1.5) -- (1.5,-0.5);
\draw[red, thick, ->-=0.5] (1.5,-0.5) -- (0.5,0.5);

\filldraw[red] (1.5,-0.5) circle (2pt);
\filldraw[red] (0.5,0.5) circle (2pt);

\node[red, left] at (1.8,0.3) {\scriptsize $\mathcal{A}$};

\node[red, below] at (-1.5,-1.5) {\scriptsize $\mathcal{A}$};
\node[red, above] at (1.5,1.5) {\scriptsize $\mathcal{A}$};
\node[red, below] at (2.5,-1.5) {\scriptsize $\mathcal{A}$};
\node[red, below] at (0.5,-1.5) {\scriptsize $\mathcal{A}$};
\node[red, right] at (1.5,-0.5) {\scriptsize $\mu$};
\node[red, left] at (0.5,0.5) {\scriptsize $\mu$};
\end{tikzpicture} \\
\Longleftrightarrow \sum_{g,h,k\in G} \varphi(g,h) \varphi(gh,k) \begin{tikzpicture}[baseline={([yshift=-.5ex]current bounding box.center)},vertex/.style={anchor=base,
    circle,fill=black!25,minimum size=18pt,inner sep=2pt},scale=0.5]
\draw[black, thick, ->-=0.5] (-1.5,-1.5) -- (-0.5,-0.5);
\draw[black, thick, ->-=0.5] (-0.5,-0.5) -- (0.5,0.5);
\draw[black, thick, ->-=0.5] (0.5,0.5) -- (1.5,1.5);

\draw[black, thick, ->-=0.5] (0.5,-1.5) -- (-0.5,-0.5);

\draw[black, thick, ->-=0.5] (2.5,-1.5) -- (0.5,0.5);

\node[black, above] at (-0.2,-0.2) {\scriptsize $gh$};

\node[black, below] at (-1.5,-1.5) {\scriptsize $g$};
\node[black, above] at (1.5,1.5) {\scriptsize $ghk$};
\node[black, below] at (2.5,-1.5) {\scriptsize $k$};
\node[black, below] at (0.5,-1.5) {\scriptsize $h$};

\end{tikzpicture} &  = \hspace{0.1 in} \sum_{g,h,k \in G} \varphi(h,k) \varphi(g,hk)
\begin{tikzpicture}[baseline={([yshift=-.5ex]current bounding box.center)},vertex/.style={anchor=base,
    circle,fill=black!25,minimum size=18pt,inner sep=2pt},scale=0.5]
\draw[black, thick, ->-=0.5] (-1.5,-1.5) -- (0.5,0.5);
\draw[black, thick, ->-=0.5] (0.5,0.5) --  (1.5,1.5);
\draw[black, thick, ->-=0.5] (0.5,-1.5) -- (1.5,-0.5);

\draw[black, thick, ->-=0.5] (2.5,-1.5) -- (1.5,-0.5);
\draw[black, thick, ->-=0.5] (1.5,-0.5) -- (0.5,0.5);

\node[black, left] at (1.8,0.3) {\scriptsize $hk$};

\node[black, below] at (-1.5,-1.5) {\scriptsize $g$};
\node[black, above] at (1.5,1.5) {\scriptsize $ghk$};
\node[black, below] at (2.5,-1.5) {\scriptsize $k$};
\node[black, below] at (0.5,-1.5) {\scriptsize $h$};
\end{tikzpicture}
\end{aligned}
\end{equation}
which in components reads
\begin{equation} \label{eq:trivialization}
    \frac{\varphi(g,h)\varphi(gh,k)}{\varphi(h,k)\varphi(g,hk)} = \omega(g,h,k).
\end{equation}
The equation \eqref{eq:trivialization} can be solved if and only if $\omega$ is cohomologically trivial in $H^3(G,U(1))$, i.e. the symmetry $G$ is anomaly-free. 
When $[\omega]$ is trivial, the inequivalent solutions are classified by $H^2(G,U(1))$ corresponding to different discrete torsions.
Additional conditions required for a consistent gauging of a general algebra object $\CA$ will be reviewed later.

For a CFT $\CT$ with a $\IZ_2$ symmetry generated by a topological line $a$, $H^2(\IZ_2,U(1))$ is trivial, and there is a unique orbifold torus partition function,
\begin{equation}\label{eq:z2orbifold}
    Z_{\CT/\IZ_2} = \frac{1}{2}\left(Z_{\CT}[\dsi,\dsi,\dsi]+Z_{\CT}[\dsi,a,a]+Z_{\CT}[a,\dsi,a]+Z_{\CT}[a,a,\dsi]\right).
\end{equation}
For a $\IZ_2\times \IZ_2$ symmetry, generated by two topological lines $a$ and $b$, $H^2(\IZ_2\times \IZ_2,U(1))=\IZ_2$, and there are two different ways to orbifold (differing by the $\pm$ sign in the last line),
\begin{align}
\begin{split}
   Z_{\CT/\IZ_2 \times \IZ_2} &= \frac{1}{4}( Z_{\CT}[\dsi,\dsi,\dsi]+Z[\dsi,a,a]+Z_{\CT}[\dsi,b,b]+Z_{\CT}[\dsi,ab,ab] \\
   &+Z_{\CT}[a,\dsi,a]+Z_{\CT}[b,\dsi,b]+Z_{\CT}[ab,\dsi,ab]+Z_{\CT}[a,a,\dsi]+Z_{\CT}[b,b,\dsi]+Z_{\CT}[ab,ab,\dsi]  \\
   &\pm Z_{\CT}[a,b,ab]\pm Z_{\CT}[a,ab,b]\pm Z_{\CT}[b,a,ab]\pm Z_{\CT}[b,ab,a]\pm Z_{\CT}[ab,a,b]\pm Z_{\CT}[ab,b,a]) \,.
\end{split}
\end{align}

\subsection{Half-space gauging and non-invertible symmetries}\label{sec:half_space_inv}

A useful way to discover a class of non-invertible, codimension-1 topological defects in general spacetime dimensions is through the ``half-space gauging'' (or ``half-gauging'' in short) 
\cite{Thorngren:2021yso,Choi:2021kmx,Choi:2022zal}.
Consider a 1+1d CFT $\CT$ with a non-anomalous global (abelian) symmetry $G$. We divide the spacetime into left and right regions, and gauge the symmetry $G$ only in the right region. 
At the codimension-1 interface between the left and right, we impose the topological Dirichlet boundary condition for the discrete $G$ gauge field.
If $\CT$ is isomorphic to the gauged theory $\CT/G$, then the interface defines a topological defect line $\CN$ in theory $\CT$:
\begin{equation}
\begin{tikzpicture}[baseline={([yshift=+.5ex]current bounding box.center)},vertex/.style={anchor=base,
    circle,fill=black!25,minimum size=18pt,inner sep=2pt},scale=0.6]
    \filldraw[grey] (-6,-2) rectangle ++(4,4);
    \filldraw[lightblue] (-2,-2) rectangle ++(4,4);
    \draw[line width=0.5mm, red] (-2,-2) -- (-2,+2);
    \draw[thick, dgrey] (-6,-2) -- (+2,-2);
    \draw[thick, dgrey] (+2,+2) -- (-6,+2);
    \node[black] at (-4,0) {$\CT$};
    \node[black] at (0,0) {$\CT/G \cong \CT$};
    \node[red,below] at (-2,-2) {$\CN$};
\end{tikzpicture} \,.
\end{equation}

A well-known example is the Kramers-Wannier duality in the Ising CFT, where the Ising CFT is self-dual under the $\IZ_2$ orbifold \eqnref{eq:z2orbifold}.
The half-gauging in this case produces the Kramers-Wannier duality defect line \cite{Oshikawa:1996ww,Oshikawa:1996dj,Petkova:2000ip,Frohlich:2004ef,Frohlich:2006ch,Frohlich:2009gb,Aasen:2016dop,Aasen:2020jwb,Seiberg:2023cdc}.
With this additional duality line $\CN$ as well as the $\IZ_2$ symmetry, the symmetry of the Ising CFT is described by the Tambara-Yamagami fusion category, $\TY(\IZ_2,\chi,+1)$.
The notation for general Tambara-Yamagami fusion categories is explained momentarily. 

More generally, if the theory is self-dual under gauging an anomaly free abelian symmetry $A$, then it admits a $\TY(A,\chi,\epsilon)$ fusion category symmetry \cite{tambara1998tensor}.\footnote{To be more precise, for there to be a $\TY(A,\chi,\epsilon)$ fusion category symmetry, the theory must be self-dual under the gauging with an appropriately chosen discrete torsion such that the gauging is of order 2 \cite{Choi:2022zal} (see also \cite{Thorngren:2021yso}).}
The simple objects of $\TY(A,\chi,\epsilon)$ category are group-like lines $g$ of the abelian group $A$, and a non-invertible line $\CN$. These simple lines satisfy the following fusion rules,
\begin{equation}\label{eq:TYfusion}
    g \otimes h = gh \,, \quad g\otimes \CN = \CN\otimes g = \CN \,, \quad \CN\otimes \CN = \bigoplus_{g\in A} g \,.
\end{equation}
The only non-trivial $F$-symbols are
\begin{align}\label{eq:TYfsymbols}
\begin{split}
    &[F^{g\CN h}_\CN]_{\CN,\CN} = [F^{\CN g \CN}_h]_{\CN,\CN} = \chi(g,h)\,,\\
    &[F^{\CN\CN \CN}_\CN]_{g,h} = \frac{\epsilon}{\sqrt{\abs{A}}}\chi(g,h)^{-1}\,,
\end{split}
\end{align}
where $\epsilon=\pm 1$ is the Frobenius-Schur indicator for $\CN$, which is classified by $\epsilon\in H^3(\IZ_2,U(1))=\IZ_2$, and $\chi:A\times A\rightarrow U(1)$ is a non-degenerate symmetric bicharacter, which satisfies
\begin{equation}
    \chi(g,h)=\chi(h,g) \,,\quad \chi(gh,k)=\chi(g,k)\chi(h,k) \,,\quad \chi(g,hk)=\chi(g,h)\chi(g,k) \,.
\end{equation}
The fusion rules \eqnref{eq:TYfusion} and the $F$-symbols \eqnref{eq:TYfsymbols} define the fusion category $\TY(A,\chi,\epsilon)$ \cite{tambara1998tensor}.
For the case of $A=\IZ_2$, the choice of $\chi$ is unique, and there are two distinct Tambara-Yamagami fusion categories based on $\IZ_2$ corresponding to $\epsilon = \pm 1$.
We will denote these two categories as $\TY(\IZ_2)_\pm \equiv \TY(\IZ_2,\chi,\pm 1)$.
The $\TY(\IZ_2)_+$ category is realized in the Ising CFT as mentioned above, whereas the $\TY(\IZ_2)_-$ category is realized in the $SU(2)_2$ WZW model, for instance.

Similar to ordinary global symmetries, fusion category symmetries can also have an 't Hooft anomaly, which obstructs the existence of a symmetric trivially gapped phase \cite{Chang:2018iay,Thorngren:2019iar}. 
For instance, the anomaly-free condition for $\TY(A,\chi,\epsilon)$ contains two parts, roughly speaking, (1) the quantum dimension of the duality line should be an integer and (2) the total Frobenius-Schur indicator should be trivial \cite{Thorngren:2019iar,Zhang:2023wlu,Tambara2000TYa,Meir2010TYa,Benini2023TYa}.

\subsection{Group extension of a fusion category}\label{sec:group_extension}
The Tambara-Yamagami fusion category is a special example of a group extended fusion category. 
In general, a fusion category $\mathcal{C}$ is a $G$-extension of the fusion category $\mathcal{C}'$ if $\mathcal{C}$ is a $G$-graded fusion category whose trivial grading component $\mathcal{C}_{\dsi} = \mathcal{C}'$. 
Namely, $\mathcal{C}$ admits a decomposition of abelian categories
\begin{equation}
    \mathcal{C} = \bigoplus_{g\in G}\mathcal{C}_g
\end{equation}
with $\mathcal{C}_{\dsi} = \mathcal{C}'$, such that the tensor product $\otimes$ maps $\mathcal{C}_g\times \mathcal{C}_h$ to $\mathcal{C}_{gh}$ for every $g,h \in G$. 

Consider the case $G= \IZ_2 = \{\dsi, \eta\}$. 
In this langauge, the Tambara-Yamagami fusion category $\TY(A,\chi,\epsilon)$ is a $\doubleZ_2$-extension of the fusion category $\VEC_A \equiv \TY(A,\chi,\epsilon)_{\dsi}$ where the non-trivial grading component $\TY(A,\chi,\epsilon)_{\eta}$ contains a unique simple object $\mathcal{N}$. 
Note that a graded fusion category does not necessarily contain non-invertible symmetries, as one can see from the simplest $G$-graded fusion category $\VEC_G$.

The trivial grading component $\mathcal{C}_{\dsi}$ is a tensor subcategory of $\mathcal{C}$, and each $\mathcal{C}_g$ is a $\mathcal{C}_{\dsi}$-bimodule category. 
This observation leads to the classification of $G$-extensions of a fusion category \cite{2009arXiv0909.3140E} which we briefly review now.
The grading components $\mathcal{C}_g$ satisfy the $G$-multiplication rule under the tensor product $\boxtimes_{\mathcal{C}_{\dsi}}$ of $\mathcal{C}_{\dsi}$-bimodule categories:
\begin{equation}
    \mathcal{C}_g \boxtimes_{\mathcal{C}_{\dsi}} \mathcal{C}_h \simeq \mathcal{C}_{gh} \,, \quad \forall g,h \in G \,.
\end{equation}
In particular, this means that $\mathcal{C}_{\dsi}$ acts as an identity under $\boxtimes_{\mathcal{C}_{\dsi}}$, $\mathcal{C}_{\dsi} \boxtimes_{\mathcal{C}_{\dsi}} \mathcal{C}_g = \mathcal{C}_g \boxtimes_{\mathcal{C}_{\dsi}} \mathcal{C}_{\dsi} = \mathcal{C}_g$, and each $\mathcal{C}_g$ admits an inverse $\mathcal{C}_{g^{-1}}$ under $\boxtimes_{\mathcal{C}_{\dsi}}$.
Namely, $\mathcal{C}_g$ is an invertible $\mathcal{C}_{\dsi}$-bimodule category. 

This implies that the $G$-extension of a fusion category $\mathcal{C}$ contains the data of a group homomorphism $\rho$ from $G$ to the so-called Brauer-Picard group $\BrPic(\mathcal{C})$ of the fusion category $\mathcal{C}$, whose elements are invertible $\mathcal{C}$-bimodule categories and the group multiplication is the tensor product $\boxtimes_{\mathcal{C}}$. 

However, not every $\rho:G\rightarrow \BrPic(\mathcal{C})$ can be made into a $G$-extension of $\mathcal{C}$, and furthermore, the extension associated to a given $\rho$ is not necessarily unique in general. 
The obstructions and additional data required to specify a $G$-extension are worked out in \cite{2009arXiv0909.3140E} and are interpreted physically in \cite{2014arXiv1402.2214B}. 
It starts with an important observation that $\BrPic(\mathcal{C})$ is isomorphic to $\EqBr(\mathcal{Z}(\mathcal{C}))$, the group of braided equivalences of the Drinfeld center $\mathcal{Z}(\mathcal{C})$. 
The latter is simply the symmetry group of the symmetry topological field theory (symTFT) of the fusion category $\mathcal{C}$ \cite{Barkeshli:2014cna}. 
The additional data contain a choice of the symmetry fractionalization class $M \in H^2_{[\rho]}(G,A)$ where $A$ is the group of Abelian anyons in the symTFT $\mathcal{Z}(\mathcal{C})$, and a choice of the discrete torsion $\epsilon \in H^3(G,U(1))$. 
The combined data $(\rho,M,\epsilon)$ must satisfy the conditions that the obstruction class $O^3_{[\rho]} \in H^3_{[\rho]}(G,A)$ to the fractionalization of the symmetry $\rho: G \rightarrow \EqBr(\mathcal{Z}(\mathcal{C}))$ as well as the 't Hooft anomaly $O^4(\rho,M) \in H^4(G,U(1))$ of the $G$ symmetry with the chosen fractionalization both vanish.

We introduce the following notation
\begin{equation}
    \mathcal{E}_{G}^{(\rho,M,\epsilon)} \mathcal{C}
\end{equation}
to denote the $G$-extension of the fusion category $\mathcal{C}$ with the data $(\rho,M,\epsilon)$. 
In this work, we focus on $G = \IZ_2= \{\dsi, \eta\}$ and also on a special class of graded extensions in which the $\eta$-component of the extension contains a unique simple object $\mathcal{D}$, and we will denote this type of extensions by adding an underline $\underline{\mathcal{E}}$. 
Then, the grading structure uniquely determines the fusion rules:
\begin{equation}\label{eq:G_extension_fusion_rule}
    \mathcal{L} \otimes \mathcal{D} = \mathcal{D} \otimes \mathcal{L} = \langle \mathcal{L}\rangle \mathcal{D} \,, \quad \mathcal{D} \otimes \mathcal{D} =  \bigoplus\limits_{\text{simple}\, \mathcal{L}} \langle \mathcal{L}\rangle  \mathcal{L}  \,,
\end{equation}
where $\CL$ lines are in the trivial grading component.
It immediately follows that such an extension can exist only if $\langle \CL \rangle \in \mathbb{Z}_{> 0}$ for every $\CL$.
As an example, the Tambara-Yamagami fusion category can be written as
\begin{equation}
    \TY(A,\chi,\epsilon) = \underline{\mathcal{E}}_{\doubleZ_2}^{(\chi,\epsilon)} \VEC_A \,.
\end{equation}
In the case where the corresponding extension data $(\rho,M,\epsilon)$ are not known explicitly,
we will replace the superscript by suitable labels which distinguish different extensions.

\section{(Half-)gauging non-invertible symmetries in 1+1d} \label{sec:gauging_noninv}
In this section, we first review how to gauge non-invertible symmetries using an algebra object \cite{Fuchs:2002cm,Bhardwaj:2017xup}, generalizing the discussion in Section \ref{sec:invert_sym_gauging}. 
We then argue that given a theory that is self-dual under gauging a non-invertible symmetry, one can obtain a new topological defect line by half-space gauging.
Finally, we work out how to gauge the non-invertible $\Rep(H_8)$ symmetry.

\subsection{Gauging non-invertible symmetries using algebra objects}
As motivated in Section \ref{sec:invert_sym_gauging}, to gauge a (finite) symmetry is to insert a mesh of the corresponding algebra object $\mathcal{A}$ across the spacetime manifold.
To begin with, an algebra object is characterized by a triple $(\CA,\mu,u)$.
Here, $\mu \in \Hom_{\mathcal{C}}(\mathcal{A}\otimes \mathcal{A}, \mathcal{A})$ is a fusion junction of $\mathcal{A}$, also known as the multiplication morphism, and $u \in \Hom_{\mathcal{C}}(\dsi,\mathcal{A})$ is the unit morphism. 
They satisfy the following consistent conditions:
\begin{equation} \label{eq:alg}
\begin{tikzpicture}[baseline={([yshift=-.5ex]current bounding box.center)},vertex/.style={anchor=base,
    circle,fill=black!25,minimum size=18pt,inner sep=2pt},scale=0.5]
\draw[red, thick, ->-=0.5] (-1.5,-1.5) -- (-0.5,-0.5);
\draw[red, thick, ->-=0.5] (-0.5,-0.5) -- (0.5,0.5);
\draw[red, thick, ->-=0.5] (0.5,0.5) -- (1.5,1.5);

\draw[red, thick, ->-=0.5] (0.5,-1.5) -- (-0.5,-0.5);

\draw[red, thick, ->-=0.5] (2.5,-1.5) -- (0.5,0.5);

\filldraw[red] (-0.5,-0.5) circle (3pt);
\filldraw[red] (0.5,0.5) circle (3pt);

\node[red, above] at (-0.2,-0.2) {\scriptsize $\mathcal{A}$};

\node[red, below] at (-1.5,-1.5) {\scriptsize $\mathcal{A}$};
\node[red, above] at (1.5,1.5) {\scriptsize $\mathcal{A}$};
\node[red, below] at (2.5,-1.5) {\scriptsize $\mathcal{A}$};
\node[red, below] at (0.5,-1.5) {\scriptsize $\mathcal{A}$};
\node[red, left] at (-0.5,-0.5) {\scriptsize $\mu$};
\node[red, right] at (0.5,0.5) {\scriptsize $\mu$};

\end{tikzpicture} = \begin{tikzpicture}[baseline={([yshift=-.5ex]current bounding box.center)},vertex/.style={anchor=base,
    circle,fill=black!25,minimum size=18pt,inner sep=2pt},scale=0.5]
\draw[red, thick, ->-=0.5] (-1.5,-1.5) -- (0.5,0.5);
\draw[red, thick, ->-=0.5] (0.5,0.5) --  (1.5,1.5);
\draw[red, thick, ->-=0.5] (0.5,-1.5) -- (1.5,-0.5);

\draw[red, thick, ->-=0.5] (2.5,-1.5) -- (1.5,-0.5);
\draw[red, thick, ->-=0.5] (1.5,-0.5) -- (0.5,0.5);

\filldraw[red] (1.5,-0.5) circle (3pt);
\filldraw[red] (0.5,0.5) circle (3pt);

\node[red, left] at (1.8,0.3) {\scriptsize $\mathcal{A}$};

\node[red, below] at (-1.5,-1.5) {\scriptsize $\mathcal{A}$};
\node[red, above] at (1.5,1.5) {\scriptsize $\mathcal{A}$};
\node[red, below] at (2.5,-1.5) {\scriptsize $\mathcal{A}$};
\node[red, below] at (0.5,-1.5) {\scriptsize $\mathcal{A}$};
\node[red, right] at (1.5,-0.5) {\scriptsize $\mu$};
\node[red, left] at (0.5,0.5) {\scriptsize $\mu$};
\end{tikzpicture}, \quad \begin{tikzpicture}[baseline={([yshift=-.5ex]current bounding box.center)},vertex/.style={anchor=base,
    circle,fill=black!25,minimum size=18pt,inner sep=2pt},scale=0.5]
    \draw[red, thick, ->-=0.5] (0,-2) -- (0,0);
    \draw[red, thick, ->-=0.5] (0,0) -- (0,+2);
    \draw[red, thick, ->-=0.5] (-1.5,-1.5) -- (0,0);
    \filldraw[red] (0,0) circle (3pt);
    \filldraw[red] (-1.5,-1.5) circle (3pt);
    \node[red, below] at (0,-2) {\scriptsize $\mathcal{A}$};
    \node[red, above] at (0,+2) {\scriptsize $\mathcal{A}$};
    \node[red, above] at (-0.75,-0.75) {\scriptsize $\mathcal{A}$};
    \node[red, below] at (-1.5,-1.5) {\scriptsize $u$};
    \node[red, right] at (0,0) {\scriptsize $\mu$};
\end{tikzpicture} = \begin{tikzpicture}[baseline={([yshift=-.5ex]current bounding box.center)},vertex/.style={anchor=base,
    circle,fill=black!25,minimum size=18pt,inner sep=2pt},scale=0.5]
    \draw[red, thick, ->-=0.5] (0,-2) -- (0,0);
    \draw[red, thick, ->-=0.5] (0,0) -- (0,+2);
    \draw[red, thick, ->-=0.5] (+1.5,-1.5) -- (0,0);
    \filldraw[red] (0,0) circle (3pt);
    \filldraw[red] (+1.5,-1.5) circle (3pt);
    \node[red, below] at (0,-2) {\scriptsize $\mathcal{A}$};
    \node[red, above] at (0,+2) {\scriptsize $\mathcal{A}$};
    \node[red, above] at (+0.75,-0.75) {\scriptsize $\mathcal{A}$};
    \node[red, below] at (+1.5,-1.5) {\scriptsize $u$};
    \node[red, left] at (0,0) {\scriptsize $\mu$};
\end{tikzpicture} = \begin{tikzpicture}[baseline={([yshift=-.5ex]current bounding box.center)},vertex/.style={anchor=base,
    circle,fill=black!25,minimum size=18pt,inner sep=2pt},scale=0.5]
    \draw[red, thick, ->-=0.5] (0,-2) -- (0,2);
    \node[red, below] at (0,-2) {\scriptsize $\mathcal{A}$};
    \node[red, above] at (0,+2) {\scriptsize $\mathcal{A}$};
\end{tikzpicture}.
\end{equation}
Throughout the paper, we focus on the case where $\mathcal{A}$ contains every simple object $\mathcal{L}_i$ of $\CC$ with multiplicities given by the quantum dimensions,
\begin{equation}\label{eq:algebra_expansion}
    \mathcal{A} = \bigoplus_{i} \langle \mathcal{L}_i\rangle  \mathcal{L}_i \,.
\end{equation}
Algebra objects $(\mathcal{A},\mu,u)$ of this form are in 1-to-1 correspondence with fiber functors of the fusion category \cite{ostrik2003module,Choi:2023xjw}. An algebra object $\mathcal{A}$ of the form \eqref{eq:algebra_expansion} is an example of a haploid algebra object, that is, the multiplicity of $\dsi$ in $\mathcal{A}$ is $1$. 
This fixes the unit morphism $u$ up to rescaling. 

To insert a mesh of $\mathcal{A}$ on a Riemann surface which implements the gauging, we first choose a triangulation, and insert $\mathcal{A}$ along the edges of the dual triangulation.
To do this, we need not only the fusion junction $\mu$ but also a splitting junction $\mu^\vee \in \Hom_{\mathcal{C}}(\mathcal{A}, \mathcal{A}\otimes \mathcal{A})$. Furthermore, in order for the gauging to be unambiguously defined, the result must be invariant under changing the triangulation of the Riemann surface. 
Any two triangulations of a Riemann surface can be transformed into each other via a sequence of the \textit{fusion} and \textit{bubble} moves \cite{chung1994tri,fukuma1994tri,karimipour1997tri},
\begin{equation}
\begin{tikzpicture}[scale=0.7, baseline={([yshift=-.5ex]current bounding box.center)},vertex/.style={anchor=base,
    circle,fill=black!25,minimum size=18pt,inner sep=2pt},scale=0.5]
\draw[black, thick] (-2,0) -- (2,0);
\draw[black, thick] (+2,0) -- (0,+3.464);
\draw[black, thick] (-2,0) -- (0,+3.464);   
\draw[black, thick] (-2,0) -- (0,-3.464);
\draw[black, thick] (+2,0) -- (0,-3.464);
\draw[red, thick] (0,-1.155) -- (0,+1.155);
\draw[red, thick] (2,+2.309) -- (0,+1.155);
\draw[red, thick] (2,-2.309) -- (0,-1.155);
\draw[red, thick] (-2,-2.309) -- (0,-1.155);
\draw[red, thick] (-2,+2.309) -- (0,+1.155);
\filldraw[black] (-2,0) circle(3pt);
\filldraw[black] (+2,0) circle(3pt);
\filldraw[black] (0,+3.464) circle(3pt);
\filldraw[black] (0,-3.464) circle(3pt);
\filldraw[red] (0,+1.155) circle (3pt);
\filldraw[red] (0,-1.155) circle (3pt);
\end{tikzpicture}  \xleftrightarrow{\,\, \textit{fusion} \,\,}  \begin{tikzpicture}[scale=0.7, baseline={([yshift=-.5ex]current bounding box.center)},vertex/.style={anchor=base,
    circle,fill=black!25,minimum size=18pt,inner sep=2pt},scale=0.5]
\draw[black, thick] (0,-3.464) -- (0,+3.464);
\draw[black, thick] (+2,0) -- (0,+3.464);
\draw[black, thick] (-2,0) -- (0,+3.464);   
\draw[black, thick] (-2,0) -- (0,-3.464);
\draw[black, thick] (+2,0) -- (0,-3.464);
\draw[red, thick] (-1,0) -- (+1,0);
\draw[red, thick] (2,+2.309) -- (+1,0);
\draw[red, thick] (2,-2.309) -- (+1,0);
\draw[red, thick] (-2,-2.309) -- (-1,0);
\draw[red, thick] (-2,+2.309) -- (-1,0);
\filldraw[black] (-2,0) circle(3pt);
\filldraw[black] (+2,0) circle(3pt);
\filldraw[black] (0,+3.464) circle(3pt);
\filldraw[black] (0,-3.464) circle(3pt);
\filldraw[red] (-1,0) circle (3pt);
\filldraw[red] (+1,0) circle (3pt);
\end{tikzpicture}\quad , \quad \begin{tikzpicture}[scale=0.7, baseline={([yshift=-.5ex]current bounding box.center)},vertex/.style={anchor=base,
    circle,fill=black!25,minimum size=18pt,inner sep=2pt},scale=0.5]
\draw[black, thick] (0,0) ellipse (2.5 and 2);
\draw[black, thick] (-2.5,0) -- (2.5,0);
\filldraw[black] (0,0) circle (3pt);
\filldraw[black] (-2.5,0) circle (3pt);
\filldraw[black] (+2.5,0) circle (3pt);
\draw[red, thick] (0,0) circle (1.5);
\draw[red, thick] (0,1.5) -- (0,4);
\draw[red, thick] (0,-1.5) -- (0,-4);
\filldraw[red] (0,+1.5) circle (3pt); 
\filldraw[red] (0,-1.5) circle (3pt);
\end{tikzpicture} \xleftrightarrow{\,\, \textit{bubble} \,\,} \begin{tikzpicture}[scale=0.7, baseline={([yshift=-.5ex]current bounding box.center)},vertex/.style={anchor=base,
    circle,fill=black!25,minimum size=18pt,inner sep=2pt},scale=0.5]
\draw[black, thick] (-2.5,0) -- (2.5,0);
\filldraw[black] (-2.5,0) circle (3pt);
\filldraw[black] (+2.5,0) circle (3pt);
\draw[red, thick] (0,-4) -- (0,4);
\end{tikzpicture}\quad .
\end{equation}
This leads to various additional constraints that the algebra object $\mathcal{A}$ and the junctions $\mu$, $\mu^\vee$ must satisfy, including \eqref{eq:finite_G_alg_res} and \eqref{eq:fintie_G_alg_constraint}.
Mathematically, these conditions are summarized by saying that $\CA$ must be a symmetric $\Delta$-separable Frobenius algebra object in the fusion category $\mathcal{C}$ \cite{Fuchs:2002cm,Bhardwaj:2017xup}. 
Some of them are listed below:
\begin{equation}\label{eq:algebracond}
\begin{tikzpicture}[baseline={([yshift=-.5ex]current bounding box.center)},vertex/.style={anchor=base,
    circle,fill=black!25,minimum size=18pt,inner sep=2pt},scale=0.5]
\draw[red, thick, ->-=0.5] (-1.5,-1.5) -- (-0.5,-0.5);
\draw[red, thick, ->-=0.5] (-0.5,-0.5) -- (0.5,0.5);
\draw[red, thick, ->-=0.5] (0.5,0.5) -- (1.5,1.5);

\draw[red, thick, ->-=0.5] (0.5,-1.5) -- (-0.5,-0.5);

\draw[red, thick, ->-=0.5] (2.5,-1.5) -- (0.5,0.5);

\filldraw[red] (-0.5,-0.5) circle (3pt);
\filldraw[red] (0.5,0.5) circle (3pt);

\node[red, above] at (-0.2,-0.2) {\scriptsize $\mathcal{A}$};

\node[red, below] at (-1.5,-1.5) {\scriptsize $\mathcal{A}$};
\node[red, above] at (1.5,1.5) {\scriptsize $\mathcal{A}$};
\node[red, below] at (2.5,-1.5) {\scriptsize $\mathcal{A}$};
\node[red, below] at (0.5,-1.5) {\scriptsize $\mathcal{A}$};
\node[red, left] at (-0.5,-0.5) {\scriptsize $\mu$};
\node[red, right] at (0.5,0.5) {\scriptsize $\mu$};

\end{tikzpicture} = \begin{tikzpicture}[baseline={([yshift=-.5ex]current bounding box.center)},vertex/.style={anchor=base,
    circle,fill=black!25,minimum size=18pt,inner sep=2pt},scale=0.5]
\draw[red, thick, ->-=0.5] (-1.5,-1.5) -- (0.5,0.5);
\draw[red, thick, ->-=0.5] (0.5,0.5) --  (1.5,1.5);
\draw[red, thick, ->-=0.5] (0.5,-1.5) -- (1.5,-0.5);

\draw[red, thick, ->-=0.5] (2.5,-1.5) -- (1.5,-0.5);
\draw[red, thick, ->-=0.5] (1.5,-0.5) -- (0.5,0.5);

\filldraw[red] (1.5,-0.5) circle (3pt);
\filldraw[red] (0.5,0.5) circle (3pt);

\node[red, left] at (1.8,0.3) {\scriptsize $\mathcal{A}$};

\node[red, below] at (-1.5,-1.5) {\scriptsize $\mathcal{A}$};
\node[red, above] at (1.5,1.5) {\scriptsize $\mathcal{A}$};
\node[red, below] at (2.5,-1.5) {\scriptsize $\mathcal{A}$};
\node[red, below] at (0.5,-1.5) {\scriptsize $\mathcal{A}$};
\node[red, right] at (1.5,-0.5) {\scriptsize $\mu$};
\node[red, left] at (0.5,0.5) {\scriptsize $\mu$};
\end{tikzpicture}, \quad \begin{tikzpicture}[baseline={([yshift=-.5ex]current bounding box.center)},vertex/.style={anchor=base,
    circle,fill=black!25,minimum size=18pt,inner sep=2pt},scale=-0.5]
\draw[red, thick, -<-=.5] (-1.5,-1.5) -- (0.5,0.5);
\draw[red, thick, -<-=.5] (0.5,0.5) -- (1.5,1.5);
\draw[red, thick, -<-=.5] (0.5,-1.5) -- (1.5,-0.5);
\draw[red, thick, -<-=.5] (2.5,-1.5)  -- (1.5,-0.5);
\draw[red, thick, -<-=.5] (1.5,-0.5) -- (0.5,0.5);

\filldraw[red] (1.5,-0.5) circle (3pt);
\filldraw[red] (0.5,0.5) circle (3pt);

\node[red, right] at (1.8,0.3) {\scriptsize $\mathcal{A}$};

\node[red, above] at (-1.5,-1.5) {\scriptsize $\mathcal{A}$};
\node[red, below] at (1.5,1.5) {\scriptsize $\mathcal{A}$};
\node[red, above] at (2.5,-1.5) {\scriptsize $\mathcal{A}$};
\node[red, above] at (0.5,-1.5) {\scriptsize $\mathcal{A}$};
\node[red, left] at (1.3,-0.4) {\scriptsize $\mu^\vee$};
\node[red, right] at (0.5,0.5) {\scriptsize $\mu^\vee$};
\end{tikzpicture} = \hspace{0.1 in} \begin{tikzpicture}[baseline={([yshift=-.5ex]current bounding box.center)},vertex/.style={anchor=base,
    circle,fill=black!25,minimum size=18pt,inner sep=2pt},scale=-0.5]
\draw[red, thick, -<-=.5] (-1.5,-1.5) -- (-0.5,-0.5);
\draw[red, thick, -<-=.5] (-0.5,-0.5)  -- (0.5,0.5);
\draw[red, thick, -<-=.5] (0.5,0.5) -- (1.5,1.5);
\draw[red, thick, -<-=.5] (0.5,-1.5) -- (-0.5,-0.5);
\draw[red, thick, -<-=.5] (2.5,-1.5) -- (0.5,0.5);

\filldraw[red] (-0.5,-0.5) circle (3pt);
\filldraw[red] (0.5,0.5) circle (3pt);

\node[red, below] at (-0.2,-0.2) {\scriptsize $\mathcal{A}$};

\node[red, above] at (-1.5,-1.5) {\scriptsize $\mathcal{A}$};
\node[red, below] at (1.5,1.5) {\scriptsize $\mathcal{A}$};
\node[red, above] at (2.5,-1.5) {\scriptsize $\mathcal{A}$};
\node[red, above] at (0.5,-1.5) {\scriptsize $\mathcal{A}$};
\node[red, right] at (-0.5,-0.5) {\scriptsize $\mu^{\vee}$};
\node[red, left] at (0.5,0.5) {\scriptsize $\mu^{\vee}$};
\end{tikzpicture},\quad \begin{tikzpicture}[scale=0.50,baseline = {(0,0)}]
    \draw[thick, red, ->-=.5] (0,-2) -- (0,-0.7);
    \draw[thick, red, ->-=.5] (0,0.7) -- (0,2);
    \draw[thick, red] (0,0) circle [radius=0.7cm];
    \draw[thick,red, ->-=1.0] (-0.7,0.0) -- (-0.7, 0.1);
    \draw[thick,red, ->-=1.0] (0.7,0.0) -- (0.7, 0.1);
    \filldraw[red] (0,0.7) circle (3pt);
    \filldraw[red] (0,-0.7) circle (3pt);
    \node[red, below] at (0,-2) {\scriptsize$\mathcal{A}$};
    \node[red, above] at (0,2) {\scriptsize$\mathcal{A}$};
    \node[red, right] at (0.7,0) {\scriptsize$\mathcal{A}$};
    \node[red, right] at (0,-1.0) {\scriptsize $\mu^\vee$};
    \node[red, right] at (0,+1.0) {\scriptsize $\mu$};
\end{tikzpicture} = \,\,   \begin{tikzpicture}[scale=0.50,baseline = {(0,0)}]
    \draw[thick, red, ->-=.5] (0,-2) -- (0,2);
    \node[red, right] at (0.,0) {\scriptsize$\mathcal{A}$};
\end{tikzpicture}.
\end{equation}
In practice, solving the above set of conditions is enough to guarantee that one has a symmetric $\Delta$-separable Frobenius algebra object, due to the fact that every semisimple haploid algebra object admits a unique coalgebra structure (namely $\mu^\vee$, the counit is unique up to rescaling) such that it becomes a symmetric $\Delta$-separable Frobenius algebra \cite{ostrik2003module,Fuchs:2002cm}.

When we gauge a finite group symmetry, we obtain a quantum (or dual) symmetry in the gauged theory \cite{Vafa:1989ih}.
Similarly, when we gauge a general algebra object $\CA$, we also get a quantum symmetry in the gauged theory \cite{Bhardwaj:2017xup,Komargodski:2020mxz}.
The quantum symmetry one gets after gauging $\mathcal{A}$ is described by the fusion category ${}_{\mathcal{A}}\mathcal{C}_{\mathcal{A}}$, that is, the category of $\mathcal{A}$-$\mathcal{A}$-bimodule objects $(\mathcal{M},\lambda,\rho)$ in $\mathcal{C}$. 
Here, $\lambda$ and $\rho$ describe left/right multiplication structures of $\mathcal{A}$ on $\CM$,
\begin{equation}
\begin{tikzpicture}[baseline={([yshift=-.5ex]current bounding box.center)},vertex/.style={anchor=base,
    circle,fill=black!25,minimum size=18pt,inner sep=2pt},scale=0.75]
\draw[red, thick, ->-=.5] (-1.5,-1) -- (-0.5,0);
\draw[dgreen, thick, ->-=.5] (-0.5,0) -- (0.5,1.0);
\draw[dgreen, thick, ->-=.5] (0.5,-1.0) -- (-0.5,0);

\node[red, below] at (-1.5,-1.5+0.5) {\footnotesize $\mathcal{A}$};
\node[dgreen, above] at (0.5,0.5+0.5) {\footnotesize $\mathcal{M}$};
\node[dgreen, below] at (0.5,-1.5+0.5) {\footnotesize $\mathcal{M}$};

\filldraw[dgreen] (-0.5,-0.5+0.5) circle (1.5pt);

\node[dgreen, left] at (-0.5,-0.5+0.5) {\footnotesize $\lambda$};
\end{tikzpicture}, \quad \begin{tikzpicture}[baseline={([yshift=-.5ex]current bounding box.center)},vertex/.style={anchor=base,
    circle,fill=black!25,minimum size=18pt,inner sep=2pt},scale=0.75]
\draw[dgreen, thick, ->-=.5] (-1.5,-1) -- (-0.5,0);
\draw[dgreen, thick, ->-=.5] (-0.5,0) -- (0.5,1.0);
\draw[red, thick, ->-=.5] (0.5,-1.0) -- (-0.5,0);

\node[dgreen, below] at (-1.5,-1.5+0.5) {\footnotesize $\mathcal{M}$};
\node[dgreen, above] at (0.5,0.5+0.5) {\footnotesize $\mathcal{M}$};
\node[red, below] at (0.5,-1.5+0.5) {\footnotesize $\mathcal{A}$};

\filldraw[dgreen] (-0.5,-0.5+0.5) circle (1.5pt);

\node[dgreen, left] at (-0.5,-0.5+0.5) {\footnotesize $\rho$};
\end{tikzpicture}\, ,
\end{equation}
satisfying
\begin{equation}\label{eq:bimodule_condition}
\begin{tikzpicture}[baseline={([yshift=-.5ex]current bounding box.center)},vertex/.style={anchor=base,
    circle,fill=black!25,minimum size=18pt,inner sep=2pt},scale=0.4]
\draw[red, thick, ->-=.5] (-1.5,-1.5) -- (-0.5,-0.5);
\draw[red, thick, ->-=.5] (-0.5,-0.5) -- (0.5,0.5);
\draw[dgreen, thick, ->-=.5] (0.5,0.5) -- (1.5,1.5);
\draw[red, thick, ->-=.5] (0.5,-1.5) -- (-0.5,-0.5);
\draw[dgreen, thick, ->-=.5] (2.5,-1.5) -- (0.5,0.5);

\filldraw[red] (-0.5,-0.5) circle (3pt);
\filldraw[dgreen] (0.5,0.5) circle (3pt);

\node[red, below] at (0.3,0.3) {\scriptsize $\mathcal{A}$};

\node[red, below] at (-1.5,-1.5) {\scriptsize $\mathcal{A}$};
\node[dgreen, above] at (1.5,1.5) {\scriptsize $\mathcal{M}$};
\node[dgreen, below] at (2.5,-1.5) {\scriptsize $\mathcal{M}$};
\node[red, below] at (0.5,-1.5) {\scriptsize $\mathcal{A}$};
\node[red, left] at (-0.3,-0.3) {\scriptsize $\mu$};
\node[dgreen, left] at (0.7,0.7) {\scriptsize $\lambda$};
\end{tikzpicture} = \begin{tikzpicture}[baseline={([yshift=-.5ex]current bounding box.center)},vertex/.style={anchor=base,
    circle,fill=black!25,minimum size=18pt,inner sep=2pt},scale=0.4]
\draw[red, thick, ->-=.5] (-1.5,-1.5) -- (0.5,0.5);
\draw[dgreen, thick, ->-=.5] (0.5,0.5) -- (1.5,1.5);
\draw[red, thick, ->-=.5] (0.5,-1.5) -- (1.5,-0.5);
\draw[dgreen, thick, ->-=.5] (2.5,-1.5) -- (1.5,-0.5);
\draw[dgreen, thick, ->-=.5] (1.5,-0.5) -- (0.5,0.5);

\filldraw[dgreen] (1.5,-0.5) circle (3pt);
\filldraw[dgreen] (0.5,0.5) circle (3pt);

\node[dgreen, below] at (0.6,0.3) {\scriptsize $\mathcal{M}$};

\node[red, below] at (-1.5,-1.5) {\scriptsize $\mathcal{A}$};
\node[dgreen, above] at (1.5,1.5) {\scriptsize $\mathcal{M}$};
\node[dgreen, below] at (2.5,-1.5) {\scriptsize $\mathcal{M}$};
\node[red, below] at (0.5,-1.5) {\scriptsize $\mathcal{A}$};
\node[dgreen, right] at (1.3,-0.3) {\scriptsize $\lambda$};
\node[dgreen, left] at (0.7,0.7) {\scriptsize $\lambda$};
\end{tikzpicture}, \begin{tikzpicture}[baseline={([yshift=-.5ex]current bounding box.center)},vertex/.style={anchor=base,
    circle,fill=black!25,minimum size=18pt,inner sep=2pt},scale=0.4]
\draw[dgreen, thick, ->-=.5] (-1.5,-1.5) -- (-0.5,-0.5);
\draw[dgreen, thick, ->-=.5] (-0.5,-0.5) -- (0.5,0.5);
\draw[dgreen, thick, ->-=.5] (0.5,0.5) -- (1.5,1.5);
\draw[red, thick, ->-=.5] (0.5,-1.5) -- (-0.5,-0.5);
\draw[red, thick, ->-=.5] (2.5,-1.5) -- (0.5,0.5);

\filldraw[dgreen] (-0.5,-0.5) circle (3pt);
\filldraw[dgreen] (0.5,0.5) circle (3pt);

\node[dgreen, below] at (0.3,0.3) {\scriptsize $\mathcal{M}$};

\node[dgreen, below] at (-1.5,-1.5) {\scriptsize $\mathcal{M}$};
\node[dgreen, above] at (1.5,1.5) {\scriptsize $\mathcal{M}$};
\node[red, below] at (2.5,-1.5) {\scriptsize $\mathcal{A}$};
\node[red, below] at (0.5,-1.5) {\scriptsize $\mathcal{A}$};
\node[dgreen, left] at (-0.3,-0.3) {\scriptsize $\rho$};
\node[dgreen, left] at (0.7,0.7) {\scriptsize $\rho$};
\end{tikzpicture} = \begin{tikzpicture}[baseline={([yshift=-.5ex]current bounding box.center)},vertex/.style={anchor=base,
    circle,fill=black!25,minimum size=18pt,inner sep=2pt},scale=0.4]
\draw[dgreen, thick, ->-=.5] (-1.5,-1.5) -- (0.5,0.5);
\draw[dgreen, thick, ->-=.5] (0.5,0.5) -- (1.5,1.5);
\draw[red, thick, ->-=.5] (0.5,-1.5) -- (1.5,-0.5);
\draw[red, thick, ->-=.5] (2.5,-1.5) -- (1.5,-0.5);
\draw[red, thick, ->-=.5] (1.5,-0.5) -- (0.5,0.5);

\filldraw[red] (1.5,-0.5) circle (3pt);
\filldraw[dgreen] (0.5,0.5) circle (3pt);

\node[red, below] at (0.6,0.3) {\scriptsize $\mathcal{A}$};

\node[dgreen, below] at (-1.5,-1.5) {\scriptsize $\mathcal{M}$};
\node[dgreen, above] at (1.5,1.5) {\scriptsize $\mathcal{M}$};
\node[red, below] at (2.5,-1.5) {\scriptsize $\mathcal{A}$};
\node[red, below] at (0.5,-1.5) {\scriptsize $\mathcal{A}$};
\node[red, right] at (1.3,-0.3) {\scriptsize $\mu$};
\node[dgreen, left] at (0.7,0.7) {\scriptsize $\rho$};
\end{tikzpicture}, \begin{tikzpicture}[baseline={([yshift=-.5ex]current bounding box.center)},vertex/.style={anchor=base,
    circle,fill=black!25,minimum size=18pt,inner sep=2pt},scale=0.4]
\draw[red, thick, ->-=.5] (-1.5,-1.5) -- (-0.5,-0.5);
\draw[dgreen, thick, ->-=.5] (-0.5,-0.5) -- (0.5,0.5);
\draw[dgreen, thick, ->-=.5] (0.5,0.5) -- (1.5,1.5);
\draw[dgreen, thick, ->-=.5] (0.5,-1.5) -- (-0.5,-0.5);
\draw[red, thick, ->-=.5] (2.5,-1.5) -- (0.5,0.5);

\filldraw[dgreen] (-0.5,-0.5) circle (3pt);
\filldraw[dgreen] (0.5,0.5) circle (3pt);

\node[dgreen, below] at (0.3,0.3) {\scriptsize $\mathcal{M}$};

\node[red, below] at (-1.5,-1.5) {\scriptsize $\mathcal{A}$};
\node[dgreen, above] at (1.5,1.5) {\scriptsize $\mathcal{M}$};
\node[red, below] at (2.5,-1.5) {\scriptsize $\mathcal{A}$};
\node[dgreen, below] at (0.5,-1.5) {\scriptsize $\mathcal{M}$};
\node[dgreen, left] at (-0.3,-0.3) {\scriptsize $\lambda$};
\node[dgreen, left] at (0.7,0.7) {\scriptsize $\rho$};
\end{tikzpicture} = \begin{tikzpicture}[baseline={([yshift=-.5ex]current bounding box.center)},vertex/.style={anchor=base,
    circle,fill=black!25,minimum size=18pt,inner sep=2pt},scale=0.4]
\draw[red, thick, -stealth] (-1.5,-1.5) -- (-0.5,-0.5);
\draw[red, thick] (-0.5,-0.5) -- (0.5,0.5);
\draw[dgreen, thick, -stealth] (0.5,0.5) -- (1.0,1.0);
\draw[dgreen, thick] (1.0,1.0) -- (1.5,1.5);

\draw[dgreen, thick, -stealth] (0.5,-1.5) -- (1.0,-1.0);
\draw[dgreen, thick] (1.0,-1.0) -- (1.5,-0.5);

\draw[red, thick, -stealth] (2.5,-1.5) -- (2,-1.0);
\draw[red, thick] (2,-1.0) -- (1.5,-0.5);
\draw[dgreen, thick, -stealth] (1.5,-0.5) -- (1.0,0);
\draw[dgreen, thick] (1.0,0) -- (0.5,0.5);

\filldraw[dgreen] (1.5,-0.5) circle (3pt);
\filldraw[dgreen] (0.5,0.5) circle (3pt);

\node[dgreen, below] at (0.6,0.3) {\scriptsize $\mathcal{M}$};

\node[red, below] at (-1.5,-1.5) {\scriptsize $\mathcal{A}$};
\node[dgreen, above] at (1.5,1.5) {\scriptsize $\mathcal{M}$};
\node[red, below] at (2.5,-1.5) {\scriptsize $\mathcal{A}$};
\node[dgreen, below] at (0.5,-1.5) {\scriptsize $\mathcal{M}$};
\node[dgreen, right] at (1.3,-0.3) {\scriptsize $\rho$};
\node[dgreen, left] at (0.7,0.7) {\scriptsize $\lambda$};
\end{tikzpicture}.
\end{equation}
Generally, a bimodule object can be decomposed as a direct sum of several other bimodule objects. The ones which can not be decomposed are called indecomposable bimodule objects, and they are the simple topological defect lines generating the quantum symmetry in the gauged theory. 
The intuition is that the conditions in \eqref{eq:bimodule_condition} allow the line $\CM$ to be inserted and deformed across the mesh of $\CA$ in a consistent way.

\subsection{Half-gauging and graded extensions}

Here, we provide a description of half-gauging non-invertible symmetries, generalizing the discussion in Section \ref{sec:half_space_inv}.
We begin by describing the topological interface obtained from gauging an algebra object $\mathcal{A}$ on half of the spacetime. 
Let us consider a CFT $\mathcal{T}$ with a fusion category symmetry $\mathcal{C}$. 
If $\mathcal{C}$ is anomaly-free, then we can find an algebra object $\mathcal{A}$ of the form \eqref{eq:algebra_expansion}. 
Gauging the algebra object $\mathcal{A}$ leads to the CFT $\mathcal{T}/\mathcal{C}$ with the quantum fusion category symmetry ${}_{\mathcal{A}}\mathcal{C}_{\mathcal{A}}$. 
It is then possible to consider gauging the symmetry $\mathcal{C}$ on only half of the spacetime to obtain the theory $\mathcal{T}/\mathcal{C}$ on half-space which is separated from $\mathcal{T}$ by a topological interface $\mathcal{I}$,
\begin{equation}
\begin{tikzpicture}[baseline={([yshift=+.5ex]current bounding box.center)},vertex/.style={anchor=base,
    circle,fill=black!25,minimum size=18pt,inner sep=2pt},scale=0.6]
    \filldraw[grey] (-8,-2) rectangle ++(6,4);
    \filldraw[lightblue] (-2,-2) rectangle ++(6,4);
    \draw[line width=0.5mm, red] (-2,-2) -- (-2,+2);
    \draw[thick, dgrey] (-8,-2) -- (+4,-2);
    \draw[thick, dgrey] (+4,+2) -- (-8,+2);
    \node[black] at (-5,0) {\footnotesize $\CT$ with $\mathcal{C}$-sym};
    \node[black] at (1,0) {\footnotesize $\CT/\mathcal{C}$ with ${}_{\mathcal{A}}\mathcal{C}_{\mathcal{A}}$-sym};
    \node[red,below] at (-2,-2) {$\mathcal{I}$};
\end{tikzpicture} \,.
\end{equation}
The topological interface $\mathcal{I}$ is a simple object in the category $\mathcal{C}_{\mathcal{A}}$ of the right $\mathcal{A}$-modules in $\mathcal{C}$, given by $\CA$ itself.
Here, $\mathcal{C}_{\mathcal{A}}$ naturally carries the structure of a $\mathcal{C}$-${}_{\mathcal{A}}\mathcal{C}_{\mathcal{A}}$-bimodule category.
Namely, the topological lines in $\CC$ acts on the interface from the left, whereas the quantum topological lines in ${}_{\mathcal{A}}\mathcal{C}_{\mathcal{A}}$ acts on the interface from the right.
Since we are interested in $\mathcal{A}$ of the form \eqref{eq:algebra_expansion}, $\mathcal{C}_{\mathcal{A}}$ contains a unique simple object, therefore the topological interface $\mathcal{I}$ is uniquely defined.
If we have an invertible symmetry $\CC = \VEC_G$, such an interface is the one coming from imposing the topological Dirichlet boundary condition for the discrete $G$ gauge field at the interface, recovering the ordinary half-gauging construction in Section \ref{sec:half_space_inv}.

In the special case where ${}_{\mathcal{A}}\mathcal{C}_{\mathcal{A}} \simeq \mathcal{C}$ and $\mathcal{T} \simeq \mathcal{T}/\mathcal{C}$, the topological interface $\mathcal{I}$ can be regarded as a topological defect line $\mathcal{D}$ in the theory $\mathcal{T}$. 
In general, such a defect $\mathcal{D}$ is not necessarily the same as its orientation reversal $\overline{\mathcal{D}}$, but in the following we will restrict to the case where $\mathcal{D} = \overline{\mathcal{D}}$.
Since the fusion category symmetry $\mathcal{C}$ (or equivalently, ${}_{\mathcal{A}}\mathcal{C}_{\mathcal{C}} \cong \mathcal{C}$) is gauged across the defect $\mathcal{D}$, any topological defect line $\mathcal{L} \in \mathcal{C}$ must become transparent up to its quantum dimension $\langle \mathcal{L}\rangle$ when it crosses $\mathcal{D}$. 
This implies the following fusion rule
\begin{equation}\label{eq:fusion_rule_general_1}
    \mathcal{L} \otimes \mathcal{D} = \mathcal{D} \otimes \mathcal{L} =  \langle\mathcal{L}\rangle \mathcal{D} \,, \quad \forall \mathcal{L} \in \mathcal{C} \,.
\end{equation}
Recall that $ \langle\mathcal{L}\rangle \in \mathbb{Z}_{>0}$ for every $\CL$ in an anomaly-free (unitary) fusion category, and the above fusion algebra is consistent.
Since $\mathcal{D} = \overline{\mathcal{D}}$, we must also have
\begin{equation}\label{eq:fusion_rule_general_2}
    \mathcal{D} \otimes \mathcal{D} = \mathcal{D} \otimes \overline{\mathcal{D}} = \bigoplus_{\text{simple}\, \mathcal{L}} \langle \mathcal{L}\rangle \mathcal{L} = \mathcal{A} \,.
\end{equation}
This is because the fusion $\mathcal{D} \otimes \overline{\mathcal{D}}$ corresponds to gauging $\CC$ inside a thin slab sandwiched by the two lines $\CD$ and $\overline{\CD}$, and such a gauging, which is given by inserting a mesh of $\CA$ inside the slab, reduces to a single $\CA$ line in the limit where the thickness of the slab goes to zero, by using the consistency conditions that $\CA$ satisfies.
Comparing with \eqref{eq:G_extension_fusion_rule}, we find that the self-duality under gauging the fusion category symmetry $\mathcal{C}$ leads to a bigger fusion category symmetry described by a $\doubleZ_2$-extension $\underline{\mathcal{E}}_{\doubleZ_2}\mathcal{C}$ of $\mathcal{C}$,
where the nontrivial grading component has the unique simple object $\CD$.
When $\CC$ is an invertible symmetry, \eqref{eq:fusion_rule_general_1} and \eqref{eq:fusion_rule_general_2} are fusion rules for the Tambara-Yamagami fusion category.

However, as discussed in Section \ref{sec:group_extension}, fusion category with the fusion rules \eqref{eq:fusion_rule_general_1} and \eqref{eq:fusion_rule_general_2} is in general not unique. 
How does the additional data arise from the above discussion? 
First, when we say that the quantum symmetry ${}_{\mathcal{A}}\mathcal{C}_{\mathcal{A}}$ is the same as the original symmetry $\mathcal{C}$, we must explicitly specify how the topological lines and the junctions between them in $\CC$ are related to the ones in ${}_{\mathcal{A}}\mathcal{C}_{\mathcal{A}}$, and vice versa. 
Such a choice is not unique in general.\footnote{\label{fit:BrPic}In the categorical language, we need to choose a tensor equivalence between the fusion categories $\mathcal{C}$ and ${}_{\mathcal{A}}\mathcal{C}_{\mathcal{A}}$, and we expect it to be classified by the group $\Aut(\mathcal{C})$ of autoequivalences of the fusion category $\mathcal{C}$. Given a choice of tensor equivalence, we can make $\mathcal{C}_{\mathcal{A}}$ either to be a $\mathcal{C}$-$\mathcal{C}$-bimodule category, or to be a ${}_{\mathcal{A}}\mathcal{C}_{\mathcal{A}}$-${}_{\mathcal{A}}\mathcal{C}_{\mathcal{A}}$-bimodule category. Both can then be used to further construct the extensions $\underline{\mathcal{E}}_{\doubleZ_2}\mathcal{C}$ and $\underline{\mathcal{E}}_{\doubleZ_2}{}_{\mathcal{A}}\mathcal{C}_{\mathcal{A}}$. Since $\mathcal{C}\simeq {}_{\mathcal{A}}\mathcal{C}_{\mathcal{A}}$, $\underline{\mathcal{E}}_{\doubleZ_2}\mathcal{C}$ has the same fusion rules as $\underline{\mathcal{E}}_{\doubleZ_2}{}_{\mathcal{A}}\mathcal{C}_{\mathcal{A}}$ but generically we do not expect them to be equivalent. Since the choice of $\rho$ is equivalent to the choice of invertible $\mathcal{C}$-$\mathcal{C}$-bimodule categories with a unique simple object, we then expect the number of the inequivalent choices of $\rho$ does not exceed $2|\Aut(\mathcal{C})|$. } Furthermore, in general there may be more than one set of consistent local junction data we can choose for the global fusion \eqref{eq:fusion_rule_general_2}. These choices must be made in order to unambiguously determine the larger symmetry $\underline{\mathcal{E}}_{\doubleZ_2}\mathcal{C}$. 

When applying this to a concrete CFT $\mathcal{T}$, the choice of the additional data mentioned above must be compatible with the other data in the $\mathcal{T}$, and we generally expect the theory $\mathcal{T}$ to realize only a subset of possible $\underline{\mathcal{E}}_{\doubleZ_2}\mathcal{C}$ symmetries. 

It is interesting to point out that one should not expect that $\underline{\mathcal{E}}_{\doubleZ_2}\mathcal{C}$ can be uniquely fixed in a theory $\mathcal{T}$ in general. An example is the critical 3-states Potts model, see \cite{Chang:2018iay} for instance. 
There, the theory is self-dual under gauging a $\doubleZ_3$ symmetry, but the theory admits two duality lines $\mathcal{N}$ and $\mathcal{N}'$ both describing this self-dual property. 
They realize two different $\doubleZ_2$-extensions of the $\doubleZ_3$ symmetry in the same theory $\mathcal{T}$.

\subsection{$\Rep(H_8)$ symmetry and its gauging}\label{sec:RepH8_gauging}

$\Rep(H_8)$ is the representation category of the Hopf algebra $H_8$ and it is also a Tambara-Yamagami fusion category $\TY(\doubleZ_2 \times \doubleZ_2, \chi_{diag},+1)$ describing self-duality under gauging $\doubleZ_2 \times \doubleZ_2$. 
It contains 5 simple topological defect lines--4 invertible lines $\dsi, a,b,ab$ generating the non-anomalous $\doubleZ_2^a \times \doubleZ_2^b$ symmetry and a non-invertible duality line $\mathcal{N}$. 
If we parameterize elements of $\doubleZ_2^a \times \doubleZ_2^b$ as $g \equiv (g_1, g_2)$ such that $a\equiv (1,0), b\equiv (0,1)$, then the bicharacter is given by $\chi_{diag}(g,h) = (-1)^{g_1 h_1 + g_2 h_2}$.
For simplicity, we will drop the subscript $diag$ and write $\chi_{diag}$ simply as $\chi$ for the rest of the paper. 
$\Rep(H_8)$ is anomaly-free and there exists a unique way of gauging it since it admits a unique fiber functor \cite{Thorngren:2021yso}.

One way to obtain a theory with the $\Rep(H_8)$ symmetry is by stacking two (potentially different) theories with 
the symmetry of the Ising CFT, namely the Tambara-Yamagami fusion category $\TY(\IZ_2)_+$. 
In this case, the $\doubleZ_2^a \times \doubleZ_2^b$ symmetry comes from the $\IZ_2$ symmetries of the two theories, and the duality line $\mathcal{N}_{\Rep(H_8)} = \mathcal{N}_{\Ising,1}\mathcal{N}_{\Ising,2}$ is the diagonal duality line.

The unique algebra object of the form $\mathcal{A} = \dsi \oplus a \oplus b\oplus ab \oplus 2\mathcal{N}$ and the corresponding junctions $\mu$, $\mu^\vee$ can be explicitly computed by solving the conditions \eqnref{eq:algebracond}.\footnote{The $\Rep(H_8)$ fusion category in total contains six (Morita classes of) algebra objects that can be gauged \cite{etingof2021tensor,Perez-Lona:2023djo,Diatlyk:2023fwf}. They are $\dsi \oplus a \oplus b\oplus ab \oplus 2\mathcal{N}$ and $\dsi \oplus ab \oplus \CN$, which include the non-invertible line, and the ones corresponding to gauging various invertible symmetries.}
We find that fusion junction $\mu$ and the splitting junction $\mu^\vee$ are given by
\begin{equation}
\begin{aligned}
\begin{tikzpicture}[scale=0.8,baseline={([yshift=-.5ex]current bounding box.center)},vertex/.style={anchor=base,
    circle,fill=black!25,minimum size=18pt,inner sep=2pt},scale=0.50]
    \draw[thick, red] (-2,-2) -- (0,0);
    \draw[thick, red] (+2,-2) -- (0,0);
    \draw[thick, red] (0,0) -- (0,2);
    \draw[thick, red, -stealth] (-2,-2) -- (-1,-1);
    \draw[thick, red, -stealth] (+2,-2) -- (1,-1);
    \draw[thick, red, -stealth] (0,0) -- (0,1);
    \filldraw[red] (0,0) circle (2pt);

    \node[red, below] at (-2,-2) {\scriptsize$\mathcal{A}$};
    \node[red, below] at (2,-2) {\scriptsize$\mathcal{A}$};
    \node[red, above] at (0,2) {\scriptsize$\mathcal{A}$};
    \node[red, left]  at (0,0) {\scriptsize$\mu$};
\end{tikzpicture} = & \sum_{g,h \in \doubleZ_2 \times \doubleZ_2} \psi(g,h) \begin{tikzpicture}[scale=0.8, baseline={([yshift=-.5ex]current bounding box.center)},vertex/.style={anchor=base,
    circle,fill=black!25,minimum size=18pt,inner sep=2pt},scale=0.50]
    \draw[thick, black] (-2,-2) -- (0,0);
    \draw[thick, black] (+2,-2) -- (0,0);
    \draw[thick, black] (0,0) -- (0,2);
    \draw[thick, black, -stealth] (-2,-2) -- (-1,-1);
    \draw[thick, black, -stealth] (+2,-2) -- (1,-1);
    \draw[thick, black, -stealth] (0,0) -- (0,1);

    \node[black, below] at (-2,-2) {\scriptsize$g$};
    \node[black, below] at (2,-2) {\scriptsize$h$};
    \node[black, above] at (0,2) {\scriptsize$gh$};
\end{tikzpicture} 
+ \sum_{g\in \doubleZ_2 \times \doubleZ_2 } [L(g)]_\mu{}^\nu \begin{tikzpicture}[scale=0.8,baseline={([yshift=-.5ex]current bounding box.center)},vertex/.style={anchor=base,
    circle,fill=black!25,minimum size=18pt,inner sep=2pt},scale=0.50]
    \draw[thick, black] (-2,-2) -- (0,0);
    \draw[thick, black] (+2,-2) -- (0,0);
    \draw[thick, black] (0,0) -- (0,2);
    \draw[thick, black, -stealth] (-2,-2) -- (-1,-1);
    \draw[thick, black, -stealth] (+2,-2) -- (1,-1);
    \draw[thick, black, -stealth] (0,0) -- (0,1);

    \node[black, below] at (-2,-2) {\scriptsize$g$};
    \node[black, below] at (2,-2) {\scriptsize$\CN_\mu$};
    \node[black, above] at (0,2) {\scriptsize$\CN_\nu$};
\end{tikzpicture} \\
& + \sum_{g\in \doubleZ_2 \times \doubleZ_2 } [R(g)]_\mu{}^\nu \begin{tikzpicture}[scale=0.8,baseline={([yshift=-.5ex]current bounding box.center)},vertex/.style={anchor=base,
    circle,fill=black!25,minimum size=18pt,inner sep=2pt},scale=0.50]
    \draw[thick, black] (-2,-2) -- (0,0);
    \draw[thick, black] (+2,-2) -- (0,0);
    \draw[thick, black] (0,0) -- (0,2);
    \draw[thick, black, -stealth] (-2,-2) -- (-1,-1);
    \draw[thick, black, -stealth] (+2,-2) -- (1,-1);
    \draw[thick, black, -stealth] (0,0) -- (0,1);

    \node[black, below] at (-2,-2) {\scriptsize$\CN_\mu$};
    \node[black, below] at (2,-2) {\scriptsize$g$};
    \node[black, above] at (0,2) {\scriptsize$\CN_\nu$};
\end{tikzpicture} + \sum_{g\in \doubleZ_2 \times \doubleZ_2} W(g)_{\mu\nu} \begin{tikzpicture}[scale=0.8,baseline={([yshift=-.5ex]current bounding box.center)},vertex/.style={anchor=base,
    circle,fill=black!25,minimum size=18pt,inner sep=2pt},scale=0.50]
    \draw[thick, black] (-2,-2) -- (0,0);
    \draw[thick, black] (+2,-2) -- (0,0);
    \draw[thick, black] (0,0) -- (0,2);
    \draw[thick, black, -stealth] (-2,-2) -- (-1,-1);
    \draw[thick, black, -stealth] (+2,-2) -- (1,-1);
    \draw[thick, black, -stealth] (0,0) -- (0,1);

    \node[black, below] at (-2,-2) {\scriptsize$\CN_\mu$};
    \node[black, below] at (2,-2) {\scriptsize$\CN_\nu$};
    \node[black, above] at (0,2) {\scriptsize$g$};
\end{tikzpicture},
\end{aligned}
\end{equation}
where 
\begin{align}
    &\psi(g,h) = \frac{1}{2\sqrt{2}}\left(
\begin{smallmatrix}
 1 & 1 & 1 & 1 \\
 1 & 1 & 1 & 1 \\
 1 & -1 & 1 & -1 \\
 1 & -1 & 1 & -1 \\
\end{smallmatrix}
\right), \quad L(g) = \frac{1}{2\sqrt{2}}\left(\sigma^0,-\sigma^3,-\sigma^1,-\ii \sigma^2\right), \nonumber\\
&R(g) = \frac{1}{2\sqrt{2}}\left(\sigma^0,\sigma^3,\sigma^1,-\ii \sigma^2\right), \quad W(g) = \frac{1}{4}\left(\sigma ^3-\sigma ^1,-\sigma ^0+\ii \sigma ^2,\sigma ^0+\ii \sigma ^2,-\sigma ^1-\sigma ^3\right),
\end{align}
and
\begin{equation}
\begin{aligned}
\begin{tikzpicture}[scale=0.8,baseline={([yshift=-.5ex]current bounding box.center)},vertex/.style={anchor=base,
    circle,fill=black!25,minimum size=18pt,inner sep=2pt},scale=0.50]
    \draw[thick, red] (-2,2) -- (0,0);
    \draw[thick, red] (+2,2) -- (0,0);
    \draw[thick, red] (0,0) -- (0,-2);
    \draw[thick, red, -stealth] (0,0) -- (-1,1);
    \draw[thick, red, -stealth] (0,0) -- (1,1);
    \draw[thick, red, -stealth] (0,-2) -- (0,-1);
    \filldraw[red] (0,0) circle (3pt);

    \node[red, above] at (-2,2) {\scriptsize$\mathcal{A}$};
    \node[red, above] at (2,2) {\scriptsize$\mathcal{A}$};
    \node[red, below] at (0,-2) {\scriptsize$\mathcal{A}$};
    \node[red, right]  at (0,0) {\scriptsize$\mu^\vee$};
\end{tikzpicture} = & \sum_{g,h \in \doubleZ_2 \times \doubleZ_2} \psi^\vee(g,h) \begin{tikzpicture}[scale=0.8,baseline={([yshift=-.5ex]current bounding box.center)},vertex/.style={anchor=base,
    circle,fill=black!25,minimum size=18pt,inner sep=2pt},scale=0.50]
    \draw[thick, black] (-2,2) -- (0,0);
    \draw[thick, black] (+2,2) -- (0,0);
    \draw[thick, black] (0,0) -- (0,-2);
    \draw[thick, black, -stealth] (0,0) -- (-1,1);
    \draw[thick, black, -stealth] (0,0) -- (1,1);
    \draw[thick, black, -stealth] (0,-2) -- (0,-1);

    \node[black, above] at (-2,2) {\scriptsize$g$};
    \node[black, above] at (2,2) {\scriptsize$h$};
    \node[black, below] at (0,-2) {\scriptsize$gh$};
\end{tikzpicture}
+ \sum_{g\in \doubleZ_2 \times \doubleZ_2 } [L^\vee(g)]_\mu{}^\nu \begin{tikzpicture}[scale=0.8,baseline={([yshift=-.5ex]current bounding box.center)},vertex/.style={anchor=base,
    circle,fill=black!25,minimum size=18pt,inner sep=2pt},scale=0.50]
    \draw[thick, black] (-2,2) -- (0,0);
    \draw[thick, black] (+2,2) -- (0,0);
    \draw[thick, black] (0,0) -- (0,-2);
    \draw[thick, black, -stealth] (0,0) -- (-1,1);
    \draw[thick, black, -stealth] (0,0) -- (1,1);
    \draw[thick, black, -stealth] (0,-2) -- (0,-1);

    \node[black, above] at (-2,2) {\scriptsize$g$};
    \node[black, above] at (2,2) {\scriptsize$\CN_\nu$};
    \node[black, below] at (0,-2) {\scriptsize$\CN_\mu$};
\end{tikzpicture} \\
& + \sum_{g\in \doubleZ_2 \times \doubleZ_2 } [R^\vee(g)]_\mu{}^\nu \begin{tikzpicture}[scale=0.8,baseline={([yshift=-.5ex]current bounding box.center)},vertex/.style={anchor=base,
    circle,fill=black!25,minimum size=18pt,inner sep=2pt},scale=0.50]
    \draw[thick, black] (-2,2) -- (0,0);
    \draw[thick, black] (+2,2) -- (0,0);
    \draw[thick, black] (0,0) -- (0,-2);
    \draw[thick, black, -stealth] (0,0) -- (-1,1);
    \draw[thick, black, -stealth] (0,0) -- (1,1);
    \draw[thick, black, -stealth] (0,-2) -- (0,-1);

    \node[black, above] at (-2,2) {\scriptsize$\CN_\nu$};
    \node[black, above] at (2,2) {\scriptsize$g$};
    \node[black, below] at (0,-2) {\scriptsize$\CN_\mu$};
\end{tikzpicture} + \sum_{g\in \doubleZ_2 \times \doubleZ_2} [W^\vee(g)]_{\mu\nu} \begin{tikzpicture}[scale=0.8,baseline={([yshift=-.5ex]current bounding box.center)},vertex/.style={anchor=base,
    circle,fill=black!25,minimum size=18pt,inner sep=2pt},scale=0.50]
    \draw[thick, black] (-2,2) -- (0,0);
    \draw[thick, black] (+2,2) -- (0,0);
    \draw[thick, black] (0,0) -- (0,-2);
    \draw[thick, black, -stealth] (0,0) -- (-1,1);
    \draw[thick, black, -stealth] (0,0) -- (1,1);
    \draw[thick, black, -stealth] (0,-2) -- (0,-1);

    \node[black, above] at (-2,2) {\scriptsize$\CN_\mu$};
    \node[black, above] at (2,2) {\scriptsize$\CN_\nu$};
    \node[black, below] at (0,-2) {\scriptsize$g$};
\end{tikzpicture},
\end{aligned}
\end{equation}
where 
\begin{align}
    &\psi^\vee(g,h) = \frac{1}{2\sqrt{2}}\left(
\begin{smallmatrix}
 1 & 1 & 1 & 1 \\
 1 & 1 & 1 & 1 \\
 1 & -1 & 1 & -1 \\
 1 & -1 & 1 & -1 \\
\end{smallmatrix}
\right), \quad L^\vee(g) = \frac{1}{2\sqrt{2}}\left(\sigma^0,-\sigma^3,-\sigma^1,\ii \sigma^2\right), \nonumber\\
&R^\vee(g) = \frac{1}{2\sqrt{2}}\left(\sigma^0,\sigma^3,\sigma^1,\ii \sigma^2\right), \quad W^\vee(g) = \frac{1}{4}\left(\sigma ^3-\sigma ^1,-\sigma ^0+\ii \sigma ^2,\sigma ^0+\ii \sigma ^2,-\sigma ^1-\sigma ^3\right).
\end{align}
More details on finding such an algebra object are given in \appref{app:algebra}. The torus partition function of the gauged theory $\CT/\Rep(H_8)$ is computed by expanding the following diagram in terms of simple topological defect lines:
\begin{equation}
\begin{tikzpicture}[scale=0.7]
    \filldraw[grey] (-2,-2) rectangle ++(4,4);
    \draw[thick, dgrey] (-2,-2) rectangle ++(4,4);
    \draw[thick, red, -stealth] (0,-2) -- (0.354,-1.354);
    \draw[thick, red] (0,-2) -- (0.707,-0.707);
    \draw[thick, red, -stealth] (2,0) -- (1.354,-0.354);
    \draw[thick, red] (2,0) -- (0.707,-0.707);
    \draw[thick, red, -stealth] (-0.707,0.707) -- (-0.354,1.354);
    \draw[thick, red] (0,2) -- (-0.707,0.707);
    \draw[thick, red, -stealth] (-0.707,0.707) -- (-1.354,0.354);
    \draw[thick, red] (-2,0) -- (-0.707,0.707);
    \draw[thick, red, -stealth] (0.707,-0.707) -- (0,0);
    \draw[thick, red] (0.707,-0.707) -- (-0.707,0.707);
    
    \filldraw[red] (+0.707,-0.707) circle (2pt);
    \filldraw[red] (-0.707,+0.707) circle (2pt);
    
    \node[red, below] at (0,-2) {$\mathcal{A}$};
    \node[red, right] at (2,0) {$\mathcal{A}$};
    \node[red, above] at (0,2) {$\mathcal{A}$};
    \node[red, left] at (-2,0) {$\mathcal{A}$};
    \node[red, above] at (0,0) {$\mathcal{A}$};

    \node[red, below] at (+0.807,-0.707) {$\mu$};
    \node[red, above] at (-0.807,+0.707) {$\mu^\vee$};
    
\end{tikzpicture} \,.
\end{equation}
We find
\begin{align} \label{eq:RepH8_gauging_general}
\begin{split}
    Z_{\mathcal{T}/\Rep(H_8)}(\tau) &= \frac{1}{8} \Big(
        Z_{\mathcal{T}}[\dsi,\dsi,\dsi](\tau) + Z_{\mathcal{T}}[\dsi,a,a](\tau) +
        Z_{\mathcal{T}}[a,\dsi,a](\tau) + Z_{\mathcal{T}}[a,a,\dsi](\tau) \\
        &\quad\quad +  Z_{\mathcal{T}}[\dsi,b,b](\tau) +
        Z_{\mathcal{T}}[b,\dsi,b](\tau) + Z_{\mathcal{T}}[b,b,\dsi](\tau) \\
         &\quad\quad +  Z_{\mathcal{T}}[\dsi,ab,ab](\tau) +
        Z_{\mathcal{T}}[ab,\dsi,ab](\tau) + Z_{\mathcal{T}}[ab,ab,\dsi](\tau) \\
        &\quad\quad -  Z_{\mathcal{T}}[a,b,ab](\tau) -
        Z_{\mathcal{T}}[b,a,ab](\tau) - Z_{\mathcal{T}}[b,ab,a](\tau) \\
        &\quad\quad -  Z_{\mathcal{T}}[ab,a,b](\tau) -
        Z_{\mathcal{T}}[a,ab,b](\tau) - Z_{\mathcal{T}}[ab,b,a](\tau) \\
        &\quad\quad + 2Z_{\mathcal{T}}[\dsi,\mathcal{N},\mathcal{N}](\tau) +
        2Z_{\mathcal{T}}[\mathcal{N},\dsi,\mathcal{N}](\tau) + 2Z_{\mathcal{T}}[\mathcal{N},\mathcal{N},\dsi](\tau) \\
        &\quad\quad + 2Z_{\mathcal{T}}[ab,\mathcal{N},\mathcal{N}](\tau) +
        2Z_{\mathcal{T}}[\mathcal{N},ab,\mathcal{N}](\tau) + 2Z_{\mathcal{T}}[\mathcal{N},\mathcal{N},ab](\tau)
    \Big) \,.
\end{split}
\end{align}
Here, the terms on the RHS that come with $-1$ coefficients have a sign ambiguity, due to the freedom to shift the counterterm associated to the nontrivial element of $H^2(\mathbb{Z}_2^a \times \mathbb{Z}_2^b,U(1)) \cong \mathbb{Z}_2$. 
On the other hand, the $\Rep(H_8)$ symmetry implies that the theory is invariant under gauging the $\doubleZ_2^a\times \doubleZ_2^b$ symmetry with an appropriate choice of such a counterterm. 
We fix the counterterm ambiguity by requiring the theory to be self-dual under gauging $\doubleZ_2^a\times \doubleZ_2^b$ with no discrete torsion.
This leads to the relation
\begin{align} \label{eq:Z2xZ2_gauging}
\begin{split}
    Z_{\mathcal{T}}[\dsi,\dsi,\dsi](\tau) &= \frac{1}{4} \Big(
        Z_{\mathcal{T}}[\dsi,\dsi,\dsi](\tau) + Z_{\mathcal{T}}[\dsi,a,a](\tau) +
        Z_{\mathcal{T}}[a,\dsi,a](\tau) + Z_{\mathcal{T}}[a,a,\dsi](\tau) \\
        &\quad\quad +  Z_{\mathcal{T}}[\dsi,b,b](\tau) +
        Z_{\mathcal{T}}[b,\dsi,b](\tau) + Z_{\mathcal{T}}[b,b,\dsi](\tau) \\
         &\quad\quad +  Z_{\mathcal{T}}[\dsi,ab,ab](\tau) +
        Z_{\mathcal{T}}[ab,\dsi,ab](\tau) + Z_{\mathcal{T}}[ab,ab,\dsi](\tau) \\
        &\quad\quad + Z_{\mathcal{T}}[a,b,ab](\tau) +
        Z_{\mathcal{T}}[b,a,ab](\tau) + Z_{\mathcal{T}}[b,ab,a](\tau) \\
        &\quad\quad +  Z_{\mathcal{T}}[ab,a,b](\tau) +
        Z_{\mathcal{T}}[a,ab,b](\tau) + Z_{\mathcal{T}}[ab,b,a](\tau)
    \Big) \,.
\end{split}
\end{align}
Using \eqref{eq:Z2xZ2_gauging} to simplify \eqref{eq:RepH8_gauging_general}, we find 
\begin{equation}\label{eq:RepH8_gauging_general_1}
\begin{aligned}
    Z_{\mathcal{T}/\Rep(H_8)}(\tau) &= \frac{1}{4} \Big(
        - Z_{\mathcal{T}}[\dsi,\dsi,\dsi](\tau) + Z_{\mathcal{T}}[\dsi,a,a](\tau) +
        Z_{\mathcal{T}}[a,\dsi,a](\tau) + Z_{\mathcal{T}}[a,a,\dsi](\tau) \\
        &\quad\quad +  Z_{\mathcal{T}}[\dsi,b,b](\tau) +
        Z_{\mathcal{T}}[b,\dsi,b](\tau) + Z_{\mathcal{T}}[b,b,\dsi](\tau) \\
         &\quad\quad +  Z_{\mathcal{T}}[\dsi,ab,ab](\tau) +
        Z_{\mathcal{T}}[ab,\dsi,ab](\tau) + Z_{\mathcal{T}}[ab,ab,\dsi](\tau) \\
        &\quad\quad + Z_{\mathcal{T}}[\dsi,\mathcal{N},\mathcal{N}](\tau) +
        Z_{\mathcal{T}}[\mathcal{N},\dsi,\mathcal{N}](\tau) + Z_{\mathcal{T}}[\mathcal{N},\mathcal{N},\dsi](\tau) \\
        &\quad\quad + Z_{\mathcal{T}}[ab,\mathcal{N},\mathcal{N}](\tau) +
        Z_{\mathcal{T}}[\mathcal{N},ab,\mathcal{N}](\tau) + Z_{\mathcal{T}}[\mathcal{N},\mathcal{N},ab](\tau)
    \Big) \,.
\end{aligned}
\end{equation}
The twisted partition functions on the RHS containing only invertible symmetries can be computed via modular transformations once their action on the Hilbert space is known. 
The twisted partition functions containing the duality line $\mathcal{N}$ can be computed using the following relation:
\begin{equation}\label{eq:RepH8_twisted_1}
    Z_{\mathcal{T}}[\mathcal{N},\dsi,\mathcal{N}](\tau) + Z_{\mathcal{T}}[\mathcal{N},ab,\mathcal{N}](\tau) =  Z_{\mathcal{T}}[\mathcal{N},\dsi,\mathcal{N}](\tau+2) + Z_{\mathcal{T}}[\mathcal{N},\dsi,\mathcal{N}](\tau-2) \,,
\end{equation}
where the RHS can be computed once we know the action of $\CN$ on the Hilbert space.
To see this relation, notice that  
\begin{equation}
\begin{aligned}
& \begin{tikzpicture}[baseline={([yshift=0]current bounding box.center)},vertex/.style={anchor=base,
    circle,fill=black!25,minimum size=18pt,inner sep=2pt},scale=0.5]
    \filldraw[grey] (-2,-2) rectangle ++(4,4);
    \draw[thick, dgrey] (-2,-2) rectangle ++(4,4);
    \draw[thick, black] (0,-2) -- (+2,-1);
    \draw[thick, black] (-2,-1) -- (+2,1);
    \draw[thick, black] (-2,+1) -- (0,+2);
    \node[black, above] at (0,0) {$\mathcal{N}$};
\end{tikzpicture}\,  = \frac{1}{2}\sum_{g} \chi(\dsi,g) \, \begin{tikzpicture}[baseline={([yshift=0]current bounding box.center)},vertex/.style={anchor=base,
    circle,fill=black!25,minimum size=18pt,inner sep=2pt},scale=0.5]
    \filldraw[grey] (-2,-2) rectangle ++(4,4);
    \draw[thick, dgrey] (-2,-2) rectangle ++(4,4);
    \draw[thick, black] (-2,-1) arc(-90:90:1);
    \draw[thick, black] (2,-1) arc(90:143:2.5);
    \draw[thick, black] (2,1) arc(-90:-143:2.5);
    \draw[thick, dashed] (-1.293, 0.707) -- (0.6,1.4);
    \node[black, below] at (-1,-0.5) {$\mathcal{N}$};
    \node[black, below] at (-0.347,1.054) {$g$};
\end{tikzpicture} = \frac{1}{2}\sum_{g} \chi(\dsi,g) \, \begin{tikzpicture}[baseline={([yshift=0]current bounding box.center)},vertex/.style={anchor=base,
    circle,fill=black!25,minimum size=18pt,inner sep=2pt},scale=0.5]
    \filldraw[grey] (-2,-2) rectangle ++(4,4);
    \draw[thick, dgrey] (-2,-2) rectangle ++(4,4);
    \draw[thick, black] (0,-2) -- (0.707,-0.707);
    \draw[thick, black, dashed] (2,0) -- (0.707,-0.707);
    \draw[thick, black] (0,2) -- (-0.707,0.707);
    \draw[thick, black, dashed] (-2,0) -- (-0.707,0.707);
    \draw[thick, black] (0.707,-0.707) -- (-0.707,0.707);
    \node[black, left] at (0.2,-0.3) {$\mathcal{N}$};
    \node[black, below] at (1.5,-0.3) {$g$};
\end{tikzpicture} \, , \\
& \begin{tikzpicture}[baseline={([yshift=0]current bounding box.center)},vertex/.style={anchor=base,
    circle,fill=black!25,minimum size=18pt,inner sep=2pt},scale=0.5]
    \filldraw[grey] (-2,-2) rectangle ++(4,4);
    \draw[thick, dgrey] (-2,-2) rectangle ++(4,4);
    \draw[thick, black] (0,-2) -- (-2,-1);
    \draw[thick, black] (2,-1) -- (-2,1);
    \draw[thick, black] (2,+1) -- (0,+2);
    \node[black, above] at (0,0) {$\mathcal{N}$};
\end{tikzpicture} \,  = \frac{1}{2}\sum_{g} \chi(\dsi,g) \, \begin{tikzpicture}[baseline={([yshift=0]current bounding box.center)},vertex/.style={anchor=base,
    circle,fill=black!25,minimum size=18pt,inner sep=2pt},scale=0.5]
    \filldraw[grey] (-2,-2) rectangle ++(4,4);
    \draw[thick, dgrey] (-2,-2) rectangle ++(4,4);
    \draw[thick, black] (2,-1) arc(-90:-270:1);
    \draw[thick, black] (-2,-1) arc(90:37:2.5);
    \draw[thick, black] (-2,1) arc(-90:-37:2.5);
    \draw[thick, dashed] (1.293, 0.707) -- (-0.6,1.4);
    \node[black, below] at (1,-0.5) {$\mathcal{N}$};
    \node[black, below] at (0.347,1.054) {$g$};
\end{tikzpicture} \, = \frac{1}{2}\sum_g \chi(\dsi,g) \chi(g,g) \, \begin{tikzpicture}[baseline={([yshift=0]current bounding box.center)},vertex/.style={anchor=base,
    circle,fill=black!25,minimum size=18pt,inner sep=2pt},scale=0.5]
    \filldraw[grey] (-2,-2) rectangle ++(4,4);
    \draw[thick, dgrey] (-2,-2) rectangle ++(4,4);
    \draw[thick, black] (0,-2) -- (0.707,-0.707);
    \draw[thick, black, dashed] (2,0) -- (0.707,-0.707);
    \draw[thick, black] (0,2) -- (-0.707,0.707);
    \draw[thick, black, dashed] (-2,0) -- (-0.707,0.707);
    \draw[thick, black] (0.707,-0.707) -- (-0.707,0.707);
    \node[black, left] at (0.2,-0.3) {$\mathcal{N}$};
    \node[black, below] at (1.5,-0.3) {$g$};
\end{tikzpicture} \, ,
\end{aligned}
\end{equation}
and use the relation $\chi(\dsi,g) (1+\chi(g,g)) = 2\delta_{g_1,g_2}$. 
The rest of the twisted partition functions can be obtained by performing a modular transformation to the LHS of \eqref{eq:RepH8_twisted_1}:
\begin{equation}\label{eq:RepH8_twisted_2}
\begin{aligned}
    & Z_{\mathcal{T}}[\mathcal{N},\mathcal{N},\dsi](\tau) + Z_{\mathcal{T}}[\mathcal{N},\mathcal{N},ab](\tau) = Z_{\mathcal{T}}[\mathcal{N},\dsi,\mathcal{N}](\tau + 1) + Z_{\mathcal{T}}[\mathcal{N},ab,\mathcal{N}](\tau + 1), \\
    & Z_{\mathcal{T}}[\dsi,\mathcal{N},\mathcal{N}](\tau) +  Z_{\mathcal{T}}[ab,\mathcal{N},\mathcal{N}](\tau) = Z_{\mathcal{T}}[\mathcal{N},\dsi,\mathcal{N}]\left(-\frac{1}{\tau}\right)+ Z_{\mathcal{T}}[\mathcal{N},ab,\mathcal{N}]\left(-\frac{1}{\tau}\right). 
\end{aligned}
\end{equation}
To summarize, the torus partition function of the gauged theory $\CT/\Rep(H_8)$ can be computed once we know the action of $\Rep(H_8)$ on the (untwisted) Hilbert space of $\CT$.

When we gauge a $\doubleZ_2 = \{\dsi,\eta\}$ symmetry, we keep the $\doubleZ_2$-invariant sector of the Hilbert space $\mathcal{H}$ as well as the $\doubleZ_2$-invariant sector of the defect Hilbert space $\mathcal{H}_\eta$.
It is natural to ask what is the analog in the case of gauging $\Rep(H_8)$. 
We will show that there is a similar interpretation. 

Let us first study the defect Hilbert space of invertible symmetries.
Different topological sectors and their charges under the $\Rep(H_8)$ symmetry are listed in Table \ref{tab:Z2Z2sector}. 
This can be obtained by solving the irreps of the algebra of \text{lasso} actions following Appendix \ref{app:lassocompose}.

\begin{table}[htbp]
    \centering
    \begin{tabular}{c|c|c|c|c|c|c|c}
    \hline \hline
         & $\dsi$ & $a$ & $b$ & $ab$ & $\mathcal{N}$ & $\mathcal{Z}(\VEC_{\doubleZ_2 \times \doubleZ_2})$ & spin $s \mod 1$  \\
         \hline
         & $1$ & $1$ & $1$ & $1$ & $2$ & $1$ & $0$ \\
         & $1$ & $1$ & $1$ & $1$ & $-2$ & $1$ & $0$\\
         $\mathcal{H}_\dsi$ & $1$ & $-1$ & $1$ & $-1$ & $0$ & $e_1$ & 0 \\
         & $1$ & $1$ & $-1$ & $-1$ & $0$ & $e_2$ & $0$ \\
         & $1$ & $-1$ & $-1$ & $1$ & $0$ & $e_1 e_2$ & $0$ \\
         \hline
         & $1$ & $-1$ & $1$ & $-1$ & $2\ii$ & $m_1 e_1$ & $\frac{1}{2}$\\
         & $1$ & $-1$ & $1$ & $-1$ & $-2\ii$ & $m_1 e_1$ & $\frac{1}{2}$\\
         $\mathcal{H}_a$ & $1$ & $1$ & $1$ & $1$ & $0$ & $m_1$ & $0$\\
         & $1$ & $1$ & $-1$ & $-1$ & $0$ & $m_1 e_2$ & $0$ \\
         & $1$ & $-1$ & $-1$ & $1$ & $0$ & $m_1 e_1 e_2$ & $\frac{1}{2}$\\
         \hline
         & $1$ & $1$ & $-1$ & $-1$ & $2\ii$ & $m_2 e_2$ & $\frac{1}{2}$\\
         & $1$ & $1$ & $-1$ & $-1$ & $-2\ii$ & $m_2 e_2$ & $\frac{1}{2}$\\
         $\mathcal{H}_b$ & $1$ & $1$ & $1$ & $1$ & $0$ & $m_2$ & $0$\\
         & $1$ & $-1$ & $1$ & $-1$ & $0$ & $m_2 e_1$ & $0$\\
         & $1$ & $-1$ & $-1$ & $1$ & $0$ & $m_2 e_1 e_2$ & $\frac{1}{2}$ \\
         \hline
         & $1$ & $-1$ & $-1$ & $1$ & $2$ & $m_1 m_2 e_1 e_2$ & $0$\\
         & $1$ & $-1$ & $-1$ & $1$ & $-2$ & $m_1 m_2 e_1 e_2$ & $0$\\
         $\mathcal{H}_{ab}$ & $1$ & $1$ & $1$ & $1$ & $0$ & $m_1 m_2$ & $0$\\
         & $1$ & $-1$ & $1$ & $-1$ & $0$ & $m_1 m_2 e_1$ & $\frac{1}{2}$ \\
         & $1$ & $1$ & $-1$ & $-1$ & $0$ & $m_1 m_2 e_2$ & $\frac{1}{2}$\\
         \hline \hline
    \end{tabular}
    \caption{Topological sectors of defect Hilbert spaces of invertible symmetries together with their charges and spins. We also use the simple anyons in the symTFT of $\doubleZ_2 \times \doubleZ_2$ to label the topological sectors. The duality line $\mathcal{N}$ corresponds to the $\doubleZ_2^{em}$ symmetry which exchanges $e_i$ with $m_i$. In the special case where the anyon is invariant under $\doubleZ_2^{em}$, the corresponding topological sector will split into two where the duality line $\mathcal{N}$ acts differently.}
    \label{tab:Z2Z2sector}
\end{table}
For $\mathcal{H} \equiv \mathcal{H}_\dsi$ and $\mathcal{H}_{ab}$, the action of the duality line $\mathcal{N}$ can be unambiguously defined, therefore we expect the gauging to keep the states where $\mathcal{N}$ acts by its quantum dimension $\langle \mathcal{N}\rangle = 2$. 
Indeed, looking at the corresponding partition functions from \eqref{eq:RepH8_gauging_general},
\begin{equation}
\begin{aligned}
    & Z_{\mathcal{T}/\Rep(H_8),\mathcal{H}} = \frac{1}{8}\left(Z_{\mathcal{T}}[\dsi,\dsi,\dsi] + Z_{\mathcal{T}}[\dsi,a,a] + Z_{\mathcal{T}}[\dsi,b,b]+Z_{\mathcal{T}}[\dsi,ab,ab] + 2Z_{\mathcal{T}}[\dsi,\mathcal{N},\mathcal{N}]\right), \\
    & Z_{\mathcal{T}/\Rep(H_8),\mathcal{H}_{ab}} = \frac{1}{8}\left(Z_{\mathcal{T}}[ab,\dsi,ab] + Z_{\mathcal{T}}[ab,ab,\dsi] - Z_{\mathcal{T}}[ab,b,a] - Z_{\mathcal{T}}[ab,b,a] + 2Z_{\mathcal{T}}[ab,\mathcal{N},\mathcal{N}] \right),
\end{aligned}
\end{equation}
we see that the gauging keeps only the states with charge $(1,1,1,1,2)$ under $(\dsi,a,b,ab,\mathcal{N})$ in $\mathcal{H}$ and the states with charge $(1,-1,-1,1,2)$ in $\mathcal{H}_{ab}$. 
In the latter case, we are keeping the states which are odd under both $\doubleZ_2^a$ and $\doubleZ_2^b$ because of the non-trivial discrete torsion of $\doubleZ_2^a \times \doubleZ_2^b$ in the algebra object.

For $\mathcal{H}_{a}$ and $\mathcal{H}_b$, the situation is more subtle, with no analog in the invertible symmetry case. 
Here, the action of the duality $\mathcal{N}$ can not be unambiguously defined and as a result it acts as $\pm 2\ii$ or $0$. 
It no longer makes sense to keep the sectors where $\mathcal{N}$ acts as $\pm 2\ii$, and this is confirmed by the fact that both sectors have fractional spins. 
From the partition functions
\begin{equation}
\begin{aligned}
    & Z_{\mathcal{T}/\Rep(H_8),\mathcal{H}_a} = \frac{1}{2}\cdot \frac{1}{4}\left(Z_{\mathcal{T}}[a,\dsi,a] + Z_{\mathcal{T}}[a,a,\dsi] - Z_{\mathcal{T}}[a,ab,b] - Z_{\mathcal{T}}[a,b,ab]\right) \,, \\
    & Z_{\mathcal{T}/\Rep(H_8),\mathcal{H}_b} = \frac{1}{2}\cdot \frac{1}{4}\left(Z_{\mathcal{T}}[b,\dsi,b] + Z_{\mathcal{T}}[b,b,\dsi] - Z_{\mathcal{T}}[b,ab,a] - Z_{\mathcal{T}}[b,a,ab]\right) \,,
\end{aligned}
\end{equation}
we find that we keep the states with charge $(1,1,-1,-1,0)$ in $\mathcal{H}_a$ and the states with charge $(1,-1,+1,-1,0)$ in $\mathcal{H}_b$. But in both cases, there is an extra undesired factor of $\frac{1}{2}$ in the partition function. This seems to be pathological, but notice that the two sectors are actually identical because of the self-duality under gauging $\doubleZ_2^a \times \doubleZ_2^b$. 
Therefore, we should really interpret this as keeping a single sector in $\mathcal{H}_a$ (which is identified with the corresponding sector in $\mathcal{H}_b$ in the gauged theory).

Finally, let us consider the defect Hilbert space $\mathcal{H}_{\mathcal{N}}$. In this case, the action of the duality line $\mathcal{N}$ splits into 4 operators which we denote as $\mathcal{U}_{\mathcal{N},g}$. 
Among these, only two of them, $\mathcal{U}_{\mathcal{N},\dsi}$ and $\mathcal{U}_{\mathcal{N},ab}$ can be unambiguously defined with charges $\pm 1$, as shown in Table \ref{tab:Nsector}. 
Meanwhile, there are only two invertible lines $\dsi$ and $ab$ acting unambiguously with charges $\pm 1$.
Therefore, it is natural to expect that the $\Rep(H_8)$-gauging will project $\mathcal{H}_{\mathcal{N}}$ to the states where these four operators acting trivially as $+1$. 
Indeed, we can confirm this from the corresponding partition function
\begin{equation}
    Z_{\mathcal{T}/\Rep(H_8),\mathcal{H}_{\mathcal{N}}} = \frac{1}{4}(\mathcal{Z}_{\mathcal{T}}[\mathcal{N},\dsi,\mathcal{N}]+\mathcal{Z}_{\mathcal{T}}[\mathcal{N},ab,\mathcal{N}]+\mathcal{Z}_{\mathcal{T}}[\mathcal{N},\mathcal{N},\dsi] + \mathcal{Z}_{\mathcal{T}}[\mathcal{N},\mathcal{N},ab]).
\end{equation}

To summarize, similar to the gauging of finite invertible symmetries, we find gauging the non-invertible $\Rep(H_8)$ symmetry can be understood as projecting to the invariant states (up to the quantum dimension) of the topological lines whose actions can be unambiguously defined on each defect Hilbert space.
\begin{table}[htbp]
    \centering
    \begin{tabular}{c|c|c|c|c|c|c|c|c|c}
    \hline \hline
         & $\dsi$ & $a$ & $b$ & $ab$ & $\mathcal{U}_{\mathcal{N},\dsi}$ & $\mathcal{U}_{\mathcal{N},a}$ & $\mathcal{U}_{\mathcal{N},b}$ & $\mathcal{U}_{\mathcal{N},ab}$ & spin $s \mod 1$  \\
         \hline 
         \multirow{8}*{$\mathcal{H}_{\mathcal{N}}$}& $1$ & $-\ii$ & $-\ii$ & $-1$ & $e^{\frac{3\pi \ii}{4}}$ & $e^{-\frac{3\pi \ii}{4}}$ & $e^{-\frac{3\pi \ii}{4}}$ & $e^{-\frac{\pi \ii}{4}}$ & $\frac{3}{8}$\\
         \cline{2-10}
         & $1$ & $-\ii$ & $-\ii$ & $-1$ & $e^{-\frac{\pi \ii}{4}}$ & $e^{\frac{\pi \ii}{4}}$ & $e^{\frac{\pi \ii}{4}}$ & $e^{\frac{3\pi \ii}{4}}$ & $\frac{7}{8}$\\
         \cline{2-10}
         & $1$ & $\ii$ & $\ii$ & $-1$ & $e^{-\frac{3\pi \ii}{4}}$ & $e^{\frac{3\pi \ii}{4}}$ & $e^{\frac{3\pi \ii}{4}}$ & $e^{\frac{\pi \ii}{4}}$ & $\frac{5}{8}$\\
         \cline{2-10}
         & $1$ & $\ii$ & $\ii$ & $-1$ & $e^{\frac{\pi \ii}{4}}$ & $e^{-\frac{\pi \ii}{4}}$ & $e^{-\frac{\pi \ii}{4}}$ & $e^{-\frac{3\pi \ii}{4}}$ & $\frac{1}{8}$\\
         \cline{2-10}
         & $1$ & $\ii$ & $-\ii$ & $1$ & $1$ & $-\ii$ & $\ii$ & $1$ & $0$\\
         \cline{2-10}
         & $1$ & $-\ii$ & $\ii$ & $1$ & $1$ & $\ii$ & $-\ii$ & $1$ & $0$\\
         \cline{2-10}
         & $1$ & $\ii$ & $-\ii$ & $1$ & $-1$ & $\ii$ & $-\ii$ & $-1$ & $\frac{1}{2}$\\
         \cline{2-10}
         & $1$ & $-\ii$ & $\ii$ & $1$ & $-1$ & $-\ii$ & $\ii$ & $-1$ & $\frac{1}{2}$\\
         \hline\hline
    \end{tabular}
    \caption{Topological sectors of $\mathcal{H}_{\mathcal{N}}$  together with their charges solved by performing similar calculations as in Appendix \ref{app:lassocompose}. Notice that in this case, the spin $s$ can be determined using the fact that $e^{2\pi \ii s}$ is the same as the $\mathcal{U}_{\mathcal{N},\dsi}$ action.}
    \label{tab:Nsector}
\end{table}

To conclude this section, we argue that the quantum symmetry after gauging is again $\Rep(H_8)$. 
As pointed out in \cite{2016arXiv160304318M}, gauging any possible algebra object (not necessarily of the form \eqref{eq:algebra_expansion}) in $\Rep(H_8)$ leads to either $\Rep(H_8)$ or $\VEC_{D_8}^\omega$ for some $[\omega] \in H^3(D_8,U(1))$. 
Obviously the two can be distinguished by the number of invertible lines. 
It is straightforward to check using \eqref{eq:bimodule_condition} that there are only $4$ indecomposable bimodule objects with quantum dimension $1$. 
Hence, the quantum symmetry must be $\Rep(H_8)$.

\section{Gauging $\Rep(H_8)$ at $c = 1$} \label{sec:gauging_c=1}

The primary example with the $\Rep(H_8)$ symmetry that we consider will be $c=1$ CFTs \cite{Thorngren:2021yso}. 
In particular, we show below that a stack of two decoupled Ising CFTs (Ising$^2$ in short) is invariant under gauging its $\Rep(H_8)$ symmetry.
This in turn implies the existence of a new topological defect line in the Ising$^2$ CFT coming from the half-space gauging of the $\Rep(H_8)$ symmetry, which will be discussed in Section \ref{sec:new_line}.

Before we begin, we briefly review the $\Rep(H_8)$ symmetry that is realized at $c=1$, following \cite{Thorngren:2021yso}.
First, recall that the moduli space of (known) $c=1$ CFTs consists of two continuous branches and three isolated points \cite{Ginsparg:1987eb}, as shown in Figure \ref{fig:moduli_space}.

One of the continuous branches, called the circle branch, corresponds to the free compact boson CFTs parameterized by the radius $R$ of the compact boson.
The other continuous branch, also parameterized by $R$, arises from gauging the charge conjugation symmetry (denoted as $\mathbb{Z}_2^C$) of the free boson CFTs, and we call it the orbifold branch.
The orbifold branch and circle branch meet at the Berezinskii–Kosterlitz–Thouless (BKT) transition point where the minimal winding vertex operator $V_{0,1}$ becomes relevant.
In our convention, the BKT point corresponds to $R=2$ of the circle branch and $R=1$ of the orbifold branch, and $\mathsf{T}$-duality acts as $R \leftrightarrow 1/R$.

As was discussed in \cite{Thorngren:2021yso}, the $\Rep(H_8)$ fusion category symmetry is realized at every point along the orbifold branch.
The simplest way to understand this is as follows.
First, at $R = \sqrt{2}$ on the orbifold branch, the theory corresponds to the Ising$^2$ CFT \cite{Ginsparg:1988ui}, which has the $\TY(\IZ_2)_+ \boxtimes \TY(\IZ_2)_+$ fusion category symmetry.
Denote the simple topological lines of the two $\TY(\IZ_2)_+$ fusion categories as $\{1,a,\mathcal{N}_1\}$ and $\{1,b,\mathcal{N}_2\}$, respectively, where $a$, $b$ are $\mathbb{Z}_2$ symmetry lines and $\mathcal{N}_1$, $\mathcal{N}_2$ are Kramers-Wannier duality lines.
The $\Rep(H_8)$ symmetry is realized as a subcategory of the $\TY(\IZ_2)_+ \boxtimes \TY(\IZ_2)_+$ fusion category, consisting of the simple lines $\{1,a,b,ab,\mathcal{N}\}$ where $\mathcal{N} \equiv \mathcal{N}_1 \mathcal{N}_2$.

To arrive at other values of $R \neq \sqrt{2}$ on the orbifold branch, one can deform the Ising$^2$ CFT by the exactly marginal operator $\epsilon_1 \epsilon_2$, where $\epsilon_i$'s stand for the energy operators with $(h,\bar{h}) = (\frac{1}{2},\frac{1}{2})$ from the two Ising factors.
Such a deformation explicitly breaks the $\TY(\IZ_2)_+ \boxtimes \TY(\IZ_2)_+$ fusion category symmetry down to its $\Rep(H_8)$ subcategory, and this shows that the $\Rep(H_8)$ symmetry is preserved across the entire orbifold branch.

Throughout the paper, we consider the gauging of symmetries only at the level of torus partition functions.
Generally, the torus partition function alone may not be enough to fully determine a 1+1d CFT.
However, at $c=1$, all the known CFTs have distinct torus partition functions (up to $\mathsf{T}$-duality) and hence computing the torus partition function will be sufficient for identifying the theory after gauging, assuming that the classification in \cite{Ginsparg:1987eb} is complete.

Finally, as a side remark, we mention that there is another $\mathbb{Z}_2 \times \mathbb{Z}_2$ Tambara-Yamagami fusion category, that is equivalent to the representation category $\Rep(D_8)$ of the dihedral group of order 8, which also exists as a symmetry along the orbifold branch \cite{Thorngren:2021yso}. 
The $\Rep(D_8)$ fusion category has the same fusion algebra as $\Rep(H_8)$, but has different $F$-symbols \cite{Bhardwaj:2017xup,Thorngren:2019iar}.
The $\Rep(D_8)$ symmetry is free of an 't Hooft anomaly, and one may also consider gauging the $\Rep(D_8)$ symmetry on the orbifold branch. 
We will leave this analysis for the future, and focus on the $\Rep(H_8)$ symmetry in this work.
It is worth mentioning, however, that there is one additional feature that appears when gauging the $\Rep(D_8)$ symmetry which does not occur in the case of $\Rep(H_8)$. 
Namely, there are three inequivalent ways to gauge the $\Rep(D_8)$ symmetry, whereas for the $\Rep(H_8)$ symmetry we do not have such multiple options.
This is due to the fact that $\Rep(D_8)$ admits three distinct fiber functors, whereas $\Rep(H_8)$ admits a unique fiber functor \cite{Thorngren:2019iar}.

\subsection{Ising$^2$ is self-dual under gauging $\Rep(H_8)$}\label{sec:Ising2}

Before we discuss the $c=1$ CFTs on the orbifold branch at a generic value of $R$, here we first show that the Ising$^2$ CFT (corresponding to $R=\sqrt{2}$) is self-dual under gauging $\Rep(H_8)$.
The computation is simpler in this case since the $\Rep(H_8)$ symmetry commutes with the fully extended chiral algebra, namely two copies of the Ising chiral algebra ($c=\frac{1}{2}$ Virasoro algebra), with respect to which the theory is rational.
With respect to the extended chiral algebra, there are 9 primary operators,
\begin{equation} \label{eq:Ising2_primaries}
    1, \quad \sigma_1, \quad \sigma_2, \quad \epsilon_1,
    \quad \epsilon_2, \quad \sigma_1 \sigma_2, \quad \sigma_1 \epsilon_2, \quad \epsilon_1 \sigma_2 , \quad \epsilon_1 \epsilon_2 \,, 
\end{equation}
where $\sigma_i$ and $\epsilon_i$ are the spin and energy operators coming from the two Ising factors, respectively.
The spin operators have conformal weights $(\frac{1}{16},\frac{1}{16})$ and the energy operators have conformal weights $(\frac{1}{2},\frac{1}{2})$.

We now proceed to gauge the $\Rep(H_8)$ symmetry of the Ising$^2$ CFT on a torus.
From the general analysis in Section \ref{sec:gauging_noninv}, the torus partition function after gauging is given by
\begin{equation}\label{eq:RepH8_gauging_Ising2}
\begin{aligned}
    Z_{\text{Ising}^2/\Rep(H_8)}(\tau) &= \frac{1}{4} \Big(
        - Z_{\text{Ising}^2}[\dsi,\dsi,\dsi](\tau) \\
        &\quad\quad+ Z_{\text{Ising}^2}[\dsi,a,a](\tau) +
        Z_{\text{Ising}^2}[a,\dsi,a](\tau) + Z_{\text{Ising}^2}[a,a,\dsi](\tau) \\
        &\quad\quad +  Z_{\text{Ising}^2}[\dsi,b,b](\tau) +
        Z_{\text{Ising}^2}[b,\dsi,b](\tau) + Z_{\text{Ising}^2}[b,b,\dsi](\tau) \\
         &\quad\quad +  Z_{\text{Ising}^2}[\dsi,ab,ab](\tau) +
        Z_{\text{Ising}^2}[ab,\dsi,ab](\tau) + Z_{\text{Ising}^2}[ab,ab,\dsi](\tau) \\
        &\quad\quad + Z_{\text{Ising}^2}[\dsi,\mathcal{N},\mathcal{N}](\tau) +
        Z_{\text{Ising}^2}[\mathcal{N},\dsi,\mathcal{N}](\tau) + Z_{\text{Ising}^2}[\mathcal{N},\mathcal{N},\dsi](\tau) \\
        &\quad\quad + Z_{\text{Ising}^2}[ab,\mathcal{N},\mathcal{N}](\tau) +
        Z_{\text{Ising}^2}[\mathcal{N},ab,\mathcal{N}](\tau) + Z_{\text{Ising}^2}[\mathcal{N},\mathcal{N},ab](\tau)
    \Big) \,.
\end{aligned}
\end{equation}
The twisted partition functions appearing on the RHS of \eqref{eq:RepH8_gauging_Ising2} can be computed straightforwardly, since they are simply products of two decoupled Ising CFT partition functions twisted by various topological lines.
Generally, torus partition functions of the diagonal minimal models (such as the Ising CFT) twisted by arbitrary topological lines are known \cite{Petkova:2000ip,Chang:2018iay}.
To begin with, we have
\begin{equation}
    Z_{\text{Ising}^2}(\tau) \equiv Z_{\text{Ising}^2}[\dsi,\dsi,\dsi](\tau) = \left(|\chi_0^\text{Ising}(\tau)|^2 + |\chi_\frac{1}{16}^\text{Ising}(\tau)|^2 + |\chi_\frac{1}{2}^\text{Ising}(\tau)|^2 \right)^2 \,,
\end{equation}
where the Ising characters are
\begin{align} \label{eq:Ising_characters}
\begin{split}
    \chi_0^\text{Ising} &= \frac{1}{2} \left( \sqrt{\frac{\vartheta_3}{\eta}} + \sqrt{\frac{\vartheta_4}{\eta}}\right) \,, \\
    \chi_{\frac{1}{2}}^\text{Ising} &= \frac{1}{2} \left( \sqrt{\frac{\vartheta_3}{\eta}} - \sqrt{\frac{\vartheta_4}{\eta}}\right) \,, \\
    \chi_{\frac{1}{16}}^\text{Ising} &= \frac{1}{\sqrt{2}} \sqrt{\frac{\vartheta_2}{\eta}} \,.
\end{split}
\end{align}

The partition functions where one of the $\mathbb{Z}_2$ symmetry lines, $a$, $b$, or $ab$, is inserted along the spatial cycle can be read off from the action of the $\mathbb{Z}_2$ symmetries on the primaries \eqref{eq:Ising2_primaries}:
\begin{align} \label{eq:Ising2_Z2_spatial}
\begin{split}
    Z_{\text{Ising}^2}[\dsi,a,a](\tau) &= Z_{\text{Ising}^2}[\dsi,b,b](\tau) \\
    &= \Big(|\chi_0^\text{Ising}(\tau)|^2 - |\chi_\frac{1}{16}^\text{Ising}(\tau)|^2 + |\chi_\frac{1}{2}^\text{Ising}(\tau)|^2 \Big) \\ &~~~~~~~~~~~~\times
    \left(|\chi_0^\text{Ising}(\tau)|^2 + |\chi_\frac{1}{16}^\text{Ising}(\tau)|^2 + |\chi_\frac{1}{2}^\text{Ising}(\tau)|^2 \right) \,, \\
    Z_{\text{Ising}^2}[\dsi,ab,ab](\tau) &= \left(|\chi_0^\text{Ising}(\tau)|^2 - |\chi_\frac{1}{16}^\text{Ising}(\tau)|^2 + |\chi_\frac{1}{2}^\text{Ising}(\tau)|^2 \right)^2 \,.
\end{split}
\end{align}
Recall that the modular $S$ and $T$ matrices for the Ising characters \eqref{eq:Ising_characters} are given by
\begin{equation} \label{eq:Ising_S_T}
    S= \frac{1}{2}\begin{pmatrix}
        1 & 1 & \sqrt{2} \\
        1 & 1 & -\sqrt{2} \\
        \sqrt{2} & -\sqrt{2} & 0
    \end{pmatrix}, \quad 
    T= e^{-\frac{\pi \ii}{24}} \begin{pmatrix}
        1 & 0 & 0 \\
        0 & -1 & 0 \\
        0 & 0 & e^{\frac{\pi \ii}{8}}
    \end{pmatrix}  \,.
\end{equation}
Applying the $S$-transformation to \eqref{eq:Ising2_Z2_spatial}, we obtain the twisted partition functions where the $\mathbb{Z}_2$ symmetry lines are now wrapping around the Euclidean time cycle:
\begin{align}
\begin{split}
    Z_{\text{Ising}^2}[a,\dsi,a](\tau) &= Z_{\text{Ising}^2}[b,\dsi,b](\tau) \\
    &= \Big(\chi_0^\text{Ising}(\tau) \overline{\chi}_{\frac{1}{2}}^\text{Ising}(\tau)  + \chi_{\frac{1}{2}}^\text{Ising}(\tau) \overline{\chi}_0^\text{Ising}(\tau)
    + |\chi_\frac{1}{16}^\text{Ising}(\tau)|^2  \Big) \\
    &~~~~~~~~~~~~\times
    \left(|\chi_0^\text{Ising}(\tau)|^2 + |\chi_\frac{1}{16}^\text{Ising}(\tau)|^2 + |\chi_\frac{1}{2}^\text{Ising}(\tau)|^2 \right) \,, \\
    Z_{\text{Ising}^2}[ab,\dsi,ab](\tau) &= \left(
        \chi_0^\text{Ising}(\tau) \overline{\chi}_{\frac{1}{2}}^\text{Ising}(\tau) + \chi_{\frac{1}{2}}^\text{Ising}(\tau) \overline{\chi}_0^\text{Ising}(\tau)
    + |\chi_\frac{1}{16}^\text{Ising}(\tau)|^2
    \right)^2 \,.
\end{split}
\end{align}
Further performing an additional $T$-transformation, we get
\begin{align}
\begin{split}
    Z_{\text{Ising}^2}[a,a,\dsi](\tau) &= Z_{\text{Ising}^2}[b,b,\dsi](\tau) \\
    &= \Big(-\chi_0^\text{Ising}(\tau) \overline{\chi}_{\frac{1}{2}}^\text{Ising}(\tau)  - \chi_{\frac{1}{2}}^\text{Ising}(\tau) \overline{\chi}_0^\text{Ising}(\tau)
    + |\chi_\frac{1}{16}^\text{Ising}(\tau)|^2  \Big) \\
    &~~~~~~~~~~~~\times
    \left(|\chi_0^\text{Ising}(\tau)|^2 + |\chi_\frac{1}{16}^\text{Ising}(\tau)|^2 + |\chi_\frac{1}{2}^\text{Ising}(\tau)|^2 \right) \,, \\
    Z_{\text{Ising}^2}[ab,ab,\dsi](\tau) &= \left(-
        \chi_0^\text{Ising}(\tau) \overline{\chi}_{\frac{1}{2}}^\text{Ising}(\tau) - \chi_{\frac{1}{2}}^\text{Ising}(\tau) \overline{\chi}_0^\text{Ising}(\tau)
    + |\chi_\frac{1}{16}^\text{Ising}(\tau)|^2
    \right)^2 \,.
\end{split}
\end{align}

Next, if we insert the $\mathcal{N} \equiv \CN_1 \CN_2$ line along the spatial cycle, we obtain
\begin{equation}
    Z_{\text{Ising}^2}[\dsi,\CN,\CN](\tau) =
    2 \left(|\chi_0^\text{Ising}(\tau)|^2 - |\chi_\frac{1}{2}^\text{Ising}(\tau)|^2 \right)^2 \,,
\end{equation}
since the spin operators map to non-local operators and do not contribute to the torus partition function, and the energy operators flip the sign under the action of $\CN$.
The factor of $2$ comes from the quantum dimension of the line, $\langle \CN \rangle = 2$.
Performing the $S$-transformation, we obtain
\begin{equation}
    Z_{\text{Ising}^2}[\CN,\dsi,\CN](\tau) =
    \left(
        \chi_0^\text{Ising}(\tau) \overline{\chi}_{\frac{1}{16}}(\tau) + \chi_{\frac{1}{2}}^\text{Ising}(\tau) \overline{\chi}_{\frac{1}{16}}(\tau) +  c.c.
    \right)^2 \,.
\end{equation}
Now, we apply the identity \eqref{eq:RepH8_twisted_1} that was derived in Section \ref{sec:gauging_noninv} to obtain
\begin{align} \label{eq:Ising2_N1N_NabN}
\begin{split}
    &Z_{\text{Ising}^2}[\mathcal{N},\dsi,\mathcal{N}](\tau) + Z_{\text{Ising}^2}[\mathcal{N},ab,\mathcal{N}](\tau) \\
    &=  Z_{\text{Ising}^2}[\mathcal{N},\dsi,\mathcal{N}](\tau+2) + Z_{\text{Ising}^2}[\mathcal{N},\dsi,\mathcal{N}](\tau-2) \\
    &= \left(
        e^{-\frac{2\pi \ii}{8}} \chi_0^\text{Ising}(\tau) \overline{\chi}_{\frac{1}{16}}(\tau) + e^{-\frac{2\pi \ii}{8}} \chi_{\frac{1}{2}}^\text{Ising}(\tau) \overline{\chi}_{\frac{1}{16}}(\tau) +  c.c.
    \right)^2 \\
    & \qquad + \left(
        e^{\frac{2\pi \ii}{8}} \chi_0^\text{Ising}(\tau) \overline{\chi}_{\frac{1}{16}}(\tau) + e^{\frac{2\pi \ii}{8}} \chi_{\frac{1}{2}}^\text{Ising}(\tau) \overline{\chi}_{\frac{1}{16}}(\tau) +  c.c.
    \right)^2 \,.
\end{split}
\end{align}
Alternatively, we can also separately obtain $Z_{\text{Ising}^2}[\mathcal{N},ab,\mathcal{N}](\tau)$ as a product of two Ising CFT partition functions, namely $Z_{\text{Ising}^2}[\mathcal{N},ab,\mathcal{N}](\tau)= Z_{\text{Ising}}[\CN_1,a,\CN_1](\tau) \times Z_{\text{Ising}}[\CN_2,b,\CN_2](\tau)$, since $\CN = \CN_1 \CN_2$.
However, to compute the RHS of \eqref{eq:RepH8_gauging_Ising2}, the combination \eqref{eq:Ising2_N1N_NabN} is sufficient.
Finally, we use \eqref{eq:RepH8_twisted_2} to obtain
\begin{align}
\begin{split}
    Z_{\text{Ising}^2}[\dsi,\CN,\CN](\tau) + Z_{\text{Ising}^2}[ab,\CN,\CN](\tau) &= Z_{\text{Ising}^2}[\mathcal{N},\dsi,\mathcal{N}](-1/\tau) + Z_{\text{Ising}^2}[\mathcal{N},ab,\mathcal{N}](-1/\tau) \,, \\
    Z_{\text{Ising}^2}[\CN,\CN,\dsi](\tau) + Z_{\text{Ising}^2}[\CN,\CN,ab](\tau) &= Z_{\text{Ising}^2}[\mathcal{N},\dsi,\mathcal{N}](\tau+1) + Z_{\text{Ising}^2}[\mathcal{N},ab,\mathcal{N}](\tau+1) \,.
\end{split}
\end{align}
The explicit expressions in terms of the Ising characters are easily obtained from \eqref{eq:Ising2_N1N_NabN} and the $S$ and $T$ matrices in \eqref{eq:Ising_S_T}.

Plugging in all the ingredients into the RHS of \eqref{eq:RepH8_gauging_Ising2}, we finally obtain
\begin{align}
\begin{split}
    Z_{\text{Ising}^2/\Rep(H_8)}(\tau) &= \left(|\chi_0^\text{Ising}(\tau)|^2 + |\chi_\frac{1}{16}^\text{Ising}(\tau)|^2 + |\chi_\frac{1}{2}^\text{Ising}(\tau)|^2 \right)^2 \\
    &= Z_{\text{Ising}^2}(\tau) \,.
\end{split}
\end{align}
We conclude that the Ising$^2$ CFT is invariant under gauging its $\Rep(H_8)$ symmetry as claimed.
To be more precise, we confirmed this fact only at the level of the torus partition function.
As was mentioned earlier, this is sufficient to prove the invariance of the full theory under the gauging, under the assumption that the classification of $c=1$ CFTs given in \cite{Ginsparg:1987eb} is complete.

\subsection{$c = 1$ orbifold branch: $R \leftrightarrow 2/R$}

Next, we move on to gauge the $\Rep(H_8)$ symmetry at a generic point on the orbifold branch.
We will find that the theory at radius $R$ is mapped to that at radius at $2/R$ and vice versa under gauging the $\Rep(H_8)$ symmetry.
The Ising$^2$ CFT at $R=\sqrt{2}$ is a fixed point under this gauging, consistent with the analysis in the previous subsection.

\subsubsection{Circle branch}
We begin by briefly reviewing various facts about the $c=1$ compact boson along the circle branch to set up the conventions.
The action is
\begin{equation}
    S = \frac{R^2}{4\pi} \int d\phi \wedge \star d\phi
      = \frac{R^2}{2\pi} \int d^2 z \partial \phi \bar{\partial} \phi \,
\end{equation}
where $\phi \sim \phi + 2\pi$.
Our convention for the radius $R$ is such that the $\mathsf{T}$-duality acts as $R \leftrightarrow 1/R$ and the self-dual radius is at $R=1$, where the theory is described by the $SU(2)_1$ WZW model.
The model has a symmetry
\begin{equation}
    \left(U(1)_m \times U(1)_w \right) \rtimes \mathbb{Z}_2^C \,.
\end{equation}
The momentum symmetry $U(1)_m$ is generated by the current $J_m = \frac{-\ii R^2}{2\pi} \star d\phi$ and the winding symmetry $U(1)_w$ is generated by the current $J_w = \frac{1}{2\pi} d\phi$.
The conservation equations are given by $dJ_m = dJ_w = 0$.
The charge conjugation symmetry $\mathbb{Z}_2^C$ acts as $\phi \rightarrow -\phi$.

At every radius $R$, there is a $u(1) \times \overline{u(1)}$ current algebra generated by $j(z) = \partial \phi (z)$ and $\bar{j}(\bar{z}) = \bar{\partial} \phi (\bar{z})$.
The current algebra primaries are the local vertex operators,
\begin{equation}
    V_{n,w} (z,\bar{z}) =  e^{\ii n\phi(z,\bar{z}) + \ii w\tilde{\phi}(z,\bar{z}) }  \,,
\end{equation}
where $n,w \in \mathbb{Z}$ label the charges under the $U(1)_m$ and $U(1)_w$ symmetries, respectively.
Here, $\tilde{\phi}$ is the dual boson satisfying $d\tilde{\phi}=-\ii R^2 \star d\phi$.
The conformal weights of the vertex operators are
\begin{equation}
    h_{n,w} = \frac{1}{4} \left( \frac{n}{R} + wR \right)^2 \,, \quad
    \bar{h}_{n,w} = \frac{1}{4} \left( \frac{n}{R} - wR \right)^2 \,.
\end{equation}
The torus partition function is given by
\begin{equation} \label{eq:circ_Z}
    Z_R^{circ} (\tau) = \frac{1}{|\eta(\tau)|^2} \sum_{n,w \in \mathbb{Z}} q^{\frac{1}{4} \left( \frac{n}{R} + wR \right)^2} \bar{q}^{\frac{1}{4} \left( \frac{n}{R} - wR \right)^2} \,,
\end{equation}
where $q = e^{2\pi \ii \tau}$ as usual. 
Each term of the torus partition function \eqref{eq:circ_Z} is a $u(1) \times \overline{u(1)}$ current algebra character.

Alternatively, we can also consider decomposing the torus partition function \eqref{eq:circ_Z} into the characters of the Virasoro algebra.
This will become useful later when we consider the $\Rep(H_8)$ symmetry and its gauging along the orbifold branch.
At $c=1$, the Virasoro characters are given as follows \cite{Kiritsis:1988et}.
For a generic conformal weight $h$, we have
\begin{equation} \label{eq:vir_char_generic}
    \chi_h(\tau) = \frac{q^h}{\eta(\tau)} \,.
\end{equation}
However, when $h=\ell^2/4$ for some integer $\ell$, there are null states in the Verma module and the character is instead given by
\begin{equation} \label{eq:vir_char_short}
    \chi_{\ell^2/4}(\tau) = \frac{1}{\eta(\tau)}\left( q^{\ell^2/4} - q^{(\ell+2)^2/4} \right) \,.
\end{equation}

The precise spectrum of Virasoro primary operators on the circle branch depends on the value of the radius $R$.
For a generic value of $R$ (such that $R \notin \mathbb{Q}$), the torus partition function \eqref{eq:circ_Z} decomposes into the Virasoro characters as follows:
\begin{equation}
    Z_R^{circ} (\tau) = \frac{1}{|\eta(\tau)|^2} \left( \sum_{\substack{n,w \in \mathbb{Z} \\ (n,w) \neq (0,0)}} q^{\frac{1}{4} \left( \frac{n}{R} + wR \right)^2} \bar{q}^{\frac{1}{4} \left( \frac{n}{R} - wR \right)^2} 
    + \sum_{\ell,m \in \mathbb{Z}_{\geq 0}} \left( q^{\ell^2} - q^{(\ell+1)^2} \right)\left( \bar{q}^{m^2} - \bar{q}^{(m+1)^2} \right) \right) \,.
\end{equation}
From this, we read off the spectrum of Virasoro primary operators at a generic value of $R$:
\begin{align} \label{eq:circle_primaries_generic}
\begin{split}
    &V_{n,w} \,, \quad n,w \in \mathbb{Z} \,, (n,w)\neq (0,0) \,, \quad (h,\bar{h}) = (\frac{1}{4} (\frac{n}{R}+wR)^2, \frac{1}{4} (\frac{n}{R}-wR )^2 ) \,, \\
    &C_{\ell,m} \,, \quad \ell,m \in \mathbb{Z}_{\geq 0} \,, \quad (h,\bar{h}) = (\ell^2 , m^2) \,.
\end{split}
\end{align}
The operators $C_{\ell,m}$ are made out of the (normal-ordered) Schur polynomials of currents $j(z)$, $\bar{j}(\bar{z})$ and their derivatives \cite{wakimoto1983irreducible,Dijkgraaf:1987vp,Dijkgraaf:1989hb}.
For instance, $C_{1,1} = j(z)\bar{j}(\bar{z})$ is the exactly marginal operator (namely the kinetic term), and $C_{0,0}$ is the identity operator.

The $\mathbb{Z}_2^C$ charge conjugation symmetry acts on the Virasoro primaries as
\begin{align} \label{eq:Z2C_action}
\begin{split}
    \mathbb{Z}_2^C: \quad & V_{n,w} \rightarrow V_{-n,-w} \,, \\
    &  C_{\ell,m} \rightarrow (-1)^{\ell+m} C_{\ell,m} \,.
\end{split}
\end{align}
Denote the topological defect line that generates the $\mathbb{Z}_2^C$ symmetry as $\eta$.
From the action of the $\mathbb{Z}_2^C$ symmetry on the primary operators, we can obtain the partition function of the free boson where the $\eta$ line is inserted along the spatial cycle:
\begin{align}
\begin{split}
    Z_R^{circ}[\dsi,\eta,\eta](\tau) &=
    \frac{1}{|\eta(\tau)|^2} \sum_{\ell,m \in \mathbb{Z}_{\geq 0}} (-1)^{\ell + m} (q^{\ell^2} - q^{(\ell+1)^2})(\bar{q}^{m^2} - \bar{q}^{(m+1)^2}) \\
    &=  \frac{|\vartheta_3(\tau)\vartheta_4(\tau)|}{|\eta(\tau)|^2} \,.
\end{split}
\end{align}
By performing the $S$-transformation, we obtain the twisted partition function
\begin{equation} \label{eq:circ_eta_twisted}
    Z^{circ}_{R}[\eta,\dsi,\eta](\tau) = \frac{1}{|\eta(\tau)|^2} \sum_{\ell,m \in \mathbb{Z}_{\geq 0}} 2 q^{\frac{1}{4}(\ell + \frac{1}{2})^2} \bar{q}^{\frac{1}{4}(m+\frac{1}{2})^2} = \frac{|\vartheta_3(\tau)\vartheta_2(\tau)|}{|\eta(\tau)|^2} \,.
\end{equation}
From \eqref{eq:circ_eta_twisted}, we read off the spectrum of Virasoro primary operators in the twisted Hilbert space associated with $\eta$ \cite{Dijkgraaf:1989hb}.
First of all, we see that the twisted sector operators are doubly-degenerate, which is a consequence of the fact that the $\mathbb{Z}_2^m \times \mathbb{Z}_2^w$ subgroup of the momentum and winding symmetries acts projectively on the $\mathbb{Z}_2^C$ twisted sector due to a mixed anomaly between the three symmetries \cite{Thorngren:2021yso}.
We denote the Virasoro primaries in the $\mathbb{Z}_2^C$ twisted sector as
\begin{equation} \label{eq:eta_twisted}
    D^{(i)}_{\ell,m} \,, \quad \ell,m \in  \mathbb{Z}_{\geq 0} \,, i=1,2 \,, \quad (h,\bar{h}) = (\frac{1}{4} (\ell + \frac{1}{2})^2, \frac{1}{4}(m + \frac{1}{2})^2 ) \,.
\end{equation}
The $\mathbb{Z}_2^C$ charge of the twisted sector operators \eqref{eq:eta_twisted} is given by $2(h-\bar{h}) = \frac{1}{2}(\ell -m)(\ell+m+1)$ \cite{Chang:2018iay,Lin:2021udi}.
Namely,
\begin{align} \label{eq:Z2C_twisted_action}
\begin{split}
    \mathbb{Z}_2^C: \quad & D^{(i)}_{\ell,m} \rightarrow
 (-1)^{\frac{1}{2}(\ell -m)(\ell+m+1)} D^{(i)}_{\ell,m} \,.
\end{split}
\end{align}

\subsubsection{Orbifold branch and its $\Rep(H_8)$ symmetry}\label{sec:OrbifoldBranch}

By gauging the $\mathbb{Z}_2^C$ charge conjugation symmetry of the compact boson, we obtain the orbifold branch theories, which we again label by the radius $R$.
The torus partition function is given by
\begin{align} \label{eq:orb1}
\begin{split}
    Z_R^{orb}(\tau) &= \frac{1}{2} \left( 
       Z_R^{circ}[\dsi,\dsi,\dsi](\tau) + Z_R^{circ}[\dsi,\eta,\eta](\tau) + Z_R^{circ}[\eta,\dsi,\eta](\tau) + Z_R^{circ}[\eta,\eta,\dsi](\tau) \right) \\
       &= \frac{1}{2} \left( 
       Z_R^{circ}(\tau) + \frac{|\vartheta_3(\tau)\vartheta_4(\tau)|}{|\eta(\tau)|^2} + \frac{|\vartheta_3(\tau)\vartheta_2(\tau)|}{|\eta(\tau)|^2} + \frac{|\vartheta_4(\tau)\vartheta_2(\tau)|}{|\eta(\tau)|^2}
    \right) \,.
\end{split}
\end{align}
Primary operators on the orbifold branch consist of the $\mathbb{Z}_2^C$-invariant operators of the circle branch, including the twisted sector operators \eqref{eq:eta_twisted}.
At a generic radius $R$, they are
\begin{align} \label{eq:orb_primaries_generic}
\begin{split}
    &V_{n,w}^+ \equiv \frac{1}{\sqrt{2}} (V_{n,w} + V_{-n,-w}) \,, \quad n, w \in \mathbb{Z}  \,, (n,w)\neq (0,0) \,, \quad (h,\bar{h}) = (\frac{1}{4} (\frac{n}{R}+wR)^2, \frac{1}{4} (\frac{n}{R}-wR)^2) \,, \\
    &C_{\ell,m} \,, \quad \ell,m \in \mathbb{Z}_{\geq 0} \,, \ell + m \in 2\mathbb{Z} \,, \quad (h,\bar{h}) = (\ell^2 , m^2) \,, \\
    &D^{(i)}_{\ell,m} \,, \quad \ell,m \in  \mathbb{Z}_{\geq 0} \,, i=1,2 \,, \frac{1}{2}(\ell -m)(\ell+m+1) \in 2\mathbb{Z} \,, \quad (h,\bar{h}) = (\frac{1}{4}(\ell + \frac{1}{2})^2, \frac{1}{4}(m + \frac{1}{2})^2 ) \,,
\end{split}
\end{align}
modulo the identification $V_{n,w}^+ = V_{-n,-w}^+$.
At the Ising$^2$ point, the exactly marginal operator $C_{1,1}$ coincides with the $\epsilon_1 \epsilon_2$ operator.

We may decompose the torus partition function on the orbifold branch \eqref{eq:orb1} into the Virasoro characters (assuming a generic value of $R$),
\begin{align} \label{eq:orb2}
\begin{split}
    Z_R^{orb}(\tau) &= \frac{1}{|\eta(\tau)|^2} \Big\{ \frac{1}{2} \sum_{\substack{n,w \in \mathbb{Z} \\ (n,w) \neq (0,0)}}  q^{\frac{1}{4} \left( \frac{n}{R} + wR \right)^2} \bar{q}^{\frac{1}{4} \left( \frac{n}{R} - wR \right)^2} 
     + \sum_{\substack{\ell,m \in \mathbb{Z}_{\geq 0} \\ \ell + m \in 2\mathbb{Z}}} \left( q^{\ell^2} - q^{(\ell+1)^2} \right)\left( \bar{q}^{m^2} - \bar{q}^{(m+1)^2} \right) 
    \\
    & \quad +\sum_{\substack{\ell,m \in \mathbb{Z}_{\geq 0} \\  \frac{1}{2}(\ell -m)(\ell+m+1) \in 2\mathbb{Z}}} 2 q^{\frac{1}{4}(\ell+\frac{1}{2})^2} \bar{q}^{\frac{1}{4}(m + \frac{1}{2})^2}
    \Big\}
     \,.
\end{split}
\end{align}
The factor of $1/2$ in front of the first summation is to account for the double counting coming from the identification $V_{n,w}^+ = V_{-n,-w}^+$.
The first two terms in \eqref{eq:orb2} add up to the first two terms of \eqref{eq:orb1}, and the last term of \eqref{eq:orb2} is the same as the sum of the last two terms of \eqref{eq:orb1}.

The action of the $\Rep(H_8)$ symmetry on the Virasoro primary operators \eqref{eq:orb_primaries_generic} along the orbifold branch is given in \cite{Thorngren:2021yso}, which we quote below.
First, the two $\mathbb{Z}_2$ symmetries (denoted as $\mathbb{Z}_{2}^a$ and $\mathbb{Z}_{2}^b$) act as
\begin{align} \label{eq:Z2a_action}
\begin{split}
    \mathbb{Z}_{2}^a: \quad & V_{n,w}^+ \rightarrow (-1)^n V_{n,w}^+ \,, \\
    & (D^{(1)}_{\ell,m},D^{(2)}_{\ell,m}) \rightarrow (-i D^{(2)}_{\ell,m},iD^{(1)}_{\ell,m}) \,,\\
    &C_{\ell,m} \rightarrow C_{\ell,m} \,,
\end{split}
\end{align}
and
\begin{align} \label{eq:Z2b_action}
\begin{split}
    \mathbb{Z}_{2}^b: \quad & V_{n,w}^+ \rightarrow (-1)^n V_{n,w}^+ \,, \\
    & (D^{(1)}_{\ell,m},D^{(2)}_{\ell,m}) \rightarrow (iD^{(2)}_{\ell,m},-iD^{(1)}_{\ell,m}) \,,\\
    &C_{\ell,m} \rightarrow C_{\ell,m} \,,
\end{split}
\end{align}
respectively.
The operators $C_{\ell,m}$ are invariant under $\mathbb{Z}_2^a \times \mathbb{Z}_2^b$.
The diagonal subgroup $\mathbb{Z}_2^{ab} \subset \mathbb{Z}_2^a \times \mathbb{Z}_2^b$ corresponds to the quantum symmetry of the charge conjugation $\mathbb{Z}_2^C$ on the circle branch, which can be seen from the fact that all the twisted sector operators $D_{\ell,m}^{(i)}$ are odd under $\mathbb{Z}_2^{ab}$, whereas the other operators $V_{n,w}^+$, $C_{\ell,m}$ are even.

On the other hand, the line $\mathcal{N}$ acts as
\begin{align} \label{eq:N_action}
\begin{split}
    \mathcal{N}: \quad & V_{n,w}^+ \rightarrow 2i^n (-1)^w V_{n,w}^+ \quad{\text{for $n\in 2\mathbb{Z}$}} \,, \\
    & C_{\ell,m} \rightarrow 2C_{\ell,m} \,,\\
    & \text{other operators} \rightarrow \text{non-local operators} \,.
\end{split}
\end{align}
The factors of 2 come from the quantum dimension of the $\mathcal{N}$ line.
The action indicates that the operators $C_{\ell,m}$ also commute with the $\mathcal{N}$ line, and hence with the entire $\Rep(H_8)$.
For instance, the exactly marginal operator $C_{1,1}$ lets us move up and down on the orbifold branch, while preserving $\Rep(H_8)$.

\subsubsection{Gauging $\Rep(H_8)$} \label{sec:gaugereph8}

We now study the $\Rep(H_8)$ gauging on the orbifold branch. 
As discussed in Section \ref{sec:RepH8_gauging}, we only need to know the action of the $\Rep(H_8)$ symmetry on the Hilbert space, or equivalently, on the local primary operators to compute the relevant twisted partition functions, which has been spelled out above.
For simplicitly, below we perform the computation at a generic irrational point.
However, the conclusion remains the same for any values of $R$.

From \eqref{eq:RepH8_gauging_general_1}, we have
\begin{align} \label{eq:RepH8_gauging_2}
\begin{split}
    Z^{orb/\Rep(H_8)}_R(\tau) &= \frac{1}{4} \Big(
        - Z_R^{orb}[\dsi,\dsi,\dsi](\tau) \\
        &\quad\quad+ Z_R^{orb}[\dsi,a,a](\tau) +
        Z_R^{orb}[a,\dsi,a](\tau) + Z_R^{orb}[a,a,\dsi](\tau) \\
        &\quad\quad +  Z_R^{orb}[\dsi,b,b](\tau) +
        Z_R^{orb}[b,\dsi,b](\tau) + Z_R^{orb}[b,b,\dsi](\tau) \\
         &\quad\quad +  Z_R^{orb}[\dsi,ab,ab](\tau) +
        Z_R^{orb}[ab,\dsi,ab](\tau) + Z_R^{orb}[ab,ab,\dsi](\tau) \\
        &\quad\quad + Z_R^{orb}[\dsi,\mathcal{N},\mathcal{N}](\tau) +
        Z_R^{orb}[\mathcal{N},\dsi,\mathcal{N}](\tau) + Z_R^{orb}[\mathcal{N},\mathcal{N},\dsi](\tau) \\
        &\quad\quad + Z_R^{orb}[ab,\mathcal{N},\mathcal{N}](\tau) +
        Z_R^{orb}[\mathcal{N},ab,\mathcal{N}](\tau) + Z_R^{orb}[\mathcal{N},\mathcal{N},ab](\tau)
    \Big) \,.
\end{split}
\end{align}
From the action of $\mathbb{Z}_2^a$, $\mathbb{Z}_2^b$ symmetries on the Virasoro primaries \eqref{eq:Z2a_action}, \eqref{eq:Z2b_action}, we read off
\begin{align}
\begin{split}
    & \quad\quad Z_R^{orb}[\dsi,a,a](\tau) = Z_R^{orb}[\dsi,b,b](\tau) \\
    &=
     \frac{1}{2} \frac{1}{|\eta(\tau)|^2} \sum_{\substack{n,w \in \mathbb{Z} \\ (n,w) \neq (0,0)}} (-1)^n q^{\frac{1}{4} \left( \frac{n}{R} + wR \right)^2} \bar{q}^{\frac{1}{4} \left( \frac{n}{R} - wR \right)^2} 
    + \frac{1}{|\eta(\tau)|^2} \sum_{\substack{\ell,m \in \mathbb{Z}_{\geq 0} \\ \ell + m \in 2\mathbb{Z}}} \left( q^{\ell^2} - q^{(\ell+1)^2} \right)\left( \bar{q}^{m^2} - \bar{q}^{(m+1)^2} \right) \\
    &= \frac{1}{2} \frac{1}{|\eta(\tau)|^2} \left( 
         \sum_{n,w \in \mathbb{Z}} (-1)^n q^{\frac{1}{4} \left( \frac{n}{R} + wR \right)^2} \bar{q}^{\frac{1}{4} \left( \frac{n}{R} - wR \right)^2} 
    +  \sum_{\ell,m \in \mathbb{Z}} (-1)^{\ell + m} q^{\ell^2}\bar{q}^{m^2}
    \right) \\
    &= \frac{1}{2} \frac{1}{|\eta(\tau)|^2} \left( 
         \sum_{n,w \in \mathbb{Z}} (-1)^n q^{\frac{1}{4} \left( \frac{n}{R} + wR \right)^2} \bar{q}^{\frac{1}{4} \left( \frac{n}{R} - wR \right)^2} 
    +  |\vartheta_3(\tau) \vartheta_4(\tau)|
    \right)
    \,,
\end{split}
\end{align}
where in the second step we have used
\begin{equation}
     \sum_{\substack{\ell,m \in \mathbb{Z}_{\geq 0} \\ \ell + m \in 2\mathbb{Z}}} \left( q^{\ell^2} - q^{(\ell+1)^2} \right)\left( \bar{q}^{m^2} - \bar{q}^{(m+1)^2} \right) = \frac{1}{2}\sum_{\ell,m \in \mathbb{Z}} (-1)^{\ell + m} q^{\ell^2}\bar{q}^{m^2} + \frac{1}{2} \,,
\end{equation}
and absorbed $\frac{1}{2}$ into the $(n,w)=(0,0)$ term of the first summation.
In the last step, we used $\sum\limits_n (-1)^n q^{n^2} = \vartheta_4(2\tau) = \sqrt{\vartheta_3(\tau) \vartheta_4(\tau)}$.

The modular $S$-transformation is computed using the Poisson resummation, and we obtain 
\begin{align}
\begin{split}
    & \quad\quad Z_R^{orb}[a,\dsi,a](\tau) = Z_R^{orb}[b,\dsi,b](\tau) \\
    &= \frac{1}{|\eta(\tau)|^2} \left( \frac{1}{2} 
         \sum_{n,w \in \mathbb{Z}} q^{\frac{1}{4} \left( \frac{w}{R} + (n+\frac{1}{2})R \right)^2} \bar{q}^{\frac{1}{4} \left( \frac{w}{R} - (n+\frac{1}{2})R \right)^2} 
    +  \frac{1}{4} \sum_{\ell,m \in \mathbb{Z}} q^{\frac{1}{4}(\ell+\frac{1}{2})^2}\bar{q}^{\frac{1}{4}(m+\frac{1}{2})^2}
    \right)  \\
    &= \frac{1}{2} \frac{1}{|\eta(\tau)|^2} \left( 
         \sum_{n,w \in \mathbb{Z}} q^{\frac{1}{4} \left( \frac{w}{R} + (n+\frac{1}{2})R \right)^2} \bar{q}^{\frac{1}{4} \left( \frac{w}{R} - (n+\frac{1}{2})R \right)^2} 
    +   |\vartheta_2(\tau)\vartheta_3(\tau)|
    \right)
    \,.
\end{split}
\end{align}
We can see that the result can be combined into a non-negative integer summation over Virasoro characters if we appropriately restrict the range of the summation variables.
The spin selection rule for a non-anomaous $\mathbb{Z}_2$ symmetry, $h- \bar{h} \in \mathbb{Z}/2$ is also satisfied \cite{Lin:2021udi}.
Further performing the $T$-transformation, we obtain
\begin{align}
\begin{split}
    & \quad\quad Z_R^{orb}[a,a,\dsi](\tau) = Z_R^{orb}[b,b,\dsi](\tau) \\
    &= \frac{1}{|\eta(\tau)|^2} \left( \frac{1}{2} 
         \sum_{n,w \in \mathbb{Z}} (-1)^w q^{\frac{1}{4} \left( \frac{w}{R} + (n+\frac{1}{2})R \right)^2} \bar{q}^{\frac{1}{4} \left( \frac{w}{R} - (n+\frac{1}{2})R \right)^2} 
    +  \frac{1}{4} \sum_{\ell,m \in \mathbb{Z}} (-1)^{\frac{1}{2}(\ell-m)(\ell+m+1)} q^{\frac{1}{4}(\ell+\frac{1}{2})^2}\bar{q}^{\frac{1}{4}(m+\frac{1}{2})^2}
    \right) \\
    &= \frac{1}{2}\frac{1}{|\eta(\tau)|^2} \left(  
         \sum_{n,w \in \mathbb{Z}} (-1)^w q^{\frac{1}{4} \left( \frac{w}{R} + (n+\frac{1}{2})R \right)^2} \bar{q}^{\frac{1}{4} \left( \frac{w}{R} - (n+\frac{1}{2})R \right)^2} 
    +   |\vartheta_2(\tau)\vartheta_4(\tau)|
    \right)
    \,.
\end{split}
\end{align}

Next, recall that the diagonal $\mathbb{Z}_2^{ab} \subset \mathbb{Z}_2^a \times \mathbb{Z}_2^b$ symmetry generated by the line $ab$ is the quantum symmetry of the $\mathbb{Z}_2^C$ charge conjugation symmetry on the circle branch \cite{Thorngren:2021yso}.
Therefore, gauging $\mathbb{Z}_2^{ab}$ brings us back to the circle branch,
\begin{equation}
    Z_R^{circ}(\tau) = \frac{1}{2} \left(
         Z_R^{orb}[\dsi,\dsi,\dsi](\tau) +  Z_R^{orb}[\dsi,ab,ab](\tau) +
        Z_R^{orb}[ab,\dsi,ab](\tau) + Z_R^{orb}[ab,ab,\dsi](\tau)
    \right) \,.
\end{equation}
From this, we obtain
\begin{align}
\begin{split}
     &\quad Z_R^{orb}[\dsi,ab,ab](\tau) +
        Z_R^{orb}[ab,\dsi,ab](\tau) + Z_R^{orb}[ab,ab,\dsi](\tau) \\
        &=
        2Z_R^{circ}(\tau) - Z_R^{orb}[\dsi,\dsi,\dsi](\tau) \\
        &= \frac{3}{2} \frac{1}{|\eta(\tau)|^2} \sum_{n,w \in \mathbb{Z}} q^{\frac{1}{4} \left( \frac{n}{R} + wR \right)^2} \bar{q}^{\frac{1}{4} \left( \frac{n}{R} - wR \right)^2} \\
        &\quad\quad\quad - \frac{1}{2|\eta(\tau)|^2} \left( 
        |\vartheta_3(\tau)\vartheta_4(\tau)| +
        |\vartheta_2(\tau)\vartheta_3(\tau)| + |\vartheta_2(\tau)\vartheta_4(\tau)|
     \right) \,.
\end{split}
\end{align}

Finally, we have \cite{Thorngren:2021yso}
\begin{align}
\begin{split}
    &\quad Z^{orb}_R[\mathcal{N},\dsi,\mathcal{N}](\tau) \\
    &= \frac{1}{2|\eta(\tau)|^2} \left(
        \sum_{m,n \in \mathbb{Z}} q^{\frac{1}{4}\left(\frac{n+1/2}{R}+ \frac{(m+1/2)R}{2}  \right)^2 }
        \bar{q}^{\frac{1}{4}\left(\frac{n+1/2}{R} - \frac{(m+1/2)R}{2}  \right)^2 }
        + \sum_{m,n \in\mathbb{Z}} q^{\frac{1}{4}(m+1/2)^2} \bar{q}^{\frac{1}{4}(n+1/2)^2}
    \right) \,.
\end{split}
\end{align}
By applying the identity \eqref{eq:RepH8_twisted_1}, we obtain
\begin{align}
\begin{split}
    Z^{orb}_R[\mathcal{N},\dsi,\mathcal{N}](\tau) +  Z^{orb}_R[\mathcal{N},ab,\mathcal{N}](\tau) &= \frac{1}{|\eta(\tau)|^2}\sum_{m,n\in\mathbb{Z}} q^{\frac{1}{4}(m+1/2)^2} \bar{q}^{\frac{1}{4}(n+1/2)^2}  \\
    &= \frac{2}{|\eta(\tau)|^2} |\vartheta_2(\tau)\vartheta_3(\tau)| \,.
\end{split}
\end{align}
We then use the identities \eqref{eq:RepH8_twisted_2} to obtain
\begin{align}
\begin{split}
    Z^{orb}_R[1,\mathcal{N},\mathcal{N}](\tau) +  Z^{orb}_R[ab,\mathcal{N},\mathcal{N}](\tau) = \frac{2}{|\eta(\tau)|^2} |\vartheta_3(\tau)\vartheta_4(\tau)| \,, \\
    Z^{orb}_R[\mathcal{N},\mathcal{N},1](\tau) +  Z^{orb}_R[\mathcal{N},\mathcal{N},ab](\tau) = \frac{2}{|\eta(\tau)|^2} |\vartheta_2(\tau)\vartheta_4(\tau)| \,.
\end{split}
\end{align}

By plugging all the twisted partition functions into the RHS of \eqref{eq:RepH8_gauging_2}, we find
\begin{equation} 
    Z^{orb/\Rep(H_8)}_R(\tau) = Z^{orb}_{2/R}(\tau) \,.
\end{equation}
That is, the orbifold theories at radii $R$ and $2/R$ are mapped into each other under gauging the $\Rep(H_8)$ symmetry as claimed.
In particular, at the Ising$^2$ point $R=\sqrt{2}$, we confirm again that the theory is self-dual under gauging the $\Rep(H_8)$ symmetry.

\section{A new topological defect line in the Ising$^2$ CFT} \label{sec:new_line}

The fact that the Ising$^2$ CFT is invariant under gauging the $\Rep(H_8)$ symmetry implies the existence of a new topological defect line $\mathcal{D}$ coming from the half-space gauging of the $\Rep(H_8)$ symmetry, accroding to the general considerations in Section \ref{sec:gauging_noninv}.
In this section, we (partially) bootstrap the action of the topological line $\mathcal{D}$ on the $c=1$ Virasoro primary operators.
Upon the modular $S$-transformation, we then also obtain the spectrum of operators in the twisted Hilbert space associated with $\mathcal{D}$.

The topological line $\mathcal{D}$ does not commute with the fully extended chiral algebra, but preserves the $c=1$ Virasoro algebra.
Moreover, there does not exist any relevant or marginal operator which is invariant under the action of $\mathcal{D}$.
The exactly marginal operator $\epsilon_1 \epsilon_2 = C_{1,1}$ anticommutes with $\mathcal{D}$.
The latter property is reminiscent of the fact that the energy operator in the Ising CFT anticommutes with the Kramers-Wannier duality line.
See \cite{Thorngren:2019iar,Thorngren:2021yso} for many other known topological defect lines in the Ising$^2$ CFT.

\subsection{Virasoro primaries}

The fact that the new topological line $\CD$ does not commute with the fully extended chiral algebra of the Ising$^2$ CFT, namely two copies of the Ising chiral algebra, is easy to check, since the Ising$^2$ CFT is a diagonal rational CFT for that extended chiral algebra.
In any diagonal rational CFT, the full spectrum of topological line operators which commute with the extended chiral algebra is known, where they are referred to as Verlinde lines \cite{Verlinde:1988sn,Petkova:2000ip}.
The Verlinde lines in the Ising$^2$ CFT form the fusion category $\TY(\mathbb{Z}_2)_+ \boxtimes \TY(\mathbb{Z}_2)_+$, and by inspecting the fusion algebra, it is straightforward to confirm that $\CD$ is not one of the Verlinde lines.

Being topological, $\CD$ commutes with the $c=1$ Virasoro algebra, and we will attempt to find the action of $\CD$ on the $c=1$ Virasoro primaries.
The spectrum of Virasoro primaries at a generic point on the orbifold branch was reviewed in Section \ref{sec:gauging_c=1} (see \eqref{eq:orb_primaries_generic}).
However, at $R = \sqrt{2}$ corresponding to the Ising$^2$ CFT, there are additional null states that do not appear at generic values of $R$, and the full spectrum of Virasoro primaries is more involved.
Below we list all the $c=1$ Virasoro primaries of the Ising$^2$ CFT and their conformal weights:
\begin{align} \label{eq:orb_primaries}
\begin{split}
    &V_{n,w}^+ \,, \quad n, w \in \mathbb{Z}  \,, n \neq \pm 2w \,, \quad (h,\bar{h}) = (\frac{1}{8} (n+2w)^2, \frac{1}{8} (n-2w)^2) \,, \\
    &A_{w,m}^+ \,, \quad w \in \mathbb{Z}_{>0} \,, m \in \mathbb{Z}_{\geq 0} \,, \quad (h,\bar{h}) = (2w^2 , m^2) \,, \\
    &B_{w,\ell}^+ \,, \quad w \in \mathbb{Z}_{>0} \,, \ell \in \mathbb{Z}_{\geq 0} \,, \quad (h,\bar{h}) = (\ell^2 , 2w^2) \,, \\
    &C_{\ell,m} \,, \quad \ell,m \in \mathbb{Z}_{\geq 0} \,, \ell + m \in 2\mathbb{Z} \,, \quad (h,\bar{h}) = (\ell^2 , m^2) \,, \\
    &D^{(i)}_{\ell,m} \,, \quad \ell,m \in  \mathbb{Z}_{\geq 0} \,, i=1,2 \,, \frac{1}{2}(\ell -m)(\ell+m+1) \in 2\mathbb{Z} \,, \quad (h,\bar{h}) = (\frac{1}{4}(\ell + \frac{1}{2})^2, \frac{1}{4}(m + \frac{1}{2})^2 ) \,.
\end{split}
\end{align}
We review \eqref{eq:orb_primaries} in more detail in Appendix \ref{app:more_Ising2}.
The operators $V_{n,w}^+$, $C_{\ell,m}$ and $D_{\ell,m}^{(i)}$ are identical to those in \eqref{eq:orb_primaries_generic} at generic $R$, whereas the operators $A_{w,m}^+$ and $B_{w,\ell}^+$ are new primary operators appearing at $R=\sqrt{2}$ due to some of the generic modules at generic $R$ splitting into infinitely-many degenerate modules at $R=\sqrt{2}$.
There is an identification $V_{n,w}^+ = V_{-n,-w}^+$, and to avoid double-counting, we may restrict to $n\geq 1, w\in \mathbb{Z}$ or $n=0, w \geq 1$.

The torus partition function of the Ising$^2$ CFT decomposes into a sum over the $c=1$ Virasoro characters given in \eqref{eq:vir_char_generic} and \eqref{eq:vir_char_short} as follows:
\begin{align}
\begin{split}
    Z_{\text{Ising}^2} (\tau) = &\frac{1}{|\eta(\tau)|^2} \Big\{
        \sum_{\substack{n \in \mathbb{Z}_{\geq 1}, w \in \mathbb{Z} \\ n \neq \pm 2w}} q^{\frac{1}{8}(n+2w)^2} \bar{q}^{\frac{1}{8}(n-2w)^2} + \sum_{w \in \mathbb{Z}_{\geq 1}} q^{\frac{1}{2} w^2} \bar{q}^{\frac{1}{2}w^2}  \\
        &+\sum_{w \in \mathbb{Z}_{\geq 1}, m \in \mathbb{Z}_{\geq 0} } q^{2w^2} \left( \bar{q}^{m^2} - \bar{q}^{(m+1)^2} \right) + \sum_{w \in \mathbb{Z}_{\geq 1}, \ell \in \mathbb{Z}_{\geq 0} } \left( q^{\ell^2} - q^{(\ell+1)^2} \right) \bar{q}^{2w^2} \\
        &+\sum_{\substack{\ell,m \in \mathbb{Z}_{\geq 0} \\ \ell +m \in 2\mathbb{Z} }} \left( q^{\ell^2} - q^{(\ell+1)^2} \right) \left( \bar{q}^{m^2} - \bar{q}^{(m+1)^2} \right) +\sum_{\substack{\ell,m \in \mathbb{Z}_{\geq 0} \\  \frac{1}{2}(\ell -m)(\ell+m+1) \in 2\mathbb{Z}}} 2 q^{\frac{1}{4}(\ell+\frac{1}{2})^2} \bar{q}^{\frac{1}{4}(m + \frac{1}{2})^2}
    \Big\} \,.
\end{split}
\end{align}
Individual terms appearing in the torus partition function can be recognized as the contributions coming from the $c=1$ Virasoro families corresponding to \eqref{eq:orb_primaries}.

\subsection{Bootstrapping the action of $\CD$}

Here, we will partially determine the action of $\CD$ on the Virasoro primary operators \eqref{eq:orb_primaries} by demanding the following set of conditions:
\begin{itemize}
    \item Consistency with the fusion algebra \eqref{eq:fusion_rule_general_2}
    \item Well-defined defect Hilbert space of $\CD$
    \item Spin selection rule \eqref{eq:spinselection}
\end{itemize}
Let us elaborate on the conditions in more detail.
First, recall that (from \eqref{eq:fusion_rule_general_2})
\begin{equation}
    \CD \otimes \CD = 1 \oplus a \oplus b \oplus ab \oplus 2\CN \,.
\end{equation}
This fusion algebra implies that $\CD$ acts non-invertibly on local operators which are not invariant under the $\Rep(H_8)$ symmetry.
That is, such operators are annihilated when surrounded by a closed loop of $\CD$, or equivalently, they are mapped to non-local operators when $\CD$ sweeps past them.
Moreover, on the local operators which are $\Rep(H_8)$-invariant, the action of $\CD$ is order 2, up to the quantum dimension $\langle \CD \rangle = \sqrt{8}$.
Specifically, we have
\begin{align} \label{eq:action_condition_1}
\begin{split}
    \CD \cdot (\CD \cdot \mathcal{O}) &= 8\mathcal{O} \quad \text{if $\mathcal{O}$ is $\Rep(H_8)$-invariant} \,,\\
    \CD \cdot \mathcal{O} &= 0 \quad \text{if $\mathcal{O}$ is not $\Rep(H_8)$-invariant} \,,
\end{split}
\end{align}
where the notation $\CD \cdot \mathcal{O}$ means the action of $\CD$ on a local operator $\mathcal{O}$ by surrounding $\mathcal{O}$ with a closed loop of $\CD$ (see Figure \ref{fig:Laction}).
By the state-operator map, such an action is related to the action of $\CD$ as an operator on the state $\ket{\mathcal{O}}$.

Second, we require the twisted Hilbert space associated with $\CD$ to decompose properly into a direct sum of $c=1$ Virasoro representations \cite{Petkova:2000ip}. 
Namely, we require the twisted torus partition function to decompose into $c=1$ Virasoro characters with non-negative integer coefficients,
\begin{equation} \label{eq:D_twisted}
    Z_{\text{Ising}^2} [\mathcal{D},\dsi,\mathcal{D}] (\tau) \stackrel{!}{=} \sum_{h,\bar{h}} n_{h,\bar{h}} \chi_h (\tau) \overline{\chi}_{\bar{h}}(\tau) \,, \quad n_{h,\bar{h}}\in \mathbb{Z}_{\geq 0} \,.
\end{equation}
Such a condition is also known as a modular bootstrap condition.
Since the twisted partition function \eqref{eq:D_twisted} is related by the $S$-transformation to the partition function where $\CD$ is inserted along the spatial cycle, $Z_{\text{Ising}^2} [\mathcal{D},\dsi,\mathcal{D}] (\tau) =  Z_{\text{Ising}^2} [\dsi,\mathcal{D},\mathcal{D}] (-1/\tau)$, the condition that $n_{h,\bar{h}}\in \mathbb{Z}_{\geq 0}$ significantly constrains the action of $\CD$ on the primaries.

Finally, the spin $s= h-\bar{h}$ of the operators in the twisted Hilbert space of $\CD$ is constrained to take only a particular set of values, shown in \eqref{eq:spinselection}.
Such conditions on the spin of twisted sector operators are commonly referred to as spin selection rules \cite{Chang:2018iay}.
The spin selection rule depends on the structure of the full fusion category formed by $\CD$ and the topological lines of $\Rep(H_8)$, namely it is determined by the fusion algebra as well as the $F$-symbols.
In Section \ref{sec:F}, by solving the pentagon equations, we show that there are 8 fusion categories with the fusion algebra \eqref{eq:fusion_intro}.
All of the 8 fusion categories are distinguished from each other by different spin selection rules.
Here, we will use the spin selection rule \eqref{eq:spinselection} but postpone its derivation until Section \ref{sec:F}.
The spin selection rule not only imposes a consistency condition on the action of $\CD$, but it also lets us to determine which of the 8 fusion categories are actually realized in the Ising$^2$ CFT once the action of $\CD$ is obtained.

We now proceed to find the action of $\CD$ on the primaries which solves the consistency conditions listed above.
In particular, we focus on the action of $\CD$ on the $\Rep(H_8)$-invariant operators.
The $\mathbb{Z}_2^a$ and $\mathbb{Z}_2^b$ symmetries act on $V_{n,w}^+$ as $V_{n,w}^+ \rightarrow (-1)^n V_{n,w}^+$, and leaves $A_{w,m}^+$, $B_{w,\ell}^+$, $C_{\ell,m}$ invariant.
They also act nontrivially on $D^{(i)}_{\ell,m}$ operators for all values of $\ell$ and $m$.
The line $\mathcal{N}$ acts on the Virasoro primaries as \cite{Thorngren:2021yso}
\begin{align}
\begin{split}
    \mathcal{N}: \quad & V_{2k,w}^+ \rightarrow 2(-1)^{k+w} V_{2k,w}^+ \,, \\
    & \text{$A_{w,m}^+$, $B_{w,\ell}^+$, $C_{\ell,m}$ are left invariant up to quantum dimension} \,, \\
    & \text{other operators $\rightarrow$ non-local operators} \,.
\end{split}
\end{align}
We see that $V_{2k,w}^+$ with $k+w \in 2\mathbb{Z}$ as well as $A_{w,m}^+$, $B_{w,\ell}^+$, $C_{\ell,m}$ operators are invariant under the $\text{Rep}(H_8)$-symmetry.
Taking into account the fact that $\mathcal{O}$ and $\CD \cdot \mathcal{O}$ should have the same conformal dimensions $(h,\bar{h})$, the general ansatz for the action of $\mathcal{D}$ we can consider is
\begin{align} \label{eq:D_ansatz}
\begin{split}
    \mathcal{D}: \quad & \begin{pmatrix}
        V_{2k,w}^+ \\
        V_{2w,k}^+
    \end{pmatrix}  \rightarrow \sqrt{8} X_{k,w} \begin{pmatrix}
        V_{2k,w}^+ \\
        V_{2w,k}^+
    \end{pmatrix} \,, \\
    &  A_{w,m}^+ \rightarrow \sqrt{8} (-1)^{f (w,m)} A_{w,m}^+ \,, \\
    &  B_{w,\ell}^+ \rightarrow \sqrt{8} (-1)^{g (w,\ell)} B_{w,\ell}^+ \,, \\
    &  C_{\ell,m} \rightarrow \sqrt{8} (-1)^{h (\ell,m)} C_{\ell,m}  \,, \\
    & \text{other operators $\rightarrow$ non-local operators} \,.
\end{split}
\end{align}
Here, $X_{k,w}$ are $2\times 2$ matrices satisfying $X_{k,w}^2 = 1$, and $f$, $g$, $h$ are $\mathbb{Z}_2$-valued functions.
Note that $V_{2k,w}^+$ and $V_{2w,k}^+$ have the same conformal dimensions, and they can mix with each other under the action of $\CD$ (also recall that $k\neq w$ from \eqref{eq:orb_primaries}).
The general ansatz \eqref{eq:D_ansatz} solves the first consistency condition \eqref{eq:action_condition_1}.

It turns out that the set of consistency conditions we impose is not sufficient to fully determine the form of the matrices $X_{k,w}$.
However, we propose that
\begin{equation}
    \text{Tr}\, X_{k,w} = 0 \quad \text{for all $k$ and $w$} \,.
\end{equation}
Under this assumption, we can write down the torus partition function with the $\CD$ line wrapping around the spatial cycle as
\begin{align}
\begin{split}
    Z_{\text{Ising}^2} [\dsi,\mathcal{D},\mathcal{D}] (\tau) = &\frac{\sqrt{8}}{|\eta(\tau)|^2} \Big\{
        \sum_{w \in \mathbb{Z}_{\geq 1}, m \in \mathbb{Z}_{\geq 0} } (-1)^{f(w,m)} q^{2w^2} \left( \bar{q}^{m^2} - \bar{q}^{(m+1)^2} \right) \\
        &+ \sum_{w \in \mathbb{Z}_{\geq 1}, \ell \in \mathbb{Z}_{\geq 0} } (-1)^{g(w,\ell)} \left( q^{\ell^2} - q^{(\ell+1)^2} \right) \bar{q}^{2w^2} \\
        &+\sum_{\substack{\ell,m \in \mathbb{Z}_{\geq 0} \\ \ell +m \in 2\mathbb{Z} }} (-1)^{h(\ell,m)} \left( q^{\ell^2} - q^{(\ell+1)^2} \right) \left( \bar{q}^{m^2} - \bar{q}^{(m+1)^2} \right)
    \Big\} \,.
\end{split}
\end{align}
The $A_{w,m}^+$, $B_{w,\ell}^+$, $C_{\ell,m}$ operators contribute to the above partition function.

Next, we proceed to constrain the $\mathbb{Z}_2$-valued functions $f$, $g$, $h$ by imposing the condition \eqref{eq:D_twisted}.
We find that the following solves \eqref{eq:D_twisted}:
\begin{align} \label{eq:Z2_functions}
\begin{split}
    f(w,m) &= m + \alpha w   \,,\\
    g(w,\ell) &= \ell + \beta w  \,,\\
    h(\ell,m) &= \ell   \,.
\end{split}   
\end{align}
Here, $\alpha, \beta = 0,1$, whose values will be further constrained below by imposing the spin selection rule.
More specifically, \eqref{eq:Z2_functions} leads to
\begin{align} \label{eq:D_twisted_2}
\begin{split}
    Z_{\text{Ising}^2} [\mathcal{D},\dsi,\mathcal{D}] (\tau) = &\frac{1}{|\eta(\tau)|^2} \Big\{
        \sum_{w \in \mathbb{Z}, m \in \mathbb{Z}_{\geq 0} } q^{\frac{1}{8}(w+ \frac{\alpha}{2})^2}  \bar{q}^{\frac{1}{4}(m+\frac{1}{2})^2} + \sum_{w \in \mathbb{Z}, \ell \in \mathbb{Z}_{\geq 0} }  q^{\frac{1}{4}(\ell+\frac{1}{2})^2} \bar{q}^{\frac{1}{8}(w+ \frac{\beta}{2})^2} 
    \Big\} \,,
\end{split}
\end{align}
and we explicitly see that the condition \eqref{eq:D_twisted} is satisfied.

The twisted torus partition function \eqref{eq:D_twisted_2} provides us with the information about the spectrum of operators in the twisted Hilbert space of $\CD$.
In particular, we see the primary operators of conformal dimensions $(\frac{1}{8}(w+ \frac{\alpha}{2})^2,\frac{1}{4}(m+\frac{1}{2})^2)$ for all $w \in \mathbb{Z}$, $m \in \mathbb{Z}_{\geq 0}$, as well as those with dimensions $(\frac{1}{4}(\ell+\frac{1}{2})^2,\frac{1}{8}(w+ \frac{\beta}{2})^2)$ for all $\ell \in \mathbb{Z}_{\geq 0}$, $w \in \mathbb{Z}$.
To further constrain the values of $\alpha$ and $\beta$, we now impose the spin selection rule \eqref{eq:spinselection}.
Namely, as explained in more detail in Section \ref{sec:F}, the spin $s = h-\bar{h}$ mod $\frac{1}{2}$ of every operator in the twisted sector of $\CD$ must take one of the 4 allowed values appearing in one of the 8 rows in \eqref{eq:spinselection}.
A priori, we do not know which of the 8 fusion categories, corresponding to the 8 rows in \eqref{eq:spinselection}, is realized by $\CD$.
We find that
\begin{align}
\begin{split}
    (\alpha,\beta) = (0,1) \quad \text{is consistent with $\underline{\mathcal{E}}^{(-,+,+)}_{\doubleZ_2}\Rep(H_8)$ in \tabref{tab:fsymbolsb}} \,, \\
    (\alpha,\beta) = (1,0) \quad \text{is consistent with $\underline{\mathcal{E}}^{(+,+,+)}_{\doubleZ_2}\Rep(H_8)$ in \tabref{tab:fsymbolsa}} \,,
\end{split}
\end{align}
whereas $(\alpha,\beta) = (0,0)$ and $(\alpha,\beta) = (1,1)$ do not satisfy any of the 8 spin selection rules in \eqref{eq:spinselection}.

Therefore, we have so far two possible candidate actions of $\CD$ corresponding to the choice of $(\alpha,\beta) = (0,1)$ or $(\alpha,\beta) = (1,0)$.
In fact, these two choices are related by another $\mathbb{Z}_2$-symmetry that is present in the Ising$^2$ CFT, namely the $\mathbb{Z}_2$ symmetry which swaps the two Ising factors.
We denote this symmetry as $\mathbb{Z}_2^{\text{swap}}$, and the corresponding topological line as $r$.
It combines with $\mathbb{Z}_2^a \times \mathbb{Z}_2^b$ to form a dihedral group of order 8.
By tracking the decomposition of the Ising$^2$ characters into the $c=1$ Virasoro characters, one can show that $\mathbb{Z}_2^{\text{swap}}$ acts on $A_{w,m}^+$ and $B_{w,\ell}^+$ as $(-1)^w$, and composing $\CD$ with the $\mathbb{Z}_2^{\text{swap}}$ line has the effect of shifting both $\alpha$ and $\beta$ by 1 mod 2, see \eqref{eq:D_ansatz} and \eqref{eq:Z2_functions}.

Without loss of generality, we may let $\CD$ to be the topological line corresponding to the solution $(\alpha,\beta) = (0,1)$.
Then, another topological line $\CD' \equiv \CD r$ obtained by composing $\CD$ with the $\mathbb{Z}_2^{\text{swap}}$-symmetry line $r$ also satisfy the same fusion algebra \eqref{eq:fusion_intro}, but it has different $F$-symbols and correspond to the other solution $(\alpha,\beta) = (1,0)$.
To conclude, we find that two of the 8 possible fusion categories, that are found in Section \ref{sec:F}, are realized in the Ising$^2$ CFT.
Namely, the two $\IZ_2$-extensions of $\Rep(H_8)$, denoted as $\underline{\mathcal{E}}^{(-,+,+)}_{\doubleZ_2}\Rep(H_8)$ and $\underline{\mathcal{E}}^{(+,+,+)}_{\doubleZ_2}\Rep(H_8)$, are realized respectively by $\CD$ and $\CD'$.
The notation is explained in more detail in Section \ref{sec:F}.
The Fronbenius-Schur indicator for the $\CD$ line is $\epsilon_{\CD} = +1$, and similarly for $\CD'$.

As claimed earlier, we see that there are no relevant operators preserving $\CD$, simply due to the fact there are no relevant operators preserving $\Rep(H_8)$ and \eqref{eq:action_condition_1}.
Furthermore, the only marginal operator $C_{1,1}$, which is $\Rep(H_8)$-invariant, anticommutes with $\CD$, namely $\CD \cdot C_{1,1} = - \sqrt{8} C_{1,1}$ where the factor of $\sqrt{8}$ comes from the quantum dimension of $\CD$.
The fact that $C_{1,1}$ should anticommute with $\CD$ is intuitively clear.
Starting from the Ising$^2$ point on the orbifold branch, one may deform the theory by $C_{1,1}$ either with a positive coefficient or with a negative coefficient.
Since $C_{1,1}$ anticommutes with $\CD$, $\CD$ now becomes a topological interface between the two deformed theories.
This is nothing but the topological interface between the theories at $R$ and $2/R$ (in the vicinity of $R=\sqrt{2}$) coming from the half-space gauging of $\Rep(H_8)$.

\section{Stacking two theories with $\TY(\mathbb{Z}_2)_+$ symmetries} \label{sec:stacking}
In this section, we study gauging the $\Rep(H_8)$ symmetry of the theories obtained by stacking two (potentially different) theories with $\TY(\mathbb{Z}_2)_+$ symmetries. 
Motivated by the previous example, i.e. Ising$^2$, of self-duality under gauging $\Rep(H_8)$, we want to see if it is true that every theory obtained in such a way is self-dual under gauging $\Rep(H_8)$. 
We will start with a generic analysis and show that this is not the case in general. 
We will provide some additional special examples which are self-dual under gauging, and then provide some examples which are not.  

\subsection{General analysis}

Here, we consider taking two generic CFTs, both with $\TY(\mathbb{Z}_2)_+$ fusion category symmetry, and gauge the $\Rep(H_8)$ subcategory. 
We will find that, in general, the theory is not self-dual under gauging $\Rep(H_8)$, and we will also see from the general result why the known example is self-dual under $\Rep(H_8)$ gauging--essentially it requires some factorization property of topological sectors. 
Below, the nontrivial simple lines of the $\TY(\mathbb{Z}_2)_+$ categories of the two CFTs are denoted as $\{\eta_1, \CN_1\}$ and $\{\eta_2, \CN_2\}$, respectively.

First, note that the topological sectors and their modular properties of a 1+1d CFT with a $\TY(\mathbb{Z}_2)_+$ symmetry can be characterized by the anyons in the symTFT $\CZ(\TY(\mathbb{Z}_2)_+) = \Ising \boxtimes \overline{\Ising}$. 
Since $\TY(\mathbb{Z}_2)_+$ fusion category itself can be equipped with a braiding structure to become a modular tensor category (MTC) $\Ising$, the bulk symTFT is simply the $\Ising$ MTC together with its orientation reversal $\overline{\Ising}$. For more details on this theory, see \cite{Barkeshli:2014cna}.

We denote the simple anyons in the 3d $\Ising$ TFT as $(\dsi,\psi,\sigma)$, and the simple anyons in its orientation reversal $\overline{\Ising}$ as $(\overline{\dsi},\overline{\psi},\overline{\sigma})$. Let us order the 9 anyons in $\Ising \times \overline{\Ising}$ as
\begin{equation}
(1,\overline{1}),(1,\overline{\psi}),(1,\overline{\sigma}),(\psi,\overline{1}),(\psi,\overline{\psi}),(\psi,\overline{\sigma}),(\sigma,\overline{1}),(\sigma,\overline{\psi}),(\sigma,\overline{\sigma}) \, .
\end{equation}
In this basis, the $S$ and $T$ matrices are given by
\begin{equation}\label{eq:IsingIsingbarST}
S = \begin{pmatrix}  \frac{1}{4} & \frac{1}{4} & \frac{1}{2 \sqrt{2}} & \frac{1}{4} & \frac{1}{4} &
   \frac{1}{2 \sqrt{2}} & \frac{1}{2 \sqrt{2}} & \frac{1}{2 \sqrt{2}} & \frac{1}{2} \\
 \frac{1}{4} & \frac{1}{4} & -\frac{1}{2 \sqrt{2}} & \frac{1}{4} & \frac{1}{4} &
   -\frac{1}{2 \sqrt{2}} & \frac{1}{2 \sqrt{2}} & \frac{1}{2 \sqrt{2}} & -\frac{1}{2} \\
 \frac{1}{2 \sqrt{2}} & -\frac{1}{2 \sqrt{2}} & 0 & \frac{1}{2 \sqrt{2}} & -\frac{1}{2
   \sqrt{2}} & 0 & \frac{1}{2} & -\frac{1}{2} & 0 \\
 \frac{1}{4} & \frac{1}{4} & \frac{1}{2 \sqrt{2}} & \frac{1}{4} & \frac{1}{4} &
   \frac{1}{2 \sqrt{2}} & -\frac{1}{2 \sqrt{2}} & -\frac{1}{2 \sqrt{2}} & -\frac{1}{2}
   \\
 \frac{1}{4} & \frac{1}{4} & -\frac{1}{2 \sqrt{2}} & \frac{1}{4} & \frac{1}{4} &
   -\frac{1}{2 \sqrt{2}} & -\frac{1}{2 \sqrt{2}} & -\frac{1}{2 \sqrt{2}} & \frac{1}{2}
   \\
 \frac{1}{2 \sqrt{2}} & -\frac{1}{2 \sqrt{2}} & 0 & \frac{1}{2 \sqrt{2}} & -\frac{1}{2
   \sqrt{2}} & 0 & -\frac{1}{2} & \frac{1}{2} & 0 \\
 \frac{1}{2 \sqrt{2}} & \frac{1}{2 \sqrt{2}} & \frac{1}{2} & -\frac{1}{2 \sqrt{2}} &
   -\frac{1}{2 \sqrt{2}} & -\frac{1}{2} & 0 & 0 & 0 \\
 \frac{1}{2 \sqrt{2}} & \frac{1}{2 \sqrt{2}} & -\frac{1}{2} & -\frac{1}{2 \sqrt{2}} &
   -\frac{1}{2 \sqrt{2}} & \frac{1}{2} & 0 & 0 & 0 \\
 \frac{1}{2} & -\frac{1}{2} & 0 & -\frac{1}{2} & \frac{1}{2} & 0 & 0 & 0 & 0 \end{pmatrix},
\end{equation}
\begin{equation}
T = \begin{pmatrix}
    1 & 0 & 0 & 0 & 0 & 0 & 0 & 0 & 0 \\
 0 & -1 & 0 & 0 & 0 & 0 & 0 & 0 & 0 \\
 0 & 0 & e^{-\frac{i \pi }{8}} & 0 & 0 & 0 & 0 & 0 & 0 \\
 0 & 0 & 0 & -1 & 0 & 0 & 0 & 0 & 0 \\
 0 & 0 & 0 & 0 & 1 & 0 & 0 & 0 & 0 \\
 0 & 0 & 0 & 0 & 0 & -e^{-\frac{i \pi }{8}} & 0 & 0 & 0 \\
 0 & 0 & 0 & 0 & 0 & 0 & e^{\frac{i \pi }{8}} & 0 & 0 \\
 0 & 0 & 0 & 0 & 0 & 0 & 0 & -e^{\frac{i \pi }{8}} & 0 \\
 0 & 0 & 0 & 0 & 0 & 0 & 0 & 0 & 1
\end{pmatrix} \, .
\end{equation}
Let us denote the torus partition function of the topological sector corresponding to an anyon $\mathbf{a}$ as $Z_{i,\mathbf{a}}(\tau)$ where $i=1,2$ labels the two stacked CFTs. 
Then, we can express the following twisted partition functions as
\begin{equation}
\begin{aligned}
    Z_i[\dsi,\dsi,\dsi](\tau) & = Z_{i,(\dsi,\overline{\dsi})}(\tau) + Z_{i,(\psi,\overline{\psi})}(\tau) + Z_{i,(\sigma,\overline{\sigma})}(\tau) \, , \\
    Z_i[\eta_i,\dsi,\eta_i](\tau) & = Z_{i,(\dsi,\overline{\psi})}(\tau) + Z_{i,(\psi,\overline{\dsi})}(\tau) + Z_{i,(\sigma,\overline{\sigma})}(\tau) \, , \\
    Z_i[\mathcal{N}_i,\dsi,\mathcal{N}_i](\tau) & = Z_{i,(\dsi,\overline{\sigma})}(\tau) + Z_{i,(\sigma,\overline{\dsi})}(\tau) + Z_{i,(\sigma,\overline{\psi})}(\tau) + Z_{i,(\psi,\overline{\sigma})}(\tau) \, . \\
\end{aligned}
\end{equation}
The twisted partition functions of the stacked theory $\mathcal{T}_1 \times \mathcal{T}_2$ can be computed from the twisted partition functions $Z_i$.
For instance,
\begin{equation}\label{eq:stacked_twisted_partition_function}
\begin{aligned}
    & Z_{\mathcal{T}_1\times \mathcal{T}_2}[\dsi,\dsi,\dsi] = Z_{1}[\dsi,\dsi,\dsi] Z_{2}[\dsi,\dsi,\dsi], \quad Z_{\mathcal{T}_1\times \mathcal{T}_2}[a,\dsi,a] = Z_{1}[\eta_1,\dsi,\eta_1] Z_{2}[\dsi,\dsi,\dsi], \\
    & Z_{\mathcal{T}_1\times \mathcal{T}_2}[b,\dsi,b] = Z_{1}[\dsi,\dsi,\dsi] Z_{2}[\eta_2,\dsi,\eta_2], \quad Z_{\mathcal{T}_1\times \mathcal{T}_2}[ab,\dsi,ab] = Z_1[\eta_1,\dsi,\eta_1]Z_2[\eta_2,\dsi,\eta_2], \\
    & Z_{\mathcal{T}_1\times \mathcal{T}_2}[\mathcal{N},\dsi,\mathcal{N}] = Z_1[\mathcal{N}_1,\dsi,\mathcal{N}_1]Z_2[\mathcal{N}_2,\dsi,\mathcal{N}_2].
\end{aligned}
\end{equation}
Using modular transformations \eqref{eq:IsingIsingbarST} and the general formula in Section \ref{sec:RepH8_gauging}, we find
\begin{equation}\label{eq:TwoCopiesGauging}
\begin{aligned}
    Z_{\mathcal{T}_1\times \mathcal{T}_2/\Rep(H_8)}(\tau) = & Z_{1,(1,\overline{1})}(\tau) Z_{2,(1,\overline{1})}(\tau) + Z_{1,(\psi,\overline{1})}(\tau) Z_{2,(1,\overline{\psi})}(\tau) + Z_{1,(1,\overline{\psi})}(\tau) Z_{2,(\psi,\overline{1})}(\tau) \\
    + & Z_{1,(\psi,\overline{\psi})}(\tau) Z_{2,(\psi,\overline{\psi})}(\tau) + Z_{1,(\psi,\overline{\sigma})}(\tau) Z_{2,(\sigma,\overline{\psi})}(\tau) + Z_{1,(\sigma,\overline{\psi})}(\tau) Z_{2,(\psi,\overline{\sigma})}(\tau) \\
    + & Z_{1,(\sigma,\overline{\sigma})}(\tau) Z_{2,(\sigma,\overline{\sigma})}(\tau) + Z_{1,(1,\overline{\sigma})}(\tau) Z_{2,(\sigma,\overline{1})}(\tau) + Z_{1,(\sigma,\overline{1})}(\tau) Z_{2,(1,\overline{\sigma})}(\tau),
\end{aligned}
\end{equation}
while
\begin{equation}\label{eq:TwoCopies}
\begin{aligned}
    Z_{\mathcal{T}_1\times \mathcal{T}_2}(\tau) = \left(Z_{1,(1,\overline{1})}(\tau) + Z_{1,(\psi,\overline{\psi})}(\tau) + Z_{1,(\sigma,\overline{\sigma})}(\tau)\right)\left(Z_{2,(1,\overline{1})}(\tau) + Z_{2,(\psi,\overline{\psi})}(\tau) + Z_{2,(\sigma,\overline{\sigma})}(\tau)\right).
\end{aligned}
\end{equation}
Thus we have shown that stacking two theories with $\TY(\IZ_2)_+$ symmetries does not necessarily lead to self-duality under gauging $\Rep(H_8)$.

\subsection{Additional examples}
Here, we provide a sufficient condition for self-duality under gauging $\Rep(H_8)$, which allows us to find additional examples which are self-dual. 

Comparing two partition functions \eqref{eq:TwoCopiesGauging} and \eqref{eq:TwoCopies}, we notice that a sufficient condition for the theory $\CT_1 \times \CT_2$ to be self-dual under gauging $\Rep(H_8)$ is
\begin{equation}\label{eq:factori_cond}
    Z_{i,(a,\overline{b})}(\tau) = Z_{i,a}(\tau) Z_{i,\overline{b}}(\tau), \quad \forall a,b,
\end{equation}
as well as
\begin{equation}\label{eq:half_equality_cond}
    Z_{1,a}(\tau) = Z_{2,a}(\tau) \quad \text{or} \quad Z_{1,\overline{a}}(\tau) = Z_{2,\overline{a}}(\tau) .
\end{equation}
Then,
\begin{equation}\label{eq:two_theories_self_dual}
\begin{aligned}
    Z_{\mathcal{T}_1\times \mathcal{T}_2/\Rep(H_8)}(\tau) = &
    \left(Z_{1,1}(\tau) Z_{2,\overline{1}}(\tau) + Z_{1,\psi}(\tau) Z_{2,\overline{\psi}}(\tau) + Z_{1,\sigma}(\tau) Z_{2,\overline{\sigma}}(\tau)\right) \\ 
    & \times \left(Z_{2,1}(\tau) Z_{1,\overline{1}}(\tau) + Z_{2,\psi}(\tau) Z_{1,\overline{\psi}}(\tau) + Z_{2,\sigma}(\tau) Z_{1,\overline{\sigma}}(\tau)\right) \\
    = & \left(Z_{1,1}(\tau) Z_{1,\overline{1}}(\tau) + Z_{1,\psi}(\tau) Z_{1,\overline{\psi}}(\tau) + Z_{1,\sigma}(\tau) Z_{1,\overline{\sigma}}(\tau)\right) \\ 
    & \times \left(Z_{2,1}(\tau) Z_{2,\overline{1}}(\tau) + Z_{2,\psi}(\tau) Z_{2,\overline{\psi}}(\tau) + Z_{2,\sigma}(\tau) Z_{2,\overline{\sigma}}(\tau)\right) 
    & = Z_{\mathcal{T}_1\times \mathcal{T}_2}(\tau),
\end{aligned}
\end{equation}
where in the first equal sign we used \eqref{eq:factori_cond} and the in the second equal sign we used \eqref{eq:half_equality_cond}.

Indeed, in the case of the Ising$^2$ CFT, the partition function for the Ising CFT does factorize as \eqref{eq:factori_cond} with
\begin{equation}
\begin{aligned}
    & \left(Z_{\text{Ising},\dsi}(\tau),Z_{\text{Ising},\psi}(\tau),Z_{\text{Ising},\sigma}(\tau)\right) = \left(\chi_{0}^{\text{Ising}}(\tau),\chi_{\frac{1}{2}}^{\text{Ising}}(\tau),\chi_{\frac{1}{16}}^{\text{Ising}}(\tau)\right), \\
    & \left(Z_{\text{Ising},\overline{\dsi}}(\tau),Z_{\text{Ising},\overline{\psi}}(\tau),Z_{\text{Ising},\overline{\sigma}}(\tau)\right) = \left(\overline{\chi_{0}^{\text{Ising}}(\tau)},\overline{\chi_{\frac{1}{2}}^{\text{Ising}}(\tau)},\overline{\chi_{\frac{1}{16}}^{\text{Ising}}(\tau)}\right).
\end{aligned}
\end{equation}
To search for additional examples, we notice that the partition function of the Monster CFT also has the property \eqref{eq:factori_cond}. 
The Monster CFT is a holomorphic CFT with central charge $c_L = c = 24$ \cite{frenkel1984natural,frenkel1989vertex}, and it enjoys a $\TY(\IZ_2)_+$ symmetry \cite{Lin:2019hks}. 
Its twisted partition functions can be expressed in terms of the Ising characters ($c=1/2$) and the Baby Monster ($c=47/2$) characters in a factorized way as \eqref{eq:factori_cond}, where
\begin{equation}
\begin{aligned}
    & \left(Z_{\text{Monster},\dsi}(\tau),Z_{\text{Monster},\psi}(\tau),Z_{\text{Monster},\sigma}(\tau)\right) = \left(\chi_{0}^{\text{Ising}}(\tau),\chi_{\frac{1}{2}}^{\text{Ising}}(\tau),\chi_{\frac{1}{16}}^{\text{Ising}}(\tau)\right), \\
    & \left(Z_{\text{Monster},\overline{\dsi}}(\tau),Z_{\text{Monster},\overline{\psi}}(\tau),Z_{\text{Monster},\overline{\sigma}}(\tau)\right) = \left(\chi_{0}^{\text{Baby}}(\tau),\chi_{\frac{3}{2}}^{\text{Baby}}(\tau),\chi_{\frac{31}{16}}^{\text{Baby}}(\tau)\right).
\end{aligned}
\end{equation}
Then, by \eqref{eq:factori_cond}, \eqref{eq:half_equality_cond}, \eqref{eq:two_theories_self_dual}, we conclude that both Monster$^2$ CFT and the Ising $\times$ Monster CFT are self-dual under gauging $\Rep(H_8)$ (at the level of the torus partition functions), and therefore it is natural to suspect that the new topological defect line $\CD$ exists in these theories as well.
The self-duality under gauging $\Rep(H_8)$ can alternatively be checked by directly calculating the twisted partition functions of the Monster CFT following \cite{Lin:2019hks}.
More detailed analysis on these new topological lines is left for the future.

\subsection{Non-examples}
In this subsection, we provide two examples which are not self-dual under gauging the $\Rep(H_8)$ symmetry. 
One of them is a nontrivial CFT and the other is a TQFT.

\subsubsection*{Two copies of $U(1)_4$}
Here, we consider a stack of 2 copies of $c = 1$ compact boson theories at radius $R = \sqrt{2}$. 
The compact boson at radius $R = \sqrt{2}$ has $\TY(\mathbb{Z}_2)_+$ symmetry \cite{Thorngren:2021yso}. 
It also enjoys the $U(1)_4$ extended chiral algebra under which the theory is rational.
We have the following twisted partition functions \cite{Thorngren:2021yso}:
\begin{equation}
\begin{aligned}
    & Z_{U(1)_4}[\dsi,\dsi,\dsi] = \sum_{m = -1}^2 |K_m^2(\tau)|^2, \\
    & Z_{U(1)_4}[\dsi,\eta,\eta] = \sum_{m = -1}^2 (-1)^m |K_m^2(\tau)|^2, \\
    & Z_{U(1)_4}[\dsi,\mathcal{N},\mathcal{N}] = \sqrt{2} K_0^2(\tau) \left(\sum_{m\in\doubleZ_{\geq 0}}(-1)^m \chi_{m^2}(\overline{\tau})\right).
\end{aligned}
\end{equation}
Here, $\chi_{m^2}(\tau)$ are the Virasoro characters with scaling dimension $m^2$ (see \eqref{eq:vir_char_short}), and $K_m^2$ are characters of the $U(1)_4$ chiral algebra and given by
\begin{equation}
    K_m^2(\tau) = \frac{1}{\eta(\tau)} \sum_{r\in\doubleZ}q^{2\left(r+\frac{m}{4}\right)^2}, \quad m = -1,0,1,2.
\end{equation}
$K^2_m(\tau)$'s have the modular properties
\begin{equation}
    K_n^2\left(-\frac{1}{\tau}\right) = \frac{1}{2} \sum_{m = -1}^{2}e^{-\frac{\pi i mn}{2}} K_m^2(\tau), \quad K_n^2(\tau+1) = e^{\frac{\pi in^2}{4} - \frac{\pi i}{12}} K_n^2(\tau).
\end{equation}
It is then straightforward to use modular transformations together with \eqref{eq:stacked_twisted_partition_function} to compute the partition function of $\left(U(1)_4 \times U(1)_4\right)/\Rep(H_8)$ following the general analysis in Section \ref{sec:RepH8_gauging}. 
We find
\begin{equation}
\begin{aligned}
    Z_{\left(U(1)_4 \times U(1)_4\right)/\Rep(H_8)}(\tau) &= \frac{1}{2} K_0^2(\tau)^2 \left(\overline{K_0^2(\tau)}^2 + A^2\right) + 4 K_0^2(\tau) K_1^2(\tau) (B^2 + C^2) + 4 K_1^2(\tau)^2 \overline{K_1^2(\tau)}^2 \\
    & \quad + 8 K_2^1(\tau) K_2^2(\tau) B C + 3 K^2_0(\tau) \overline{K^2_0(\tau)} K^2_2(\tau) \overline{K^2_2(\tau)} + \frac{1}{2} K^2_2(\tau)^2 \overline{K^2_2(\tau)}^2 \\
    &= q^{\frac{1}{12}}\oq^{\frac{1}{12}} + 4 q^{\frac{5}{24}}\oq^{\frac{5}{24}} + 4 q^{\frac{1}{3}} \oq^{\frac{1}{3}} + 12 q^{\frac{7}{12}}\oq^{\frac{7}{12}} + \cdots,
\end{aligned}
\end{equation}
where
\begin{equation}
    A = \sum_{m\in\doubleZ_{\geq 0}}(-1)^m\chi_{m^2}(\overline{\tau}), \,\, B = \sum_{\substack{m\in\doubleZ_{\geq 0}, \\ m = 0,3 \mod 4}} \chi_{\frac{(m+1/2)^2}{4}}(\otau), \,\, C = \sum_{\substack{m\in\doubleZ_{\geq 0}, \\ m = 1,2 \mod 4}} \chi_{\frac{(m+1/2)^2}{4}}(\otau).
\end{equation}
On the other hand,
\begin{equation}
    Z_{U(1)_4 \times U(1)_4}(\tau) = \left(\sum_{m = -1}^2 |K_m^2(\tau)|^2\right)^2 = q^{\frac{1}{12}}\oq^{\frac{1}{12}} + 4 q^{\frac{5}{24}}\oq^{\frac{5}{24}} + 4 q^{\frac{1}{3}} \oq^{\frac{1}{3}} + 8 q^{\frac{7}{12}}\oq^{\frac{7}{12}} + \cdots.
\end{equation}
Hence, we conclude that $U(1)_4 \times U(1)_4$ is not self-dual under gauging $\Rep(H_8)$.

\subsubsection*{Regular Ising$^2$ TQFT}
Given a fusion category $\mathcal{C}$, 1+1d $\mathcal{C}$-symmetric TQFTs are classified by the module categories over $\mathcal{C}$ \cite{Huang:2021zvu}.
Here we consider a 1+1d TQFT which has the $\mathcal{C} = \TY(\mathbb{Z}_2)_+ \boxtimes \TY(\mathbb{Z}_2)_+$ symmetry.
The nontrivial simple objects in the first $\TY(\mathbb{Z}_2)_+$ factor are denoted as $\eta_1$, $\mathcal{N}_1$, and those in the second factor as $\eta_2$, $\mathcal{N}_2$.
We will consider the TQFT which corresponds to the regular module category of $\mathcal{C}$, that is, $\mathcal{C}$ viewed as a module category over itself.
The untwisted (closed) Hilbert space $\mathcal{H}$ is spanned by the states labeled by the simple objects in $\mathcal{C}$,
\begin{equation} \label{eq:TQFT_basis}
    \mathcal{H} = \text{span} \left\{ \ket{\dsi}, \ket{\eta_1} , \ket{\eta_2} , \ket{\eta_1 \eta_2} , \ket{\mathcal{N}_1} , \ket{\mathcal{N}_2} , \ket{\eta_2 \mathcal{N}_1} , \ket{\eta_1 \mathcal{N}_2} , \ket{\mathcal{N}} \right\} \,,
\end{equation}
where $\mathcal{N} \equiv \mathcal{N}_1\mathcal{N}_2$.
In particular, the theory has 9 degenerate states, and it represents a phase where the $\mathcal{C}$-symmetry is spontaneously broken completely.
More generally, the twisted Hilbert space of a simple topological line $\mathcal{L}_i$ is given by
\begin{equation}
    \mathcal{H}_{\mathcal{L}_i} \cong \bigoplus_j \text{Hom}(\mathcal{L}_j , \mathcal{L}_i \otimes \mathcal{L}_j) \,.
\end{equation}

We now consider gauging the $\Rep(H_8)$ subcategory of $\mathcal{C} = \TY(\mathbb{Z}_2)_+ \boxtimes \TY(\mathbb{Z}_2)_+$.
The action of $\mathcal{C}$ on $\mathcal{H}$ is determined by the fusion coefficients.
Namely, when a simple topological line $\mathcal{L}_i$ in $\mathcal{C}$ acts on a basis state $\ket{\mathcal{L}_j}$, we have $\mathcal{L}_i \ket{\mathcal{L}_j} = \sum\limits_k N_{ij}^k \ket{\mathcal{L}_k}$, and then the action is linearly extended to arbitrary states in $\mathcal{H}$.
The action on the twisted sectors are determined from the lasso actions.
We can obtain the following twisted torus partition functions which are needed for the $\Rep(H_8)$ gauging:
\begin{align}
\begin{split}
    &Z_{TQFT}[\dsi,\dsi,\dsi] = 9 \,,\\
    &Z_{TQFT}[\dsi,\eta_1,\eta_1] = Z_{TQFT}[\eta_1,\dsi,\eta_1] = Z_{TQFT}[\eta_1,\eta_1,\dsi] = 3 \,, \\
    &Z_{TQFT}[\dsi,\eta_2,\eta_2] = Z_{TQFT}[\eta_2,\dsi,\eta_2] = Z_{TQFT}[\eta_2,\eta_2,\dsi] = 3 \,,\\
    &Z_{TQFT}[\dsi,\eta_1 \eta_2,\eta_1 \eta_2] = Z_{TQFT}[\eta_1 \eta_2,\dsi,\eta_1 \eta_2] = Z_{TQFT}[\eta_1 \eta_2,\eta_1 \eta_2,\dsi] = 1 \,,\\
    &Z_{TQFT}[\eta_1,\eta_2,\eta_1 \eta_2] = Z_{TQFT}[\eta_2,\eta_1,\eta_1 \eta_2] = Z_{TQFT}[\eta_2,\eta_1 b,\eta_1] =1 \,, \\
    &Z_{TQFT}[\eta_1 \eta_2,\eta_1,\eta_2] = Z_{TQFT}[\eta_1,\eta_1 \eta_2,\eta_2] = Z_{TQFT}[\eta_1 \eta_2,\eta_2,\eta_1] = 1 \,,\\
    &Z_{TQFT}[\dsi,\mathcal{N},\mathcal{N}] = Z_{TQFT}[\mathcal{N},\dsi,\mathcal{N}] = Z_{TQFT}[\mathcal{N},\mathcal{N},\dsi] = 0 \,.
\end{split}
\end{align}
The torus partition function after gauging $\Rep(H_8)$ is then given by
\begin{equation}
    Z_{TQFT/\Rep(H_8)} = 3 
\end{equation}
which differs from the original torus partition function $Z_{TQFT} = 9$.\footnote{The resulting theory after gauging $\Rep(H_8)$ would correspond to a module category of $\mathcal{C}$ with 3 simple objects.}
Therefore, this TQFT serves as another non-example, which has the Ising$^2$ fusion category symmetry and yet is not invariant under gauging the $\Rep(H_8)$ subcategory.

\section{Fusion categories including the new topological line: $\underline{\mathcal{E}}^{(i,\fsb,\fsa)}_{\doubleZ_2}\Rep(H_8)$} \label{sec:F}

The fusion rules involving the new topological defect line $\CD$ are given by \eqref{eq:fusion_intro}.
The fusion rules and $F$-symbols only involving $1,a,b,ab,\CN$ are inherited from $\Rep(H_8)$. 
In this section, we solve the pentagon equations with the additional $\CD$ line, and calculate the spin selection rules using the lasso action of topological defect lines on defect operators. 
There are 8 gauge-inequivalent solutions to the pentagon equations, which differ by three $\IZ_2$-valued phases. 
The spin selection rules can unambiguously distinguish all 8 fusion categories. 
In particular, the Ising$^2$ CFT realizes 2 of the 8 solutions in \tabref{tab:fsymbolsa} and \tabref{tab:fsymbolsb}. 
More details on calculations are presented in \appref{app:F_details}.

\subsection{$F$-symbols}

Recall that non-trivial $F$-symbols of $\Rep(H_8)$ are
\begin{align}
\begin{split}
    &[F_{\CN}^{g\CN h}]_{(\CN,1,1),(\CN,1,1)} = \chi(g,h)\,,\quad [F_{h}^{\CN g\CN}]_{(\CN,1,1),(\CN,1,1)} = \chi(g,h)\,,\\
    &[F_{\CN}^{\CN \CN \CN}]_{(g,1,1),(h,1,1)} = \frac{1}{2}\chi^{-1}(g,h) \,.
\end{split}
\end{align}
All other $F$-symbols of $\Rep(H_8)$ are trivial. 
With the additional $\CD$ line, we explicitly solve the pentagon equations.
The 8 inequivalent solutions for the $F$-symbols are listed in \tabref{tab:fsymbolsa} and \tabref{tab:fsymbolsb}.\footnote{$\mathsf{Mathematica}$ files of the solutions for the $F$-symbols are uploaded as Ancillary files on the arXiv.}
We denote these 8 fusion categories as $\underline{\mathcal{E}}^{(i,\fsb,\fsa)}_{\doubleZ_2}\Rep(H_8)$ with $i=\pm,\fsb=\pm,\fsa=\pm$.

\begin{landscape}

\begin{table}[htbp]
    \centering
    \begin{tabular}[t]{c|c}
         \hline\hline
    \makecell{$[F_{\CD}^{gh\CD}]_{(gh,1,1),(\CD,1,1)}$\\$[F_{\CD}^{\CD gh}]_{(\CD,1,1),(gh,1,1)}$} & $\left(\begin{smallmatrix}
             1 & 1 & 1 & 1 \\
 1 & 1 & 1 & 1 \\
 1 & -1 & 1 & -1 \\
 1 & -1 & 1 & -1
         \end{smallmatrix}\right)$\\ \hline
         \makecell{$[F_{\CD}^{g\CN \CD}]_{(\CN,1,\mu),(\CD,\nu,1)}$\\ $[F_{\CD}^{\CD \CN g}]_{(\CD,\mu,1),(\CN,1,\nu)}$}   & $(\sigma^0,\sigma^1,\sigma^3,-\ii \sigma^2)$\\ \hline
          \makecell{$[F_\CD^{\CN g \CD}]_{(\CN,1,\mu),(\CD,1,\nu)}$\\$[F_\CD^{\CD g \CN}]_{(\CD,1,\mu),(\CN,1,\nu)}$} & $(\sigma^0,-\sigma^3,-\sigma^1,-\ii \sigma^2)$ \\ \hline
         \makecell{$[F_\CD^{\CN \CN \CD}]_{(g,1,1),(\CD,\mu,\nu)}$\\$[F_\CD^{\CD \CN \CN}]_{(\CD,\mu,\nu),(g,1,1)}$} & $(\frac{\sigma^1-\sigma^3}{2},\frac{\sigma^0-\ii\sigma^2}{2},-\frac{\sigma^0+\ii \sigma^2}{2},\mp\frac{\sigma^1+\sigma^3}{2})$\\ \hline\hline
         $[F_h^{\CD \CD g}]_{(gh,1,1),(\CD,1,1)}$ &  $\left(
\begin{smallmatrix}
 1 & 1 & 1 & 1 \\
 1 & 1 & 1 & 1 \\
 1 & -1 & 1 & -1 \\
 1 & -1 & 1 & -1 \\
\end{smallmatrix}
\right)$\\\hline
         $[F_\CN^{\CD \CD g}]_{(\CN,\mu,1),(\CD,1,\nu)}$ & $(\sigma^0,\sigma^1,\sigma^3,-\ii \sigma^2)$\\\hline
         $[F_g^{\CD \CD \CN}]_{(\CN,\mu,1),(\CD,\nu,1)}$ &  $(\frac{-\sigma^1+\sigma^3}{\sqrt{2}},\frac{-\sigma^0+\ii \sigma^2}{\sqrt{2}},\frac{\sigma^0+\ii \sigma^2}{\sqrt{2}},\frac{\sigma^1+\sigma^3}{\sqrt{2}})$\\\hline
         $[F_\CN^{\CD \CD \CN}]_{(g,1,1),(\CD,\mu,\nu)}$ &  $(-\frac{\sigma^0}{\sqrt{2}},\frac{\sigma^3}{\sqrt{2}},\frac{\sigma^1}{\sqrt{2}},-\frac{\ii \sigma^2}{\sqrt{2}})$ \\\hline \hline

         $[F_h^{g\CD \CD }]_{(\CD,1,1),(gh,1,1)}$ & $\left(
\begin{smallmatrix}
 1 & 1 & 1 & 1 \\
 1 & 1 & -1 & -1 \\
 1 & 1 & 1 & 1 \\
 1 & 1 & -1 & -1 \\
\end{smallmatrix}\right)$\\\hline
         $[F_\CN^{g \CD \CD }]_{(\CD,1,\mu),(\CN,\nu,1)}$ & $(\sigma^0,-\sigma^3,-\sigma^1,\ii \sigma^2)$\\\hline
         $[F_g^{\CN \CD \CD}]_{(\CD,\mu,1),(\CN,\nu,1)}$ & $(\sigma^0,\sigma^1,\sigma^3,\ii \sigma^2)$\\\hline
         $[F_\CN^{\CN \CD \CD}]_{(\CD,\mu,\nu),(g,1,1)}$ & $(\frac{\sigma^1-\sigma^3}{2},\frac{\sigma^0-\ii \sigma^2}{2},-\frac{\sigma^0+\ii \sigma^2}{2},-\frac{\sigma^1+\sigma^3}{2})$ \\ \hline \hline
    
    \end{tabular}
    \quad
    \begin{tabular}[t]{c|c}
         \hline\hline
         $[F_{\CD}^{g \CD h}]_{(\CD,1,1),(\CD,1,1)}$ & $\left(
\begin{smallmatrix}
 1 & 1 & 1 & 1 \\
 1 & -1 & -1 & 1 \\
 1 & -1 & -1 & 1 \\
 1 & 1 & 1 & 1 \\
\end{smallmatrix}
\right)$\\ \hline
        $[F_\CD^{\CN \CD g}]_{(\CD,\mu,1),(\CD,1,\nu)}$ &  $(\sigma^0,\sigma^2,-\sigma^2, -\sigma^0)$ \\ \hline
        $[F_\CD^{g \CD \CN}]_{(\CD,1,\mu),(\CD,\nu,1)}$ &  $(\sigma^0,-\sigma^2,\sigma^2,- \sigma^0)$\\ \hline
        $[F_\CD^{\CN \CD \CN}]_{(\CD,\mu,\nu),(\CD,\rho,\sigma)}$ & $\frac{\fsb}{2\sqrt{2}}\left(
\begin{smallmatrix}
 \left(
\begin{smallmatrix}
 1+\ii & -1+\ii \\
 -1+\ii & 1+\ii \\
\end{smallmatrix}
\right) & \left(
\begin{smallmatrix}
 -1+\ii & 1+\ii \\
 -1-\ii & 1-\ii \\
\end{smallmatrix}
\right) \\
 \left(
\begin{smallmatrix}
 -1+\ii & -1-\ii \\
 1+\ii & 1-\ii \\
\end{smallmatrix}
\right) & \left(
\begin{smallmatrix}
 1+\ii & 1-\ii \\
 1-\ii & 1+\ii \\
\end{smallmatrix}
\right) \\
\end{smallmatrix}
\right)$ \\
    \hline\hline
    $[F_h^{\CD g \CD}]_{(\CD,1,1),(\CD,1,1)}$ & $\left(
\begin{smallmatrix}
 1 & 1 & 1 & 1 \\
 1 & -1 & -1 & 1 \\
 1 & -1 & -1 & 1 \\
 1 & 1 & 1 & 1 \\
\end{smallmatrix}
\right)$\\\hline
         $[F_\CN^{\CD g \CD}]_{(\CD,1,\mu),(\CD,1,\nu)}$ & $(\sigma^0,\sigma^2,-\sigma^2, -\sigma^0)$\\\hline
         $[F_g^{\CD \CN \CD}]_{(\CD,\mu,1),(\CD,\nu,1)}$ & $(\frac{-\sigma^1+\sigma^3}{\sqrt{2}},-\ii\frac{\sigma^1+\sigma^3}{\sqrt{2}},\ii\frac{\sigma^1+\sigma^3}{\sqrt{2}}, \frac{\sigma^1-\sigma^3}{\sqrt{2}})$\\\hline
         $[F_\CN^{\CD \CN \CD}]_{(\CD,\mu,\nu),(\CD,\rho,\sigma)}$ & $\fsb\left(
\begin{smallmatrix}
 \left(
\begin{smallmatrix}
 \frac{1}{2} & -\frac{1}{2} \\
 \frac{\ii}{2} & \frac{\ii}{2} \\
\end{smallmatrix}
\right) & \left(
\begin{smallmatrix}
 -\frac{1}{2} & -\frac{1}{2} \\
 \frac{\ii}{2} & -\frac{\ii}{2} \\
\end{smallmatrix}
\right) \\
 \left(
\begin{smallmatrix}
 -\frac{\ii}{2} & -\frac{\ii}{2} \\
 \frac{1}{2} & -\frac{1}{2} \\
\end{smallmatrix}
\right) & \left(
\begin{smallmatrix}
 -\frac{\ii}{2} & \frac{\ii}{2} \\
 -\frac{1}{2} & -\frac{1}{2} \\
\end{smallmatrix}
\right) \\
\end{smallmatrix}
\right)$ \\\hline\hline
$[F_\CD^{\CD \CD \CD}]_{(g,1,1),(h,1,1)}$ & $\frac{\fsa }{2\sqrt{2}} \left(
\begin{smallmatrix}
 1 & 1 & 1 & -1 \\
 1 & -1 & -1 & -1 \\
 1 & -1 & -1 & -1 \\
 -1 & -1 & -1 & 1 \\
\end{smallmatrix}
\right)$\\\hline
    $[F_\CD^{\CD \CD \CD}]_{(g,1,1),(\CN,\mu,\nu)}$ & $\fsa (\frac{-\sigma^0}{2},\frac{\sigma^2}{2},\frac{-\sigma^2}{2},\frac{-\sigma^0}{2})$\\\hline
    $[F_\CD^{\CD \CD \CD}]_{(\CN,\mu,\nu),(g,1,1)}$ &  $\fsa  (\frac{\sigma^1-\sigma^3}{2 \sqrt{2}},\frac{-\ii (\sigma^1+\sigma^3)}{2 \sqrt{2}},\frac{\ii (\sigma^1+\sigma^3)}{2 \sqrt{2}},\frac{\sigma^1-\sigma^3}{2 \sqrt{2}})$ \\\hline
    $[F_\CD^{\CD \CD \CD}]_{(\CN,\mu,\nu),(\CN,\rho,\sigma)}$ & $\frac{\fsa\fsb}{2\sqrt{2}}\left(
\begin{smallmatrix}
 \left(
\begin{smallmatrix}
 1 & -1 \\
 -1 & -1 \\
\end{smallmatrix}
\right) & \left(
\begin{smallmatrix}
 -\ii & -\ii \\
 -\ii & \ii \\
\end{smallmatrix}
\right) \\
 \left(
\begin{smallmatrix}
 \ii & \ii \\
 \ii & -\ii \\
\end{smallmatrix}
\right) & \left(
\begin{smallmatrix}
 1 & -1 \\
 -1 & -1 \\
\end{smallmatrix}
\right) \\
\end{smallmatrix}
\right)$\\\hline\hline
    \end{tabular}
    \caption{First 4 sets of $F$-symbols for $\underline{\mathcal{E}}^{(+,\fsb,\fsa)}_{\doubleZ_2}\Rep(H_8)$ involving the $\CD$ line, specified by $\fsb=\pm 1,\fsa=\pm 1$. $g,h=(1,a,b,ab)$, Greek letters $=1,2$ due to the multiplicity $N_{\CD \CD}^\CN=N_{\CD \CN}^\CD=N_{\CN \CD}^\CD=2$. }
    \label{tab:fsymbolsa}
\end{table}

\end{landscape}
\begin{table}[htbp]
    \centering
    
    \begin{tabular}{c|c}
         \hline\hline
     $[F_{\CD}^{g \CD h}]_{(\CD,1,1),(\CD,1,1)}$ & $\left(
\begin{smallmatrix}
 1 & 1 & 1 & 1 \\
 1 & -1 & -1 & 1 \\
 1 & -1 & -1 & 1 \\
 1 & 1 & 1 & 1 \\
\end{smallmatrix}
\right)$\\ \hline
     $[F_\CD^{\CN \CD g}]_{(\CD,\mu,1),(\CD,1,\nu)}$ & $(\sigma^0,-\sigma^2,\sigma^2, -\sigma^0)$\\ \hline
     $[F_\CD^{g \CD \CN}]_{(\CD,1,\mu),(\CD,\nu,1)}$ & $(\sigma^0,\sigma^2,-\sigma^2,- \sigma^0)$ \\ \hline
     $[F_\CD^{\CN \CD \CN}]_{(\CD,\mu,\nu),(\CD,\rho,\sigma)}$ & $\frac{\fsb}{2\sqrt{2}}\left(
\begin{smallmatrix}
 \left(
\begin{smallmatrix}
 1-\ii & -1-\ii \\
 -1-\ii & 1-\ii \\
\end{smallmatrix}
\right) & \left(
\begin{smallmatrix}
 -1-\ii & 1-\ii \\
 -1+\ii & 1+\ii \\
\end{smallmatrix}
\right) \\
 \left(
\begin{smallmatrix}
 -1-\ii & -1+\ii \\
 1-\ii & 1+\ii \\
\end{smallmatrix}
\right) & \left(
\begin{smallmatrix}
 1-\ii & 1+\ii \\
 1+\ii & 1-\ii \\
\end{smallmatrix}
\right) \\
\end{smallmatrix}
\right)$\\
         \hline\hline
         $[F_h^{\CD g \CD}]_{(\CD,1,1),(\CD,1,1)}$ & $\left(
\begin{smallmatrix}
 1 & 1 & 1 & 1 \\
 1 & -1 & -1 & 1 \\
 1 & -1 & -1 & 1 \\
 1 & 1 & 1 & 1 \\
\end{smallmatrix}
\right)$\\ \hline
         $[F_\CN^{\CD g \CD}]_{(\CD,1,\mu),(\CD,1,\nu)}$ & $(\sigma^0,-\sigma^2,\sigma^2, -\sigma^0)$\\ \hline
         $[F_g^{\CD \CN \CD}]_{(\CD,\mu,1),(\CD,\nu,1)}$ & $(\frac{-\sigma^1+\sigma^3}{\sqrt{2}},\ii\frac{\sigma^1+\sigma^3}{\sqrt{2}},-\ii\frac{\sigma^1+\sigma^3}{\sqrt{2}}, \frac{\sigma^1-\sigma^3}{\sqrt{2}})$ \\ \hline
         $[F_\CN^{\CD \CN \CD}]_{(\CD,\mu,\nu),(\CD,\rho,\sigma)}$  & $\fsb\left(
\begin{smallmatrix}
 \left(
\begin{smallmatrix}
 \frac{1}{2} & -\frac{1}{2} \\
 -\frac{\ii}{2} & -\frac{\ii}{2} \\
\end{smallmatrix}
\right) & \left(
\begin{smallmatrix}
 -\frac{1}{2} & -\frac{1}{2} \\
 -\frac{\ii}{2} & \frac{\ii}{2} \\
\end{smallmatrix}
\right) \\
 \left(
\begin{smallmatrix}
 \frac{\ii}{2} & \frac{\ii}{2} \\
 \frac{1}{2} & -\frac{1}{2} \\
\end{smallmatrix}
\right) & \left(
\begin{smallmatrix}
 \frac{\ii}{2} & -\frac{\ii}{2} \\
 -\frac{1}{2} & -\frac{1}{2} \\
\end{smallmatrix}
\right) \\
\end{smallmatrix}
\right)$\\ \hline\hline
$[F_\CD^{\CD \CD \CD}]_{(g,1,1),(h,1,1)}$ & $\frac{\fsa }{2\sqrt{2}} \left(
\begin{smallmatrix}
 1 & 1 & 1 & -1 \\
 1 & -1 & -1 & -1 \\
 1 & -1 & -1 & -1 \\
 -1 & -1 & -1 & 1 \\
\end{smallmatrix}
\right)$ \\ \hline
$[F_\CD^{\CD \CD \CD}]_{(g,1,1),(\CN,\mu,\nu)}$ & $\fsa  (\frac{-\sigma^0}{2},\frac{-\sigma^2}{2},\frac{\sigma^2}{2},\frac{-\sigma^0}{2})$\\ \hline
$[F_\CD^{\CD \CD \CD}]_{(\CN,\mu,\nu),(g,1,1)}$ & $\fsa (\frac{\sigma^1-\sigma^3}{2 \sqrt{2}},\frac{\ii (\sigma^1+\sigma^3)}{2 \sqrt{2}},-\frac{\ii (\sigma^1+\sigma^3)}{2 \sqrt{2}},\frac{\sigma^1-\sigma^3}{2 \sqrt{2}})$\\ \hline
$[F_\CD^{\CD \CD \CD}]_{(\CN,\mu,\nu),(\CN,\rho,\sigma)}$ & $\frac{\fsa \fsb}{2\sqrt{2}}\left(
\begin{smallmatrix}
 \left(
\begin{smallmatrix}
 1 & -1 \\
 -1 & -1 \\
\end{smallmatrix}
\right) & \left(
\begin{smallmatrix}
 \ii & \ii \\
 \ii & -\ii \\
\end{smallmatrix}
\right) \\
 \left(
\begin{smallmatrix}
 -\ii & -\ii \\
 -\ii & \ii \\
\end{smallmatrix}
\right) & \left(
\begin{smallmatrix}
 1 & -1 \\
 -1 & -1 \\
\end{smallmatrix}
\right) \\
\end{smallmatrix}
\right)$\\ \hline\hline
    \end{tabular}
    \caption{The other 4 sets of $F$-symbols for $\underline{\mathcal{E}}^{(-,\fsb,\fsa)}_{\doubleZ_2}\Rep(H_8)$ involving the $\CD$ line, specified by $\fsb=\pm 1,\fsa=\pm 1$. Omitted $F$-symbols (left part in \tabref{tab:fsymbolsa}) are the same as \tabref{tab:fsymbolsa}. The two sets of $F$-symbols cannot be nicely grouped together because the difference is not an overall phase. }
    \label{tab:fsymbolsb}
\end{table}

The 8 solutions differ by the three $\IZ_2$-valued phases $i=\pm,\fsb=\pm,\fsa=\pm$ appearing in $F^{L \CD R}_{\CD}, F^{\CD L \CD}_R, [F^{\CD \CD \CD}_\CD]_{L,R}$, where $L,R \in \{1,a,b,ab,\CN\}$. 
In particular, $\fsa\in H^3(\IZ_2,U(1))$ is the Frobenius-Schur indicator for the new topological line $\CD$.

We briefly discuss how to interpret the above $8$ solutions in terms of the classification data $(\rho,M,\epsilon)$. 
First, specifying $\rho:\doubleZ_2 \rightarrow \BrPic(\Rep(H_8))$ is equivalent to specifying a bimodule category $\mathcal{C}_\eta$ with a unique simple object $\mathcal{D}$ and the corresponding $F$-symbols are $F^{r_1 r_2 \mathcal{D}}_{\mathcal{D}},F^{r_1 \mathcal{D} r_2 }_{\mathcal{D}},F^{\mathcal{D} r_1 r_2}_{\mathcal{D}}$'s. 
Next, we need to verify that indeed $\mathcal{C}_\eta$ is an \textit{invertible} $\Rep(H_8)$-bimodule category, and if so, we make a choice of the $\Rep(H_8)$-bimodule equivalence functor between $\mathcal{C}_\eta \boxtimes_{\Rep(H_8)} \mathcal{C}_\eta$ and $\Rep(H_8)$. This data is equivalent to the choice of the symmetry fractionalization class $M \in H^2_{[\rho]}(\doubleZ_2,A)$ in the symTFT point of view and the corresponding $F$-symbols are $F^{\mathcal{D}\mathcal{D} r_1}_{r_2},F^{\mathcal{D}r_1 \mathcal{D}}_{r_2},F^{r_1 \mathcal{D}\mathcal{D}}_{r_2}$. 
Finally, the choice of the Frobenius-Schur indicator $\epsilon_{\CD}$ is the choice of the discrete torsion of $\doubleZ_2$ in the symTFT picture, and the corresponding $F$-symbol data is the $\pm$ sign in $\left[F^{\mathcal{D}\mathcal{D}\mathcal{D}}_{\mathcal{D}}\right]_{\dsi\dsi}$. We summarize the above discussion in Table \ref{tab:clsf_F_symbol}.
\begin{table}[h]
    \centering
    \begin{tabular}{c|c|c}
    \hline \hline
        Abstract structure & $F$-symbols & Classification data  \\
    \hline
        Left $\Rep(H_8)$-module category structure & $F^{r_1 r_2 \mathcal{D}}_{\mathcal{D}}$ & \multirow{3}*{$\rho:\doubleZ_2 \rightarrow \BrPic(\Rep(H_8))$} \\
    \cline{1-2}
        Right $\Rep(H_8)$-module category structure & $F^{\mathcal{D}r_1 r_2}_{\mathcal{D}}$ & \\
    \cline{1-2}
        \makecell{Bimodule structure which glues \\ the left/right module structure together} & $F^{r_1\mathcal{D}r_2}_{\mathcal{D}}$ &  \\
    \hline
        \makecell{A choice of the equivalence functor \\ $\mathcal{C}_\eta \boxtimes_{\Rep(H_8)} \mathcal{C}_\eta \simeq \Rep(H_8)$} & $F^{\mathcal{D}\mathcal{D} r_1}_{r_2}, F^{\mathcal{D}r_1 \mathcal{D}}_{r_2}, F^{r_1 \mathcal{D}\mathcal{D}}_{r_2}$ & $M \in H^2_{[\rho]}(\doubleZ_2, A)$ \\
    \hline
        A choice of the FS indicator & the $\pm$ sign in  $\left[F^{\mathcal{D}\mathcal{D}\mathcal{D}}_{\mathcal{D}}\right]_{\dsi \dsi}$ & $\fsa \in H^3(\doubleZ_2, U(1))$ \\
    \hline\hline
    
    \end{tabular}
    \caption{The correspondence between the abstract structure appearing in the classification analysis and the concrete $F$-symbols, where $r_i \in \Rep(H_8)$. We drop the labels for the internal channel of the $F$-symbol for simplicity.}
    \label{tab:clsf_F_symbol}
\end{table}

As one can see in Tables \ref{tab:fsymbolsa} and \ref{tab:fsymbolsb}, there are $4$ distinct sets of $(F^{r_1 r_2 \mathcal{D}}_{\CD}, F^{r_1 \mathcal{D} r_2}_{\CD}, F^{\mathcal{D} r_1 r_2}_{\CD})$'s. This implies there are $4$ distinct choices of $\rho$.
For each given $\rho$, there is a unique choice of $M$, and two choices for the Frobenius-Schur indicator.\footnote{On the other hand, $\Aut(\TY(A,\chi,\epsilon))$ is given by $\Aut(A,\chi)$ which is the group of automorphisms of $A$ preserving $\chi$, see \cite[Section 4.2]{2020arXiv201000847G}.
We can then compute $\Aut(\Rep(H_8)) = \doubleZ_2$, and the discussion in Footnote \ref{fit:BrPic} suggests there should be no more than $4$ distinct choices of $\rho$, which is consistent with the above analysis.} 

\subsection{Lasso actions and spin selection rules}\label{sec:spin_selection}
Here, we derive the spin selection rules for the defect Hilbert space $\mathcal{H}_{\mathcal{D}}$ following \cite{Lu:2022ver, Chang:2018iay}.
By using the lasso actions of the topological defect lines on the defect Hilbert space $\mathcal{H}_{\CD}$, the allowed values of the spin $s$ of operators in $\mathcal{H}_{\CD}$ are constrained.

The lasso actions of topological defect lines $\{\dsi,a,b,ab,\scN\}$ define maps acting on the defect Hilbert space of $\CD$. 
The corresponding defect line configurations are shown both on the cylinder (left) and on the plane (right) below:
\begin{equation}\label{eq:tubealgs}
    \begin{tikzpicture}[baseline={([yshift=0]current bounding box.center)},vertex/.style={anchor=base,
    circle,fill=black!25,minimum size=18pt,inner sep=2pt},scale=0.5]
    \filldraw[grey] (-2,-2) rectangle ++(4,4);
    \draw[thick, dgrey] (-2,-2) rectangle ++(4,4);
    \draw[thick, black] (0,-2) -- (0.707,-0.707);
    \draw[thick, black, dashed] (2,0) -- (0.707,-0.707);
    \draw[thick, black] (0,2) -- (-0.707,0.707);
    \draw[thick, black, dashed] (-2,0) -- (-0.707,0.707);
    \draw[thick, black] (0.707,-0.707) -- (-0.707,0.707);
    \node[black, left] at (0.2,-0.3) {$\mathcal{D}$};
    \node[black, below] at (1.5,-0.3) {$g$};
\end{tikzpicture}\quad \text{or} \quad
    \begin{gathered}
    \begin{tikzpicture}[scale=0.5,rotate=270]
    \draw [fill=black] (0.6,0) circle (.1);
    \draw [line] (-3,0) -- (0.6,0);
    \draw (-3,0) node [right] {$\CD$};
    \draw (0.6,0) node [below] {$\scO^{\CD}$};
    \draw [scale=1,domain=-3.141:3.141,smooth,variable=\t]
        plot ({(1.9+0.2*\t)*cos(\t r)+0.6},{-(1.9+0.2*\t)*sin(\t r)});
    \draw (2.5,0) node [below] {$g$};
    \draw [fill=black] (-0.672,0) circle (.1);
    \draw (-0.672,0) node [left] {\scriptsize $ $};
    \draw [fill=black] (-1.928,0) circle (.1);
    \draw (-1.928,0) node [right] {\scriptsize $ $};
    \end{tikzpicture}
\end{gathered} \equiv \scU_g ,\quad     \begin{tikzpicture}[baseline={([yshift=0]current bounding box.center)},vertex/.style={anchor=base,
    circle,fill=black!25,minimum size=18pt,inner sep=2pt},scale=0.5]
    \filldraw[grey] (-2,-2) rectangle ++(4,4);
    \draw[thick, dgrey] (-2,-2) rectangle ++(4,4);
    \draw[thick, black] (0,-2) -- (0.707,-0.707);
    \draw[thin, black] (2,0) -- (0.707,-0.707);
    \draw[thick, black] (0,2) -- (-0.707,0.707);
    \draw[thin, black] (-2,0) -- (-0.707,0.707);
    \draw[thick, black] (0.707,-0.707) -- (-0.707,0.707);
    \node[black, left] at (0.2,-0.3) {$\mathcal{D}$};
    \node[black, below] at (1.5,-0.3) {\footnotesize $\CN$};
    \draw [fill=black] (0.707,-0.707) circle (.08);
    \node[black, above] at (0.707,-0.707) {\scriptsize$\mu$};
    \draw [fill=black] (-0.707,0.707) circle (.08);
    \node[black, below] at (-0.707,0.707) {\scriptsize$\nu$};
\end{tikzpicture} \quad \text{or} \quad \begin{gathered}
    \begin{tikzpicture}[scale=0.5,rotate=270]
    \draw [fill=black] (0.6,0) circle (.1);
    \draw [line] (-3,0) -- (0.6,0);
    \draw (-3,0) node [right] {$\CD$};
    \draw (0.6,0) node [below] {$\scO^{\CD}$};
    \draw [scale=1,domain=-3.141:3.141,smooth,variable=\t]
        plot ({(1.9+0.2*\t)*cos(\t r)+0.6},{-(1.9+0.2*\t)*sin(\t r)});
    \draw (2.5,0.) node [below] {$\scN$};
    \draw [fill=black] (-0.672,0) circle (.1);
    \draw (-0.672,0) node [left] {\scriptsize $\mu$};
    \draw [fill=black] (-1.928,0) circle (.1);
    \draw (-1.928,0) node [right] {\scriptsize $\nu$};
    \end{tikzpicture}
\end{gathered} \equiv \scU_{\CN,\mu\nu} .
\end{equation}
Both $\scU_{g},\scU_{\CN,\mu\nu}$ maps $\scH_\scD \rightarrow \scH_\scD$. 
The compositions of the lasso actions in \eqnref{eq:tubealgs} satisfy
\begin{align}
\begin{split}
    & \scU_g \cdot \scU_h = [F_{\CD}^{h \CD g}]_{(\CD,1,1),(\CD,1,1)} [F_{\CD}^{\CD hg}]_{(\CD,1,1),(hg,1,1)} [F_{\CD}^{gh\CD}]^{-1}_{(\CD,1,1),(gh,1,1)} \scU_{gh} \,, \\
    & \scU_g \cdot \scU_{\scN,\mu\nu} = [F_\CD^{\CN \CD g}]_{(\CD,\mu,1),(\CD,1,\rho)} [F_{\CD}^{g\CN \CD}]^{-1}_{(\CD,\rho,1),(\CN,1,\sigma)} [F_{\CD}^{\CD \CN g}]_{(\CD,\nu,1),(\CN,1,\lambda)} \scU_{\scN,\sigma \lambda}\,, \\
    &\scU_{\scN,\mu\nu} \cdot  \scU_g =[F_\CD^{g \CD \CN}]_{(\CD,1,\nu),(\CD,\rho,1)} [F_\CD^{\CN g \CD}]^{-1}_{(\CD,1,\mu),(\CN,1,\sigma)} [F_\CD^{\CD g \CN}]_{(\CD,1,\rho),(\CN,1,\lambda)} \scU_{\scN,\sigma \lambda}\,,\\
    & \scU_{\scN,\mu\nu} \cdot \scU_{\scN,\rho\sigma} =2 [F_\CN^{\CD \CN \CD}]_{(\CD,\rho,\nu),(\CD,\rho',\nu')}\sum_g [F_\CD^{\CN \CN \CD}]^{-1}_{(\CD,\nu',\mu),(g,1,1)} [F_\CD^{\CD \CN \CN}]_{(\CD,\sigma,\rho'),(g,1,1)}\scU_g \,.
\end{split}
\end{align}
The composition of more general lasso actions are listed in \appref{app:lassocompose}. There are 8 1-dimensional representations for this algebra of $\scU_g,\scU_{\scN,\mu\nu}$, for each set of $F$-symbols. 
The spin mod $\frac{1}{2}$ of the operators in the defect Hilbert space is obtained by
\begin{equation}
    e^{4\pi \ii s} = \sum_{g} F_{\CD,(1,1,1),(g,1,1)}^{\CD\CD\CD}\scU_g+\sum_{\mu\nu}F_{\CD,(1,1,1),(\CN,\mu,\nu)}^{\CD\CD\CD} \scU_{\CN,\mu\nu},
\end{equation}
which is derived using the following diagram,
\begin{align}
    &\begin{tikzpicture}[baseline={([yshift=0]current bounding box.center)},vertex/.style={anchor=base,
    circle,fill=black!25,minimum size=18pt,inner sep=2pt},scale=0.5]
    \filldraw[grey] (-2,-2) rectangle ++(4,4);
    \draw[thick, dgrey] (-2,-2) rectangle ++(4,4);
    \draw[thick, black] (0,-2) -- (+2,-1);
    \draw[thick, black] (-2,-1) -- (+2,1);
    \draw[thick, black] (-2,+1) -- (0,+2);
    \node[black, above] at (0,0) {$\mathcal{D}$};
\end{tikzpicture}\,  = \sum_{g} F_{\CD,(1,1,1),(g,1,1)}^{\CD\CD\CD} \, \begin{tikzpicture}[baseline={([yshift=0]current bounding box.center)},vertex/.style={anchor=base,
    circle,fill=black!25,minimum size=18pt,inner sep=2pt},scale=0.5]
    \filldraw[grey] (-2,-2) rectangle ++(4,4);
    \draw[thick, dgrey] (-2,-2) rectangle ++(4,4);
    \draw[thick, black] (-2,-1) arc(-90:90:1);
    \draw[thick, black] (2,-1) arc(90:143:2.5);
    \draw[thick, black] (2,1) arc(-90:-143:2.5);
    \draw[thick, dashed] (-1.293, 0.707) -- (0.6,1.4);
    \node[black, below] at (-1,-0.5) {$\mathcal{D}$};
    \node[black, below] at (-0.347,1.054) {$g$};
\end{tikzpicture}+\sum_{\mu\nu}F_{\CD,(1,1,1),(\CN,\mu,\nu)}^{\CD\CD\CD} \, \begin{tikzpicture}[baseline={([yshift=0]current bounding box.center)},vertex/.style={anchor=base,
    circle,fill=black!25,minimum size=18pt,inner sep=2pt},scale=0.5]
    \filldraw[grey] (-2,-2) rectangle ++(4,4);
    \draw[thick, dgrey] (-2,-2) rectangle ++(4,4);
    \draw[thick, black] (-2,-1) arc(-90:90:1);
    \draw[thick, black] (2,-1) arc(90:143:2.5);
    \draw[thick, black] (2,1) arc(-90:-143:2.5);
    \draw[thin] (-1.293, 0.707) -- (0.6,1.4);
    \node[black, below] at (-1,-0.5) {$\mathcal{D}$};
    \node[black, below] at (-0.347,1.054) {$\CN$};
    \draw [fill=black] (-1.293, 0.707) circle (.08);
    \node[black, above] at (-1.293, 0.707) {\scriptsize$\mu$};
    \draw [fill=black] (0.6,1.4) circle (.08);
    \node[black, below] at (0.6,1.4) {\scriptsize$\nu$};
\end{tikzpicture}\\ \nonumber
&= \sum_{g} F_{\CD,(1,1,1),(g,1,1)}^{\CD\CD\CD} \, \begin{tikzpicture}[baseline={([yshift=0]current bounding box.center)},vertex/.style={anchor=base,
    circle,fill=black!25,minimum size=18pt,inner sep=2pt},scale=0.5]
    \filldraw[grey] (-2,-2) rectangle ++(4,4);
    \draw[thick, dgrey] (-2,-2) rectangle ++(4,4);
    \draw[thick, black] (0,-2) -- (0.707,-0.707);
    \draw[thick, black, dashed] (2,0) -- (0.707,-0.707);
    \draw[thick, black] (0,2) -- (-0.707,0.707);
    \draw[thick, black, dashed] (-2,0) -- (-0.707,0.707);
    \draw[thick, black] (0.707,-0.707) -- (-0.707,0.707);
    \node[black, left] at (0.2,-0.3) {$\mathcal{D}$};
    \node[black, below] at (1.5,-0.3) {$g$};
\end{tikzpicture}+ \sum_{\mu\nu}F_{\CD,(1,1,1),(\CN,\mu,\nu)}^{\CD\CD\CD} \, \begin{tikzpicture}[baseline={([yshift=0]current bounding box.center)},vertex/.style={anchor=base,
    circle,fill=black!25,minimum size=18pt,inner sep=2pt},scale=0.5]
    \filldraw[grey] (-2,-2) rectangle ++(4,4);
    \draw[thick, dgrey] (-2,-2) rectangle ++(4,4);
    \draw[thick, black] (0,-2) -- (0.707,-0.707);
    \draw[thin, black] (2,0) -- (0.707,-0.707);
    \draw[thick, black] (0,2) -- (-0.707,0.707);
    \draw[thin, black] (-2,0) -- (-0.707,0.707);
    \draw[thick, black] (0.707,-0.707) -- (-0.707,0.707);
    \node[black, left] at (0.2,-0.3) {$\mathcal{D}$};
    \node[black, below] at (1.5,-0.3) {$\CN$};
    \draw [fill=black] (0.707,-0.707) circle (.08);
    \node[black, above] at (0.707,-0.707) {\scriptsize$\mu$};
    \draw [fill=black] (-0.707,0.707) circle (.08);
    \node[black, below] at (-0.707,0.707) {\scriptsize$\nu$};
\end{tikzpicture}.
\end{align}
The spin selection rules $s \text{ mod }\frac{1}{2}$ for the 8 different solutions of $F$-symbols in \tabref{tab:fsymbolsa} and \tabref{tab:fsymbolsb} are given by
\begin{equation}\label{eq:spinselection}
\begin{array}{c|cccc}
\hline\hline
\underline{\mathcal{E}}^{(+,+,+)}_{\doubleZ_2}\Rep(H_8);\mathrm{\tabref{tab:fsymbolsa}},\fsb=+, \fsa=+&\frac{1}{16} & \frac{7}{32} & \frac{7}{16} & \frac{15}{32} \\\hline
 \underline{\mathcal{E}}^{(+,+,-)}_{\doubleZ_2}\Rep(H_8);\mathrm{\tabref{tab:fsymbolsa}},\fsb=+, \fsa=-&\frac{3}{16} & \frac{7}{32} & \frac{5}{16} & \frac{15}{32}  \\\hline
\underline{\mathcal{E}}^{(+,-,+)}_{\doubleZ_2}\Rep(H_8);\mathrm{\tabref{tab:fsymbolsa}},\fsb=-, \fsa=+&\frac{1}{16} & \frac{3}{32} & \frac{11}{32} & \frac{7}{16} \\\hline
 \underline{\mathcal{E}}^{(+,-,-)}_{\doubleZ_2}\Rep(H_8);\mathrm{\tabref{tab:fsymbolsa}},\fsb=-, \fsa=-&\frac{3}{32} & \frac{3}{16} & \frac{5}{16} & \frac{11}{32} \\ \hline
 \underline{\mathcal{E}}^{(-,+,+)}_{\doubleZ_2}\Rep(H_8);\mathrm{\tabref{tab:fsymbolsb}},\fsb=+, \fsa=+&\frac{1}{32} & \frac{1}{16} & \frac{9}{32} & \frac{7}{16} \\\hline
 \underline{\mathcal{E}}^{(-,+,-)}_{\doubleZ_2}\Rep(H_8); \mathrm{\tabref{tab:fsymbolsb}},\fsb=+, \fsa=-&\frac{1}{32}  & \frac{3}{16} & \frac{9}{32} & \frac{5}{16}  \\\hline
 \underline{\mathcal{E}}^{(-,-,+)}_{\doubleZ_2}\Rep(H_8);\mathrm{\tabref{tab:fsymbolsb}},\fsb=-, \fsa=+&\frac{1}{16} & \frac{5}{32} & \frac{13}{32} & \frac{7}{16}  \\\hline
 \underline{\mathcal{E}}^{(-,-,-)}_{\doubleZ_2}\Rep(H_8);\mathrm{\tabref{tab:fsymbolsb}},\fsb=-, \fsa=-&\frac{5}{32} & \frac{3}{16} & \frac{5}{16} & \frac{13}{32} \\\hline\hline
\end{array}
\end{equation}
The spin selection rules for $\underline{\mathcal{E}}^{(+,\fsb,\fsa)}_{\doubleZ_2}\Rep(H_8)$ and $\underline{\mathcal{E}}^{(-,\fsb,\fsa)}_{\doubleZ_2}\Rep(H_8)$ are related by $s\leftrightarrow -s$.

\section*{Acknowledgements}
We are grateful to P.-S.\ Hsin, T.\ Jacobson, B.\ C.\ Rayhaun, Y.\ Sanghavi, S.\ Seifnashri, S.-H.\ Shao, Y.\ Wang, Y.-Z.\ You, C.\ Zhang, and Y.\ Zheng for interesting discussions. We also thank TASI 2023: Aspects of Symmetry, PiTP 2023: Understanding Confinement, and PSSCMP 2023: Fractionalization, Criticality and Unconventional Quantum Materials  summer schools for providing stimulating environments, where this work was initiated. D.C.L. is supported by the National Science Foundation (NSF) Grant No. DMR-2238360. Z.S. is partially supported by the US Department of Energy (DOE) under cooperative research agreement DE-SC0009919, Simons Foundation award No. 568420 (K.I.), and the Simons Collaboration on Global Categorical Symmetries.

\appendix

\section{Details on solving the algebra object in $\Rep(H_8)$}\label{app:algebra}
The fusion and splitting junctions satisfy the conditions in \eqnref{eq:algebracond}. The first two conditions can fix the form of the junctions and the last one fixes the normalization. From the first condition in \eqnref{eq:algebracond}, we have
\begin{align}
    &\psi(g,h)\psi(gh,k) = \psi(g,hk)\psi(h,k), \label{eq:b1}\\
    &L(h) L(g) = \psi(g,h)L(gh),\quad  R(g)  R(h) = \psi(g,h) R(gh), \label{eq:b2}\\
    &L(g)R(h) = \chi(g,h) R(h)L(g),\\
    & W(gh) R(h)^\intercal = \psi(g,h) W(g),\quad  L(g) W(gh)= \psi(g,h)W(h), \\ 
    &R(g) W(h) = \chi(g,h) W(h) L(g)^\intercal, \\
    & [W(g)]_{ab} [L(g)]_{cd}=\sum_h \frac{1}{2}\chi(g,h)^{-1} [W(h)]_{bc} [R(h)]_{ad},
\end{align}
where $L(g),R(g),W(g)$ are $2\times 2$ matrices and their multiplication is matrix multiplication. $\chi$ is the bicharacter in $\Rep(H_8)$. \eqnref{eq:b1} states that $\psi(g,h)$ is a 2-cocycle of $\IZ_2\times \IZ_2$, and is classified by $H^2(\IZ_2\times \IZ_2,U(1))=\IZ_2$. Only the non-trivial one will solve the following equations. \eqnref{eq:b2} states $L(g),R(g)$ furnish the projective representation of $\IZ_2\times \IZ_2$, and $W(g)$ can be solved accordingly. The splitting junctions can be solved similarly. The junctions are normalized such that the last condition in \eqnref{eq:algebracond} is satisfied.

\section{More on Virasoro primaries of Ising$^2$} \label{app:more_Ising2}

Here we work out the spectrum of $c=1$ Virasoro primary operators at the Ising$^2$ point.
To do so, we begin from the $R= \sqrt{2}$ point on the circle branch.
The conformal weights of the vertex operators $V_{n,w}$ are
\begin{align}
\begin{split}
    h_{n,w} &= \frac{1}{4} \left( \frac{n}{\sqrt{2}} + w \sqrt{2} \right)^2 = \frac{1}{8} \left( n + 2w \right)^2 \,,\\
    \bar{h}_{n,w} &= \frac{1}{4} \left( \frac{n}{\sqrt{2}} - w \sqrt{2} \right)^2 = \frac{1}{8} \left( n - 2w \right)^2 \,.
\end{split}
\end{align}
When $n=2w \neq 0$, there are null states since $\bar{h}_{2w,w} =0$, and the Verma module corresponding to the vertex operator $V_{2w,w}$ breaks into infinitely many irreducible $c=1$ Virasoro modules.
We denote the corresponding $c=1$ Virasoro primaries as $A_{w,m}$, where $w \in \mathbb{Z} \setminus \{0\}$ and $m \in \mathbb{Z}_{\geq 0}$.
The conformal weights of $A_{w,m}$ are $(h,\bar{h}) = (2w^2 , m^2)$.
Similarly, when $n= -2w \neq 0$, we have null states since $h_{-2w,w} = 0$, and the Verma module decomposes into infinitely many irreducible $c=1$ Virasoro modules.
The corresponding Virasoro primaries are denoted as $B_{w,\ell}$ with $w \in \mathbb{Z} \setminus \{0\}$ and $\ell \in \mathbb{Z}_{\geq 0}$, whose conformal weights are $(h,\bar{h}) = (\ell^2 , 2w^2)$.
Finally, the $(n,w)=(0,0)$ module decomposes into the irreducible $c=1$ Virasoro modules with the primary operators denoted as $C_{\ell,m}$ with $\ell,m \in \mathbb{Z}_{\geq 0}$, whose conformal weights are $(h,\bar{h}) = (\ell^2 , m^2)$.

To summarize, the $c=1$ Virasoro primaries at $R= \sqrt{2}$ on the circle branch are
\begin{align} \label{eq:circle_primaries}
\begin{split}
    &V_{n,w} \,, \quad n,w \in \mathbb{Z} \,, n \neq \pm 2w \,, \quad (h,\bar{h}) = (\frac{1}{8} (n+2w)^2, \frac{1}{8} (n-2w)^2) \,, \\
    &A_{w,m} \,, \quad w \in \mathbb{Z} \setminus \{0\} \,, m \in \mathbb{Z}_{\geq 0} \,, \quad (h,\bar{h}) = (2w^2 , m^2) \,, \\
    &B_{w,\ell} \,, \quad w \in \mathbb{Z} \setminus \{0\} \,, \ell \in \mathbb{Z}_{\geq 0} \,, \quad (h,\bar{h}) = (\ell^2 , 2w^2) \,, \\
    &C_{\ell,m} \,, \quad \ell,m \in \mathbb{Z}_{\geq 0} \,, \quad (h,\bar{h}) = (\ell^2 , m^2) \,.
\end{split}
\end{align}
Correspondingly, the torus partition function can be decomposed into the $c=1$ Virasoro characters:
\begin{align}
\begin{split}
    Z^{circ}_{R=\sqrt{2}}(\tau) &= \frac{1}{|\eta(\tau)|^2} \sum_{n,w \in \mathbb{Z}} q^{\frac{1}{8}(n+2w)^2} \bar{q}^{\frac{1}{8}(n-2w)^2} \\
    &= \frac{1}{|\eta(\tau)|^2} \sum_{\substack{n,w \in \mathbb{Z} \\ n \neq \pm 2w}} q^{\frac{1}{8}(n+2w)^2} \bar{q}^{\frac{1}{8}(n-2w)^2}
    + \frac{1}{|\eta(\tau)|^2} \sum_{\substack{w \in \mathbb{Z} \setminus \{ 0 \} \\ m \in \mathbb{Z}_{\geq 0}}} q^{2w^2} \left( \bar{q}^{m^2} - \bar{q}^{(m+1)^2} \right) \\
    &\quad + \frac{1}{|\eta(\tau)|^2} \sum_{\substack{w \in \mathbb{Z} \setminus \{ 0 \} \\ \ell \in \mathbb{Z}_{\geq 0}}} \left( q^{\ell^2} - q^{(\ell+1)^2} \right) \bar{q}^{2w^2} \\
    &\quad + \frac{1}{|\eta(\tau)|^2} \sum_{ \ell,m \in \mathbb{Z}_{\geq 0}} \left( q^{\ell^2} - q^{(\ell+1)^2} \right) \left( \bar{q}^{m^2} - \bar{q}^{(m+1)^2} \right) \,.
\end{split}
\end{align}
The $\mathbb{Z}_2^C$ charge conjugation symmetry acts on the Virasoro primaries \eqref{eq:circle_primaries} as
\begin{align}
\begin{split}
    \mathbb{Z}_2^C: \quad & V_{n,w} \rightarrow V_{-n,-w} \,, \\
    & A_{w,m} \rightarrow (-1)^m A_{-w,m} \,,\\
    & B_{w,\ell} \rightarrow (-1)^\ell B_{-w,\ell} \,,\\
    & C_{\ell,m} \rightarrow (-1)^{\ell + m} C_{\ell,m} \,.
\end{split}
\end{align}

The $\mathbb{Z}_2^C$ twisted sector spectrum can be read off from the twisted partition function (which actually does not depend on the value of $R$)
\begin{equation}
    Z^{circ}_{R=\sqrt{2}}[\eta,\dsi,\eta](\tau) = \frac{1}{|\eta(\tau)|^2} \sum_{\ell,m \in \mathbb{Z}_{\geq 0}} 2 q^{\frac{1}{4}(\ell + \frac{1}{2})^2} \bar{q}^{\frac{1}{4}(m+\frac{1}{2})^2} \,.
\end{equation}
We see that the twisted sector operators are doubly-degenerate, which is a consequence of the fact that the $\mathbb{Z}_2^m \times \mathbb{Z}_2^w$ subgroup of the momentum and winding symmetries acts projectively on the $\mathbb{Z}_2^C$ twisted sector due to a mixed anomaly between the three symmetries.
We denote the Virasoro primaries in the $\mathbb{Z}_2^C$ twisted sector as
\begin{equation}
    D^{(i)}_{\ell,m} \,, \quad \ell,m \in  \mathbb{Z}_{\geq 0} \,, i=1,2 \,, \quad (h,\bar{h}) = (\frac{1}{4}(\ell + \frac{1}{2})^2, \frac{1}{4}(m + \frac{1}{2})^2 ) \,.
\end{equation}
The $\mathbb{Z}_2^C$ charge of a twisted sector operator is given by $2(h-\bar{h}) = \frac{1}{2}(\ell -m)(\ell+m+1)$.
Namely,
\begin{align}
\begin{split}
    \mathbb{Z}_2^C: \quad & D^{(i)}_{\ell,m} \rightarrow
 (-1)^{\frac{1}{2}(\ell -m)(\ell+m+1)} D^{(i)}_{\ell,m} \,.
\end{split}
\end{align}

We are ready to write down all the $c=1$ Virasoro primaries at the Ising$^2$ point, namely at $R=\sqrt{2}$ on the orbifold branch.
These are the operators on the circle branch which are $\mathbb{Z}_2^C$ invariant:
\begin{align}
\begin{split}
    &V_{n,w}^+ \equiv \frac{1}{\sqrt{2}} (V_{n,w} + V_{-n,-w}) \,, \quad n, w \in \mathbb{Z}  \,, n \neq \pm 2w \,, \quad (h,\bar{h}) = (\frac{1}{8} (n+2w)^2, \frac{1}{8} (n-2w)^2) \,, \\
    &A_{w,m}^+ \equiv \frac{1}{\sqrt{2}} (A_{w,m} + (-1)^m A_{-w,m}) \,, \quad w \in \mathbb{Z}_{>0} \,, m \in \mathbb{Z}_{\geq 0} \,, \quad (h,\bar{h}) = (2w^2 , m^2) \,, \\
    &B_{w,\ell}^+ \equiv \frac{1}{\sqrt{2}} (B_{w,\ell} + (-1)^\ell B_{-w,\ell}) \,, \quad w \in \mathbb{Z}_{>0} \,, \ell \in \mathbb{Z}_{\geq 0} \,, \quad (h,\bar{h}) = (\ell^2 , 2w^2) \,, \\
    &C_{\ell,m} \,, \quad \ell,m \in \mathbb{Z}_{\geq 0} \,, \ell + m \in 2\mathbb{Z} \,, \quad (h,\bar{h}) = (\ell^2 , m^2) \,, \\
    &D^{(i)}_{\ell,m} \,, \quad \ell,m \in  \mathbb{Z}_{\geq 0} \,, i=1,2 \,, \frac{1}{2}(\ell -m)(\ell+m+1) \in 2\mathbb{Z} \,, \quad (h,\bar{h}) = (\frac{1}{4}(\ell + \frac{1}{2})^2, \frac{1}{4}(m + \frac{1}{2})^2 ) \,,
\end{split}
\end{align}
modulo the identification $V_{n,w}^+ = V_{-n,-w}^+$.

\section{Calculation details for $\underline{\mathcal{E}}^{(i,\fsb,\fsa)}_{\doubleZ_2}\Rep(H_8)$} \label{app:F_details}

We provide some details on the pentagon equations for $\underline{\mathcal{E}}^{(i,\fsb,\fsa)}_{\doubleZ_2}\Rep(H_8)$ and lasso actions in the defect Hilbert space of $\CD$.
\subsection{Gauge fixing} \label{app:gaugefixing}
The fusion vertices generally can be transformed by unitary matrices, and the $F$-symbols related by these transformations lead to equivalent fusion categories. 
For the purpose of classifying inequivalent fusion categories, this introduces tremendous redundancies when solving the pentagon equations. 
We will follow the physics convention to refer these redundancies as gauge redundancies in this context.

The finite number of inequivalent solutions can be obtained by gauge fixing. To solve the pentagon equations for $\underline{\mathcal{E}}^{(i,\fsb,\fsa)}_{\doubleZ_2}\Rep(H_8)$, the fusion vertices involving the new defect $\CD$ need to be gauge fixed. In particular, we set $F$-symbols with at least one of $a,b,c=\dsi$ to be 1, and also fix the expressions for 
\begin{align}
\begin{split}
    &[F_{\CD}^{ab\CD}]_{(ab,1,1),(\CD,1,1)},\quad[F_{\CD}^{a\CN \CD}]_{(\CN,1,\mu),(\CD,\nu,1)},\quad [F_{\CD}^{\CD ab}]_{(\CD,1,1),(ab,1,1)},\quad [F_{\CD}^{\CD \CN a}]_{(\CD,\mu,1),(\CN,1,\nu)}, \\
    &[F_b^{\CD \CD a}]_{(b a^{-1},1,1),(\CD,1,1)},\quad [F_\CN^{\CD \CD a}]_{(\CN,\mu,1),(\CD,1,\nu)}.
\end{split}
\end{align}
There will be residual gauge transformations which will further eliminate gauge-equivalent solutions. The explicit expressions for the choices and gauge-inequivalent solutions are listed in \tabref{tab:fsymbolsa} and \tabref{tab:fsymbolsb}.

\subsection{General lasso actions in the defect Hilbert space of $\CD$}\label{app:lassocompose}
In this section, we derive the composition of general lasso actions in the defect Hilbert space of $\CD$. We define the notation,
\begin{equation}
\begin{tikzpicture}[baseline={([yshift=0]current bounding box.center)},vertex/.style={anchor=base,
    circle,fill=black!25,minimum size=18pt,inner sep=2pt},scale=0.5]
    \filldraw[grey] (-2,-2) rectangle ++(4,4);
    \draw[thick, dgrey] (-2,-2) rectangle ++(4,4);
    \draw[thick, black] (0,-2) -- (0.707,-0.707);
    \draw[thick, red] (2,0) -- (0.707,-0.707);
    \draw[thick, black] (0,2) -- (-0.707,0.707);
    \draw[thick, red] (-2,0) -- (-0.707,0.707);
    \draw[thick, blue] (0.707,-0.707) -- (-0.707,0.707);
    \node[black] at (0.0,-1.3) {\scriptsize$\mathcal{D}$};
    \node[red, below] at (1.5,-0.3) {\scriptsize$a$};
    \filldraw[red] (0.707,-0.707) circle (.08);
    \node[red, above] at (0.707,-0.707) {\scriptsize$\mu$};
    \filldraw[red] (-0.707,0.707) circle (.08);
    \node[red, below] at (-0.707,0.707) {\scriptsize$\nu$};
    \node[blue, above] at (0.3,-0.2) {\scriptsize$b$};
\end{tikzpicture}\text{  or  }
\begin{tikzpicture}[scale=0.8,baseline={(0,0.75)}]
\node[above] at (0,2.5) {\scriptsize$\CD$};
\draw[thick] (0,0) -- (0,2.5);
\node[below] at (0,0) {\scriptsize$\mathcal{O}$};
\draw [red,thick,domain=-90:90] plot ({1*cos(\x)}, {1*sin(\x)});
\draw [red,thick,domain=90:270] plot ({1.5*cos(\x)}, {0.5+1.5*sin(\x)});
\draw[thick,blue] (0,1) -- (0,2);
\filldraw[] (0,0) circle (1.5pt);
\filldraw[red] (0,1.0) circle (1.5pt);
\filldraw[red] (0,2.0) circle (1.5pt);
\node[red,left] at (0.,1.) {\scriptsize$\mu$};
\node[red,right] at (0.,2.) {\scriptsize$\nu$};
\node[red,right] at (1.,0.) {\scriptsize$a$};
\node[blue,right] at (0.,1.5) {\scriptsize$b$};
\end{tikzpicture} \equiv \scU[a,b][\mu,\nu] \equiv
\begin{tikzpicture}[scale=0.8,baseline={(0,1)}]
\node[above] at (0,2.5) {\scriptsize$\CD$};
\draw[thick] (0,0) -- (0,2.5);
\node[below] at (0,0) {\scriptsize$\mathcal{O}$};
\draw[thick, red] (0,2) -- (-0.5,1.5);
\draw[thick, red] (0,1.) -- (0.5,0.5);
\draw[thick,blue] (0,1) -- (0,2);

\filldraw[] (0,0) circle (1.5pt);
\filldraw[red] (0,1.0) circle (1.5pt);
\filldraw[red] (0,2.0) circle (1.5pt);
\node[red,left] at (0.,1.) {\scriptsize$\mu$};
\node[red,right] at (0.,2.) {\scriptsize$\nu$};
\node[red,right] at (0.25,0.75) {\scriptsize$a$};
\node[red,left] at (-0.25,1.75) {\scriptsize$a$};
\node[blue,right] at (0.,1.5) {\scriptsize$b$};
\end{tikzpicture},
\end{equation}
where $a,b\in \mathcal{E}^{(i,\fsb,\fsa)}_{\doubleZ_2}\Rep(H_8)$. We only keep the start and end points of the $a$ line for simplicity. 
The multiplication is given by,
\begin{equation}
\begin{aligned}
    &\begin{tikzpicture}[scale=0.6,baseline={(0,1)}]
\node[above] at (0,4) {\scriptsize$\CD$};
\draw[thick] (0,0) -- (0,4);
\node[below] at (0,0) {\scriptsize$\mathcal{O}$};
\filldraw[] (0,0) circle (1.5pt);
\draw[thick, red] (0,0.5) -- (0.5,0.0);\node[red,right] at (0.25,0.25) {\scriptsize$c$};
\draw[thick, red] (0,1.5) -- (-0.5,1.0);\node[red,left] at (-0.25,1.25) {\scriptsize$c$};
\draw[thick,blue] (0,0.5) -- (0,1.5);\node[blue,right] at (0.,1.0) {\scriptsize$d$};
\filldraw[red] (0,0.5) circle (1.5pt);\node[red,left] at (0.,0.5) {\scriptsize$\rho$};
\filldraw[red] (0,1.5) circle (1.5pt);\node[red,right] at (0.,1.5) {\scriptsize$\sigma$};

\draw[thick, red] (0,2.5) -- (0.5,2.0);\node[red,right] at (0.25,2.25) {\scriptsize$a$};
\draw[thick, red] (0,3.5) -- (-0.5,3.0);\node[red,left] at (-0.25,3.25) {\scriptsize$a$};
\draw[thick,blue] (0,2.5) -- (0,3.5);\node[blue,right] at (0.,3.0) {\scriptsize$b$};
\filldraw[red] (0,2.5) circle (1.5pt);\node[red,left] at (0.,2.5) {\scriptsize$\mu$};
\filldraw[red] (0,3.5) circle (1.5pt);\node[red,right] at (0.,3.5) {\scriptsize$\nu$};
\node[left] at (0.,2) {\scriptsize$\CD$};
\end{tikzpicture} =\sum_{f',\sigma',\mu'} F_{b,(\CD,\sigma,\mu)(f',\sigma',\mu')}^{cda}\begin{tikzpicture}[scale=0.6,baseline={(0,1)}]
\node[above] at (0,4) {\scriptsize$\CD$};
\draw[thick] (0,0) -- (0,4);
\node[below] at (0,0) {\scriptsize$\mathcal{O}$};
\filldraw[] (0,0) circle (1.5pt);
\draw[thick, red] (0,0.5) -- (0.5,0.0);\node[red,right] at (0.25,0.25) {\scriptsize$c$};
\draw[thick, red] (0,1.5) -- (0.5,1.0);\node[red,right] at (0.25,1.25) {\scriptsize$a$};
\draw[thick,blue] (0,0.5) -- (0,1.5);\node[blue,left] at (0.,1.0) {\scriptsize$d$};
\filldraw[red] (0,0.5) circle (1.5pt);\node[red,left] at (0.,0.5) {\scriptsize$\rho$};
\filldraw[red] (0,1.5) circle (1.5pt);\node[red,left] at (0.,1.5) {\scriptsize$\sigma'$};

\draw[thick, red] (0,2.5) -- (-0.5,2.0);\node[red,left] at (-0.25,2.25) {\scriptsize$c$};
\draw[thick, red] (0,3.5) -- (-0.5,3.0);\node[red,left] at (-0.25,3.25) {\scriptsize$a$};
\draw[thick,blue] (0,2.5) -- (0,3.5);\node[blue,right] at (0.,3.0) {\scriptsize$b$};
\filldraw[red] (0,2.5) circle (1.5pt);\node[red,right] at (0.,2.5) {\scriptsize$\mu'$};
\filldraw[red] (0,3.5) circle (1.5pt);\node[red,right] at (0.,3.5) {\scriptsize$\nu$};
\node[right] at (0.,2) {\scriptsize$f'$};
\end{tikzpicture}\\
&=\sum_{\substack{f',\mu',\nu'\\f'',\mu'',\nu''\\e,\alpha,\beta}} F_{b,(\CD,\sigma,\mu)(f',\sigma',\mu')}^{cda} F^{\CD c a}_{f',(d,\rho,\mu'),(f'',\mu'',\nu'')} (F^{acf'}_{\CD})^{-1}_{(b,\nu',\nu)(e,\alpha,\beta)}\  \begin{tikzpicture}[scale=0.6,baseline={(0,0.0)}]
\node[above] at (0,2) {\scriptsize$\CD$};
\draw[thick] (0,0) -- (0,2);
\node[below] at (0,0) {\scriptsize$\mathcal{O}$};
\filldraw[] (0,0) circle (1.5pt);
\draw [red,thick,domain=90:-48] plot ({1*cos(\x)}, {-0.5+1*sin(\x)});
\draw [red,thick,domain=90:242] plot ({1.5*cos(\x)}, {+1.5*sin(\x)});
\draw[red, thick] (0,-1.5) circle (0.75);
\filldraw[red] (-0.72618,-1.3125) circle (1.5pt);\node[red,left] at (-0.72618,-1.4125) {\scriptsize$\alpha$};
\filldraw[red] (0.695269,-1.21875) circle (1.5pt);\node[red,right] at (0.695269,-1.31875) {\scriptsize$\mu''$};
\filldraw[red] (0,0.5) circle (1.5pt);
\filldraw[red] (0,1.5) circle (1.5pt);
\node[red,right] at (0,1.5) {\scriptsize$\beta$};
\node[red,left] at (0,0.5) {\scriptsize$\nu''$};
\node[red,right] at (1,0) {\scriptsize$f''$};
\node[red,right] at (-1.5,0.1) {\scriptsize$e$};
\node[red,below] at (0,-0.8) {\scriptsize$c$};\node[red,above] at (0,-2.2) {\scriptsize$a$};
\node[right] at (0.0,1.0) {\scriptsize$f'$};
\end{tikzpicture}
\\
&=\sum_{\substack{f',\mu',\nu',\nu''\\e,\alpha,\beta}} \sqrt{\frac{d_a d_c}{d_e}}F_{b,(\CD,\sigma,\mu)(f',\sigma',\mu')}^{cda} F^{\CD c a}_{f',(d,\rho,\mu'),(e,\alpha,\nu'')} (F^{acf'}_{\CD})^{-1}_{(b,\nu',\nu)(e,\alpha,\beta)} \begin{tikzpicture}[scale=0.6,baseline={(0,1)}]
\node[above] at (0,2.5) {\scriptsize$\CD$};
\draw[thick] (0,0) -- (0,2.5);
\node[below] at (0,0) {\scriptsize$\mathcal{O}$};
\draw[thick, red] (0,2) -- (-0.5,1.5);
\draw[thick, red] (0,1.) -- (0.5,0.5);
\draw[thick,blue] (0,1) -- (0,2);

\filldraw[] (0,0) circle (1.5pt);
\filldraw[red] (0,1.0) circle (1.5pt);
\filldraw[red] (0,2.0) circle (1.5pt);
\node[red,left] at (0.,1.) {\scriptsize$\nu''$};
\node[red,right] at (0.,2.) {\scriptsize$\beta$};
\node[red,right] at (0.25,0.75) {\scriptsize$e$};
\node[red,left] at (-0.25,1.75) {\scriptsize$e$};
\node[blue,right] at (0.,1.5) {\scriptsize$f'$};
\end{tikzpicture}.
\end{aligned}
\end{equation}

\bibliographystyle{JHEP}
\bibliography{ising2}

\end{document}